\pdfoutput=1
\documentclass[pdflatex,sn-nature, referee]{sn-jnl}

\usepackage{array}
\usepackage{graphicx}%
\usepackage{multirow}%
\usepackage{amsmath,amssymb,amsfonts}%
\usepackage{amsthm}%
\usepackage{mathrsfs}%
\usepackage[title]{appendix}%
\usepackage{xcolor}%
\usepackage{textcomp}%
\usepackage{manyfoot}%
\usepackage{booktabs}%
\usepackage{algorithm}%
\usepackage{algorithmicx}%
\usepackage{algpseudocode}%
\usepackage{listings}%
\usepackage{amsmath}
\usepackage{esint}
\usepackage{graphicx}
\usepackage{mathrsfs}
\usepackage{subcaption}
\usepackage{caption}
\usepackage{multibib}
\usepackage{hyperref}

\theoremstyle{thmstyleone}%
\theoremstyle{thmstyletwo}%

\theoremstyle{thmstylethree}%
\hypersetup{
    colorlinks=true,
    linkcolor=blue,
    filecolor=magenta,      
    urlcolor=cyan,
}

\begin{document}

\title[Article Title]{Differentiable high-fidelity vectorial propagator}


\author[1]{\fnm{Dohyun} \sur{Kim}}\email{photonics@kaist.ac.kr}
\author*[1]{\fnm{Jonghwa} \sur{Shin}}\email{qubit@kaist.ac.kr}

\affil*[1]{\orgdiv{Department of Materials Science and Engineering}, \orgname{KAIST}, \orgaddress{\city{Daejeon}, \country{Republic of Korea}}}


\abstract{The numerical accuracy and computational efficiency of free-space field propagation directly determine what physics can be predicted and what devices can be designed, from nanophotonic metasurfaces to optically addressed quantum processors. Existing methods face a fundamental accuracy–speed trade-off, which is further compounded in the vectorial regime, where major formalisms remain fragmented with no unified computational framework. Here we address both challenges: a theory that unifies the vectorial diffraction framework, and an algorithm that resolves the accuracy–speed trade-off, each reinforcing the other. We prove that all major vectorial diffraction formalisms and vector-potential seeding methods arise from a single surface equivalence principle and reduce, without approximation, to a finite sequence of scalar angular-spectrum operations. Any advance in scalar propagation can therefore be exploited across the entire vectorial ecosystem. The error-compensating angular spectrum method (E-ASM) serves as a fast and accurate scalar solver, delivering peak signal-to-noise-ratio gains exceeding 60 dB over the state-of-the-art band-extended angular spectrum method at order-of-magnitude faster speed, with a freely configurable observation window. Their combination, implemented in a GPU-accelerated, fully differentiable form, turns rigorous vectorial field computation—previously prohibitive for higher-order beams over macroscopic distances—into a practical, inverse-design-ready tool. We demonstrate the framework on optically addressed local qubit gates and on vectorial Hermite–Gaussian and Laguerre–Gaussian beam propagation, resolving field structures inaccessible to scalar or approximate vectorial methods.
}



\maketitle

\section{Introduction}\label{sec1}
Accurate and efficient computation of free-space electromagnetic 
field propagation underpins a broad spectrum of optical science and 
engineering, yet it presents a formidable challenge due to an 
inherent trade-off between numerical accuracy and execution speed. 
The demand for such wave-optics solvers has escalated with the emergence of paradigms beyond traditional ray-tracing limits. These range from differentiable wave-optics pipelines for bulk optical systems \cite{ho2025differentiable, yang2024end} to metaform optics—the integration of bulk compound optics with metasurfaces—which demands multi-scale solvers spanning macroscopic propagation and microscopic scattering \cite{nikolov2021metaform}. Crucially, as these emerging applications push deeper into non-paraxial and sub-wavelength regimes, conventional scalar treatments become fundamentally insufficient. The precise vectorial structure of propagating fields dictates entire system performance, spanning from high-NA focusing \cite{richards1959electromagnetic, sheppard1987imaging} and polarization-dependent metasurfaces\cite{rubin2019matrix, chang2022universal} to optically addressed quantum computing platforms \cite{thompson2013coherence, noek2013high, todaro2021state}, where unintended cross-polarized or longitudinal coupling directly limits qubit gate fidelity.\\
When confined strictly to the scalar regime, the angular spectrum 
method (ASM) is the standard approach. As an exact solution to the 
scalar Helmholtz equation implemented via the Fast Fourier Transform 
(FFT), it enjoys an efficient $O(N^2 \log N)$ complexity, where N us the number of samples per side of the $N\times N$ computational grid. Yet this 
exactness rarely translates to numerical robustness: at large 
propagation distances relative to the field extent, the ASM transfer 
function oscillates rapidly, violating the Nyquist condition on a 
discrete grid and triggering severe aliasing. Existing variants—
band-limited \cite{matsushima2009band, matsushima2010shifted}, 
band-extended \cite{zhang2020band, zhang2021shifted}, and 
least-sampling \cite{wei2023modeling} methods—mitigate this under 
restrictive spatial constraints but do not eliminate it 
fundamentally, particularly in long-distance, highly non-paraxial 
regimes. The difficulty compounds in the vectorial case, where 
coupling among the electromagnetic components turns propagation into 
an interdependent boundary-value problem that resists a unified 
planar treatment.\\
Here we resolve both limitations through a unified theory and a new algorithm that strengthen each other. On the theoretical side, we demonstrate that five historically separate formalisms—the Luneburg integral, the Stratton–Chu integral, the vectorial angular spectrum method, vector-potential seeding, and spherical vectorial diffraction—all trace back to the surface equivalence principle, each realized through a spectral matrix multiplication combined with a small number of scalar angular-spectrum evaluations. While elements of this picture may have been implicitly appreciated, an explicit proof of this reduction has not been reported. The consequence is that vectorial diffraction no longer requires a dedicated algorithm: any improvement made to a scalar solver is immediately available to every vectorial formalism.\\
The error-compensating angular spectrum method (E-ASM) breaks the accuracy–speed trade-off of scalar diffraction solvers. The key idea is to reformulate exact Helmholtz propagation by introducing a subtle phase-compensation term that corrects the dispersion error between the parabolic and spherical phases, followed by a real-space convolution that bypasses conventional single- or two-step FFT-based Fresnel propagation. By evaluating the subsequent Fresnel step directly in real space, all remaining discretization error is confined to the pre-compensation stage, where the slowly varying aberration correction fundamentally suppresses aliasing.\\
As a result, E-ASM approaches the precision of direct Riemann-sum evaluation of the Rayleigh–Sommerfeld integral from the near field to the far field, while leaving the observation plane free in position, extent, and sampling pitch. It reduces the propagation error by more than three orders of magnitude relative to the state-of-the-art band-extended angular spectrum method at substantially faster speed, and admits an $O(N^2 \log N)$ implementation. The solver is differentiable and runs on GPU, so these gains can be delivered inside gradient-based inverse design frameworks without any reformulation or loss of accuracy. By the reduction established above, these advantages transfer unchanged to full vectorial propagation, opening a computational regime that has so far been inaccessible. We validate the framework on real-world problems, including optically addressed local qubit gates and the propagation of high-order vector beams.\\
\begin{figure}[htbp]
  \centering
  \includegraphics[width=\textwidth, page=1]{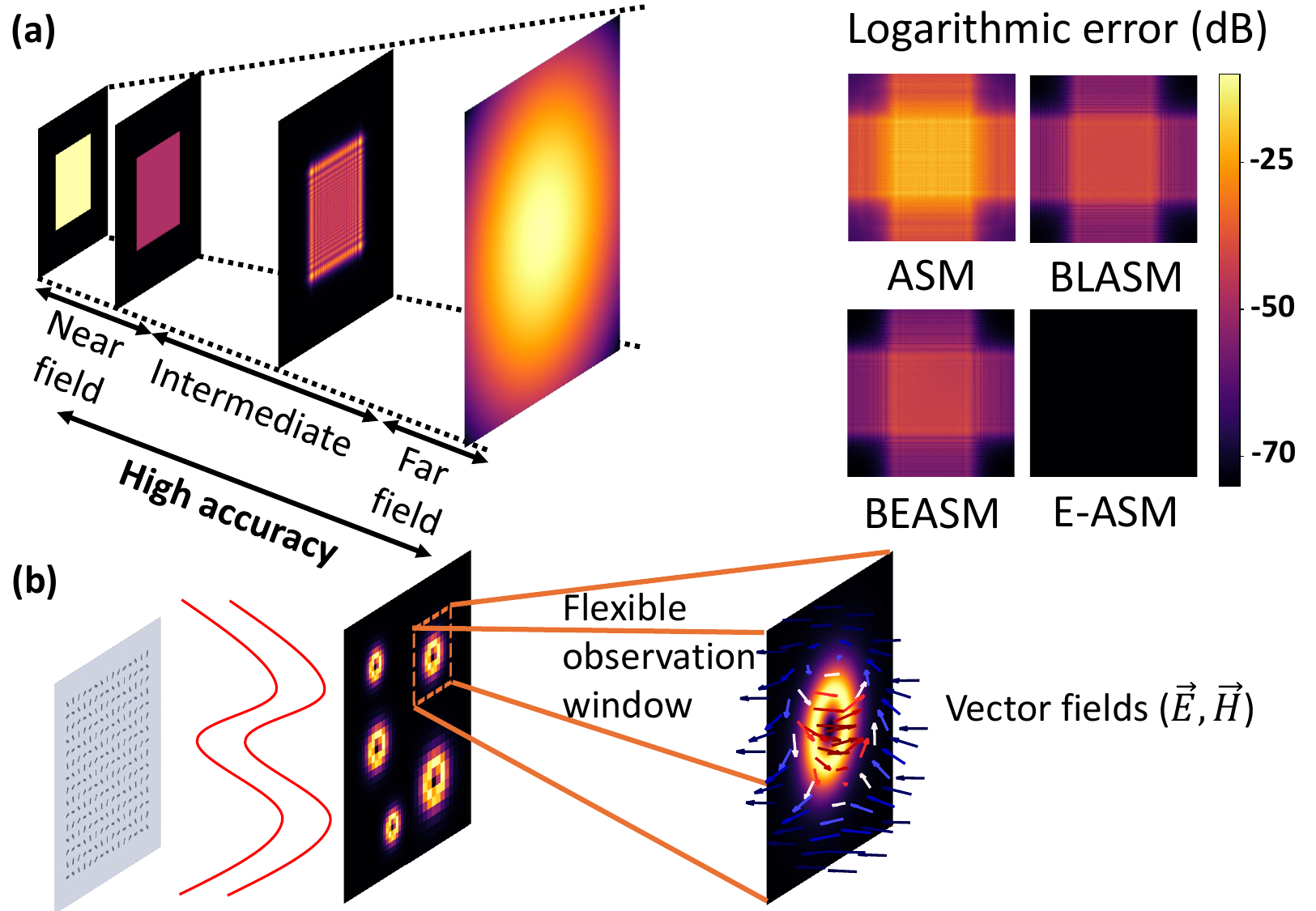}
  \caption{Conceptual schematic of the E-ASM. \protect\linebreak(a) applicable propagation distance range and accuracy of the E-ASM compared with existing methods at a propagation distance of 24984 $\lambda$. ASM, BLASM, and BEASM denote the angular spectrum method, the band-limited angular spectrum method, and the band-extended angular spectrum method, respectively. \protect\linebreak(b) Flexible observation window of the E-ASM and its extension to vectorial diffraction.}
  \label{fig1}
\end{figure}

\section{Theoretical Analysis}\label{sec2}
\subsection{The current scalar diffraction theory and its limitations}
The free-space propagation of a scalar field is governed by the 
Helmholtz equation. In the parallel-plane geometry, three well-known 
solutions exist depending on the choice of Green's function: the 
Helmholtz--Kirchhoff integral and the first and second 
Rayleigh--Sommerfeld integrals. The most widely used is the first RS 
integral, derived from the Dirichlet Green's function \cite{Goodman},
 \begin{equation}
     U(x,y;z)=-2\iint_{\Sigma} U(x',y';0)\frac{\partial G}{\partial z} ds'
     \label{1st RS}
 \end{equation}
which requires only the field value on the aperture $\Sigma$ rather than its normal derivative. Here, $U$ denotes the complex scalar field, $ds'=dx'dy'$ is the area element on the source plane, and $G$ is the free-space Green's function with a point source at the observation point. Direct Riemann-sum evaluation of this convolution carries $O(N^4)$ complexity, making it prohibitively expensive for large grids. The angular spectrum method (ASM) provides an equivalent but far more efficient formulation as a multiplication in the spatial frequency domain,
\begin{equation}
    U(x,y;z)=\mathcal{F}^{-1}\left[H_{AS}\mathcal{F}[U(x',y';0)]\right]
    \label{ASM}
\end{equation}
where $\mathcal{F}$ and $\mathcal{F}^{-1}$ denote the forward and inverse two-dimensional Fourier transforms, $k=2n\pi/\lambda$ is the wavenumber, $(k_x,k_y)$ are the transverse components of the wavevector, and $H_{AS}=\exp{\left(i\sqrt{k^2-k_x^2-k_y^2}z\right)}$ is the free-space transfer function in the spectral domain. Requiring only two sequential FFTs, the ASM achieves $O(N^2\log N)$ complexity. Unlike the Fresnel and Fraunhofer approximations, which fail in the non-paraxial and near-field regimes, respectively, the ASM is exact and has become the canonical method for 
free-space scalar propagation.\\
However, the conventional ASM based on FFT suffers from two inherent limitations. First, 
at large propagation distances the transfer function oscillates rapidly 
in the frequency domain, causing aliasing unless prohibitively large 
zero-padding is applied \cite{liu2012controlling}. Second, because the 
method relies on two sequential FFTs, the output window size and 
sampling are rigidly tied to those of the input plane \cite{voelz2011computational}.\\
To suppress aliasing, the band-limited ASM (BLASM) \cite{matsushima2009band, matsushima2010shifted}, band-extended ASM (BEASM) \cite{zhang2020band, zhang2021shifted}, and least-sampling ASM (LSASM) \cite{wei2023modeling} restrict the transfer function's spectral support under physical bounds. However, these bandwidth constraints tighten with propagation distance, so their accuracy degrades sharply as the distance increases, particularly under highly non-paraxial conditions. Separately, flexible observation windows have been achieved through the scaled DFT \cite{abedi2024improvement}, chirp-Z transform \cite{hu2020efficient}, fractional Fourier transform \cite{ozaktas2011fundamental}, and matrix triple product (MTP) \cite{zhao2020flexible}. Yet no existing method simultaneously achieves 
high accuracy and a freely configurable observation window.\\
\subsection{The current vectorial diffraction theory and its limitations}
In non-paraxial geometries such as high-NA focusing and off-axis propagation, cross-polarization and longitudinal field components become non-negligible, necessitating a rigorous vectorial treatment beyond the uncoupled propagation of individual field components. Over the past decades, various vectorial diffraction formalisms have been developed under different physical assumptions. The classical Luneburg integral \cite{baker2003mathematical} uses the scalar angular spectrum method (ASM) for transverse fields and recovers the longitudinal component via the divergence-free condition ($\nabla \cdot \mathbf{E} = 0$). For an exact description of electromagnetic propagation derived from fundamental principles, the Stratton--Chu integral—rooted in the surface equivalence principle—presents a rigorous boundary-value formulation \cite{stratton1939diffraction}. More recently, computationally focused variants have been introduced to improve numerical tractability: the generic vectorial angular spectrum method (VASM) \cite{song2025generic}, inspired by Debye--Wolf polarization projections, and the spherical vectorial diffraction (SVD) method \cite{liu2026vectorial}, derived from the Franz formula under the paraxial approximation.\\
Concurrently, the vector beam community has independently advanced the vector potential seeding framework—including magnetic ($\mathbf{A}$) and electric ($\mathbf{F}$) potential formulations—to construct exact solutions to Maxwell's equations from scalar Helmholtz seed functions. To balance asymmetries inherent to individual seeding choices, symmetrization techniques averaging $\mathbf{A}$- and $\mathbf{F}$-seeding solutions are frequently employed to model complex, free-space vector beams. Historically, this spectral approach, pioneered by Agrawal and Pattanayak \cite{levy2019mathematics, agrawal1979gaussian}, relies on the plane-wave-spectrum representation to propagate individual modes. Crucially, rather than utilizing fast transform-based grid operations, this formalism requires the numerical evaluation of highly oscillating spectral integrals via point-by-point Gaussian quadrature at each spatial location.\\
Despite their respective physical insights, existing vectorial diffraction theories and potential seeding methods face two fundamental bottlenecks. First, from a theoretical perspective, a unified formal framework that analytically bridges these disparate representations remains obscure. Because formalisms like Stratton--Chu, VASM, SVD, and potential seeding have been derived from distinct physical intuition, their rigorous exact relationships and equivalence conditions have not been fully elucidated, leaving the field of vector diffraction highly fragmented. Second, from a computational perspective, these methods are severely constrained by a rigid trade-off between numerical accuracy and computational efficiency. On one hand, fundamental formulations like the Stratton--Chu integral and Agrawal and Pattanayak's plane-wave spectrum seeding approach offer rigorous solutions. However, their direct evaluation requires dense, multi-dimensional linear convolution or Gaussian quadrature. This becomes computationally prohibitive over extended propagation ranges, particularly for higher-order beams (e.g., Hermite--Gaussian or Laguerre--Gaussian modes) that lack azimuthal symmetry \cite{levy2019mathematics}. On the other hand, computationally centered transforms such as VASM and SVD attempt to leverage fast transform-based architectures, but they inherently lack robust error-compensation mechanisms. Consequently, they remain highly susceptible to severe aliasing and wrap-around artifacts unless prohibitive grid padding and memory overhead are incurred.\\
To address both limitations, the following section introduces a unified vectorial diffraction framework based on the surface equivalence principle. We further show that all major formalisms reduce to a finite sequence of standard ASM operations, enabling an efficient, high-fidelity pipeline that bridges scalar and vectorial diffraction.\\
\subsection{A unified framework for scalar and vectorial diffraction}
The theoretical unification of major vectorial diffraction formalisms is established by unveiling a hidden link within the surface equivalence principle. Based on the uniqueness theorem, the tangential fields at a boundary can be formulated via equivalent surface currents. Love's 
equivalence principle models propagation under a null-field condition using coexisting electric and magnetic currents, whereas Schelkunoff's PEC and PMC principles isolate these boundaries by replacing the internal region with perfect conductors, where image currents either double or vanish for a planar geometry \cite{PhysRev.56.308}. Crucially, we show explicitly that by taking the direct analytical average of Schelkunoff’s PEC and PMC real-space integral equations, the vector image sources exactly cancel out to reproduce Love’s equivalence, immediately bridging to the rigorous Stratton–Chu formula. In the following, we demonstrate how this real-space integral averaging serves as the foundational mathematical bridge to derive, interface, and unify all conventional vectorial diffraction formalisms within a single spectral matrix framework.\\
As a foundational baseline, the simplest approach to vectorial propagation is the Luneburg integral Eq. \eqref{Luneburg RS}:
\begin{align}
    E_{p} = -2 \iint_{\Sigma} E^\text{inc}_{p} \left( \mathbf{r}' \right) \frac{\partial G}{\partial z} \, ds' \qquad & E_z = 2 \iint_{\Sigma} \left[ E_x^\text{inc} (\mathbf{r}') \frac{\partial G}{\partial x} + E_y^\text{inc} (\mathbf{r}') \frac{\partial G}{\partial y} \right] \, ds'
    \label{Luneburg RS}
\end{align}
where $p\in \{x,y\}$ denotes the transverse polarization, the superscript $i$ labels the incident field at the source plane, and $\mathbf{r}'$ is the source-point position on $\Sigma$; the left-hand sides are the propagated components at distance $z$ \cite{baker2003mathematical}.
Here, the transverse components are independently propagated via the first RS integral—efficiently evaluated via the ASM—while the longitudinal component $E_z$ is recovered from the divergence-free condition. Because the transverse fields appear uncoupled in Eq.~\eqref{Luneburg RS}, the Luneburg formalism has often been regarded as unable to account for cross-polarization coupling \cite{song2025generic, liu2026vectorial}. Historically, however, Schelkunoff’s early work implicitly demonstrated that if the input boundary field satisfies Maxwell's equations, this formulation is mathematically identical to the exact electric field obtained from Schelkunoff's perfect electric conductor (PEC) equivalence principle \cite{PhysRev.56.308} (Eq. \eqref{Schelkunoff PEC E}):
\begin{equation}
    \mathbf{E} = -\nabla\times\iint_{\Sigma}2\left(\mathbf{E}^\text{inc}\times\hat{\mathbf{z}}'\right)Gds'=-2\iint_{\Sigma}\left(\hat{\mathbf{z}}'\times\mathbf{E}^\text{inc}\right)\times\nabla Gds'
    \label{Schelkunoff PEC E}
\end{equation}
where $\mathbf{E}^\text{inc}$ is the source field.
Revisiting this equivalence (Supplementary Information, Section 4) makes clear that the Luneburg integral is an exact Maxwell solution, provided the incident boundary field is physically valid. To interface this real-space representation with modern numerical frameworks, these boundary integrals can be transformed into the Fourier domain. Evaluating Schelkunoff's PEC formulation in the spectral domain yields the exact matrix representations for the propagated fields:
\begin{equation}
    \begin{pmatrix} \tilde{E}_x \\ \tilde{E}_y \\ \tilde{E}_z \end{pmatrix} = 
    \begin{pmatrix} 
    1 & 0 & 0 \\ 
    0 & 1 & 0 \\ 
    -\frac{k_x}{k_z} & -\frac{k_y}{k_z} & 0 
    \end{pmatrix} 
    \begin{pmatrix} \tilde{E}_x^\text{inc} \\ \tilde{E}_x^\text{inc} \\ \tilde{E}_x^\text{inc} \end{pmatrix} e^{ik_z z}
     = 
    Z
    \begin{pmatrix} 
    0 & \sigma_z & -\sigma_y \\ 
    -\sigma_z & 0 & \sigma_x \\ 
    \sigma_y & -\sigma_x & 0 
    \end{pmatrix} 
    \begin{pmatrix} \tilde{H}_x^\text{inc} \\ \tilde{H}_y^\text{inc} \\ \tilde{H}_z^\text{inc} \end{pmatrix} e^{ik_z z}
    \label{Matrix Schelkunoff PEC E}
\end{equation}
Here the tilde denotes the Fourier-domain amplitude, $\sigma_{x,y,z}=k_{x,y,z}/k$ are the direction cosines and $Z$ is the impedance of the medium. 
\begin{flalign}
    &\begin{pmatrix} \tilde{H}_x \\ \tilde{H}_y \\ \tilde{H}_z \end{pmatrix} = 
    \frac{1}{Z} \begin{pmatrix} 
        -\frac{\sigma_y \sigma_x}{\sigma_z} & -\sigma_z - \frac{\sigma_y^2}{\sigma_z} & 0 \\ 
        \sigma_z + \frac{\sigma_x^2}{\sigma_z} & \frac{\sigma_y \sigma_x}{\sigma_z} & 0 \\ 
        -\sigma_y & \sigma_x & 0 
    \end{pmatrix} 
    \begin{pmatrix} \tilde{E}_x^\text{inc} \\ \tilde{E}_y^\text{inc} \\ \tilde{E}_z^\text{inc} \end{pmatrix}e^{ik_z z} \notag \\
    &=\begin{pmatrix}
        (1-\sigma_x^2) & -\sigma_x \sigma_y & -\sigma_x \sigma_z \\
        -\sigma_x \sigma_y & (1-\sigma_y^2) & -\sigma_y \sigma_z \\
        -\sigma_x \sigma_z & -\sigma_y \sigma_z & (1-\sigma_z^2)
    \end{pmatrix}
    \begin{pmatrix} \tilde{H}_x^\text{inc} \\ \tilde{H}_y^\text{inc} \\ \tilde{H}_z^\text{inc} \end{pmatrix}e^{ik_z z}.
    \label{Matrix Schelkunoff PEC H}
\end{flalign}
Similarly, Schelkunoff's PMC real-space integral Eq. \eqref{Schelkunoff PMC E} 
\begin{equation}
    \mathbf{E}=-\frac{2i}{\omega\varepsilon}\nabla \times \iint_{\Sigma}\left(\hat{\mathbf{z}}'\times\mathbf{H}^\text{inc}\right)\times\nabla Gds'=\frac{2i}{\omega\mu\varepsilon} \nabla\times\nabla\times\mathbf{A}
    \label{Schelkunoff PMC E}
\end{equation}
where $\mathbf{H}^\text{inc}$ is the incident magnetic field, can be mapped to a compact matrix representation in the spectral domain:
\begin{flalign}
    &\begin{pmatrix} \tilde{E}_x \\ \tilde{E}_y \\ \tilde{E}_{z} \end{pmatrix} = 
    Z \begin{pmatrix} 
        \frac{\sigma_y \sigma_x}{\sigma_z} & \sigma_z + \frac{\sigma_y^2}{\sigma_z} & 0 \\ 
        -\sigma_z - \frac{\sigma_x^2}{\sigma_z} & -\frac{\sigma_y \sigma_x}{\sigma_z} & 0 \\ 
        \sigma_y & -\sigma_x & 0 
    \end{pmatrix} 
    \begin{pmatrix} \tilde{H}_x^\text{inc} \\ \tilde{H}_y^\text{inc} \\ \tilde{H}_z^\text{inc} \end{pmatrix}e^{ik_z z} \notag \\
    &=\begin{pmatrix}
        (1-\sigma_x^2) & -\sigma_x \sigma_y & -\sigma_x \sigma_z \\
        -\sigma_x \sigma_y & (1-\sigma_y^2) & -\sigma_y \sigma_z \\
        -\sigma_x \sigma_z & -\sigma_y \sigma_z & (1-\sigma_z^2)
    \end{pmatrix}
    \begin{pmatrix} \tilde{E}_x^\text{inc} \\ \tilde{E}_y^\text{inc} \\ \tilde{E}_z^\text{inc} \end{pmatrix}e^{ik_z z}.
    \label{Matrix Schelkunoff PMC E}
\end{flalign}\\
Notably, the resulting electric field operator in Eq. \eqref{Matrix Schelkunoff PMC E} is mathematically identical to the generic VASM formulation \cite{song2025generic}. Although VASM was originally introduced via $p$/$s$ polarization projections inspired by the Debye–Wolf integral, this correspondence rigorously proves that VASM is inherently equivalent to Schelkunoff's PMC formulation for planar boundaries, while the PEC solution Eq. \eqref{Matrix Schelkunoff PEC H} serves as its exact magnetic field counterpart.
PMC solution of the $\mathbf{H}$ field is
\begin{equation}
    \begin{pmatrix} \tilde{H}_x \\ \tilde{H}_y \\ \tilde{H}_z \end{pmatrix} = 
    \begin{pmatrix} 
        1 & 0 & 0 \\ 
        0 & 1 & 0 \\ 
        -\frac{k_x}{k_z} & -\frac{k_y}{k_z} & 0 
    \end{pmatrix} 
    \begin{pmatrix} \tilde{H}_x^\text{inc} \\ \tilde{H}_y^\text{inc} \\ \tilde{H}_z^\text{inc} \end{pmatrix}e^{ik_z z} = 
    \frac{1}{Z} \begin{pmatrix} 
        0 & -\sigma_z & \sigma_y \\ 
        \sigma_z & 0 & -\sigma_x \\ 
        -\sigma_y & \sigma_x & 0 
    \end{pmatrix} 
    \begin{pmatrix} \tilde{E}_x^\text{inc} \\ \tilde{E}_y^\text{inc} \\ \tilde{E}_z^\text{inc} \end{pmatrix}e^{ik_z z}.
    \label{Matrix Schelkunoff PMC H}
\end{equation}
\noindent In addition, by taking the average of Eqs. \eqref{Schelkunoff PEC E} and \eqref{Schelkunoff PMC E}, we obtain
\begin{equation}
    \mathbf{E} = \iint_{\Sigma} \left\{ i \omega \mu \left[ \mathbf{\hat{z}}' \times \mathbf{H}^\text{inc} \left( \mathbf{r}' \right) \right] G + \left[ \mathbf{\hat{z}}' \times \mathbf{E}^\text{inc} \left( \mathbf{r}' \right) \right] \times \nabla' G + \left[ \mathbf{\hat{z}}' \cdot \mathbf{E}^\text{inc} \right] \nabla' G \right\} ds'
    \label{Stratton-Chu E planar}
\end{equation}
which is precisely the Stratton--Chu integral under a planar geometry. This argument is typically restricted to planar geometries, where an analytical closed-form Green's function under PEC and PMC conditions is available (Supplementary Information, Section 4). Following the same steps, the 
Stratton--Chu integral admits a compact matrix representation:
\begin{equation}
    \begin{pmatrix}
    \tilde{E}_x \\
    \tilde{E}_y \\
    \tilde{E}_z
    \end{pmatrix} = 
    \begin{pmatrix}
    1 - \frac{\sigma_x^2}{2} & -\frac{\sigma_x \sigma_y}{2} & -\frac{\sigma_x \sigma_z}{2} \\
    -\frac{\sigma_x \sigma_y}{2} & 1 - \frac{\sigma_y^2}{2} & -\frac{\sigma_y \sigma_z}{2} \\
    -\frac{1}{2} \left( \sigma_x \sigma_z + \frac{\sigma_x}{\sigma_z} \right) & -\frac{1}{2} \left( \sigma_y \sigma_z + \frac{\sigma_y}{\sigma_z} \right) & \frac{1-\sigma_z^2}{2}
    \end{pmatrix}
    \begin{pmatrix}
    \tilde{E}_x^\text{inc} \\
    \tilde{E}_x^\text{inc} \\
    \tilde{E}_x^\text{inc}
\end{pmatrix}e^{ik_z z}.
\label{Matrix Stratton-Chu E}
\end{equation}
Applying the paraxial approximation to the Eq. \eqref{Matrix Stratton-Chu E} and assuming a transversely polarized incident field $(\tilde{E}_{z}^\text{inc}=0)$, we get
\begin{equation}
    \begin{pmatrix}
    \tilde{E}_x \\
    \tilde{E}_y \\
    \tilde{E}_z
    \end{pmatrix} = 
    \begin{pmatrix}
    1 - \frac{\lambda^2 f_x^2}{2} & -\frac{\lambda^2 f_x f_y}{2} \\
    -\frac{\lambda^2 f_x f_y}{2} & 1 - \frac{\lambda^2 f_y^2}{2} \\
    -\lambda f_x & -\lambda f_y
    \end{pmatrix}
    \begin{pmatrix}
    \tilde{E}_x^\text{inc} \\
    \tilde{E}_x^\text{inc} \\
    \end{pmatrix}e^{ik_z z}
    \label{SVD E}
\end{equation}
where $f_{x,y}$ are the transverse spatial frequencies. This is the exact SVD formulation \cite{liu2026vectorial} (Supplementary Information, Section 7). The SVD formula is thus the paraxial approximation of the surface equivalence principle. Ultimately, the Luneburg integral, the Stratton–Chu formula, the VASM, and the SVD framework---all of which can be rigorously derived from Schelkunoff's PEC and PMC equivalences---can be universally expressed in a matrix form integrated with the angular spectrum propagation factor.\\
This theoretical unification extends further into the vector beam community's independent methodologies. By applying the Lorenz gauge condition, we establish that the magnetic potential ($\mathbf{A}$-seeding) and electric potential ($\mathbf{F}$-seeding) formulations are rigorously gauge-equivalent to Schelkunoff's PMC and PEC boundary solutions, respectively (see Section 8 of the Supplementary Information for complete mathematical proofs). Consequently, the widely used symmetrization technique—averaging $\mathbf{A}$- and $\mathbf{F}$-seeding solutions to model complex vector beams—is physically validated as an alternative representation of the exact Stratton–Chu integral.\\ 
While these formulations are mathematically consistent under ideal physical inputs, their numerical implementations diverge critically due to coordinate singularities. In practice, evaluating the PEC operator via the magnetic field ($\mathbf{H}$) and the PMC operator via the electric field ($\mathbf{E}$) is highly recommended; this specific cross-representation mathematically bypasses the numerical poles ($\sigma_z \rightarrow 0$) in the Fourier domain that otherwise destabilize computing pipelines near evanescent boundaries (see Section 10 of the Supplementary Information).\\
Crucially, because this unified framework maps all major historical vectorial formalisms into a finite sequence of ASM, any computational acceleration achieved in the scalar domain directly empowers the entire vectorial diffraction ecosystem. By utilizing our newly developed E-ASM in the following section, these unified vectorial fields can be evaluated with unprecedented numerical accuracy and efficiency, far surpassing existing state-of-the-art limitations.\\
\clearpage
\subsection{E-ASM}
\begin{figure}[htbp]
  \centering
  \includegraphics[width=0.8\textwidth, page=2, trim = 0cm 1cm 0cm 0cm, clip]{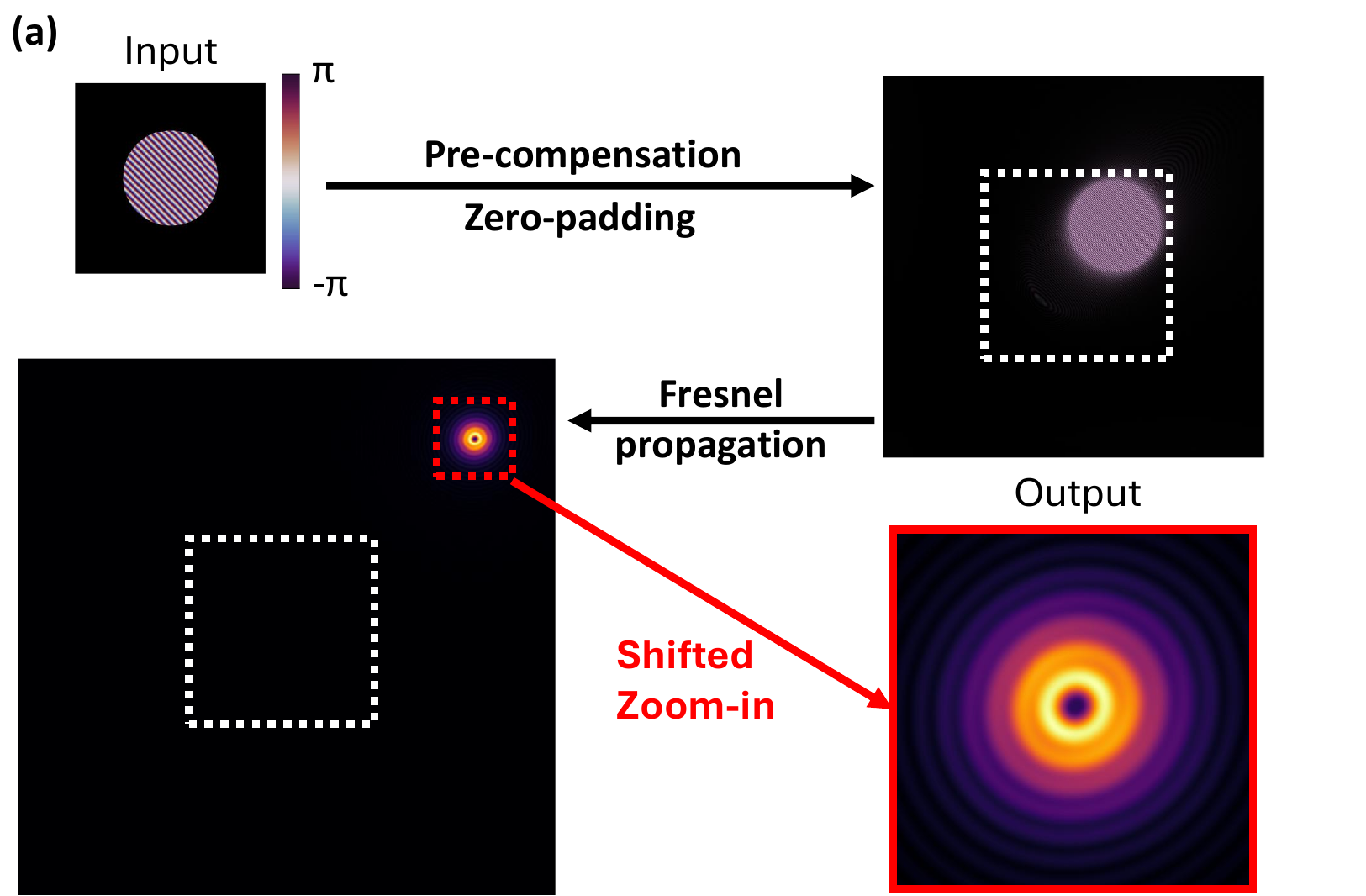}
  \includegraphics[width=0.8\textwidth, page=3, trim = 0cm 2cm 0cm 0cm, clip]{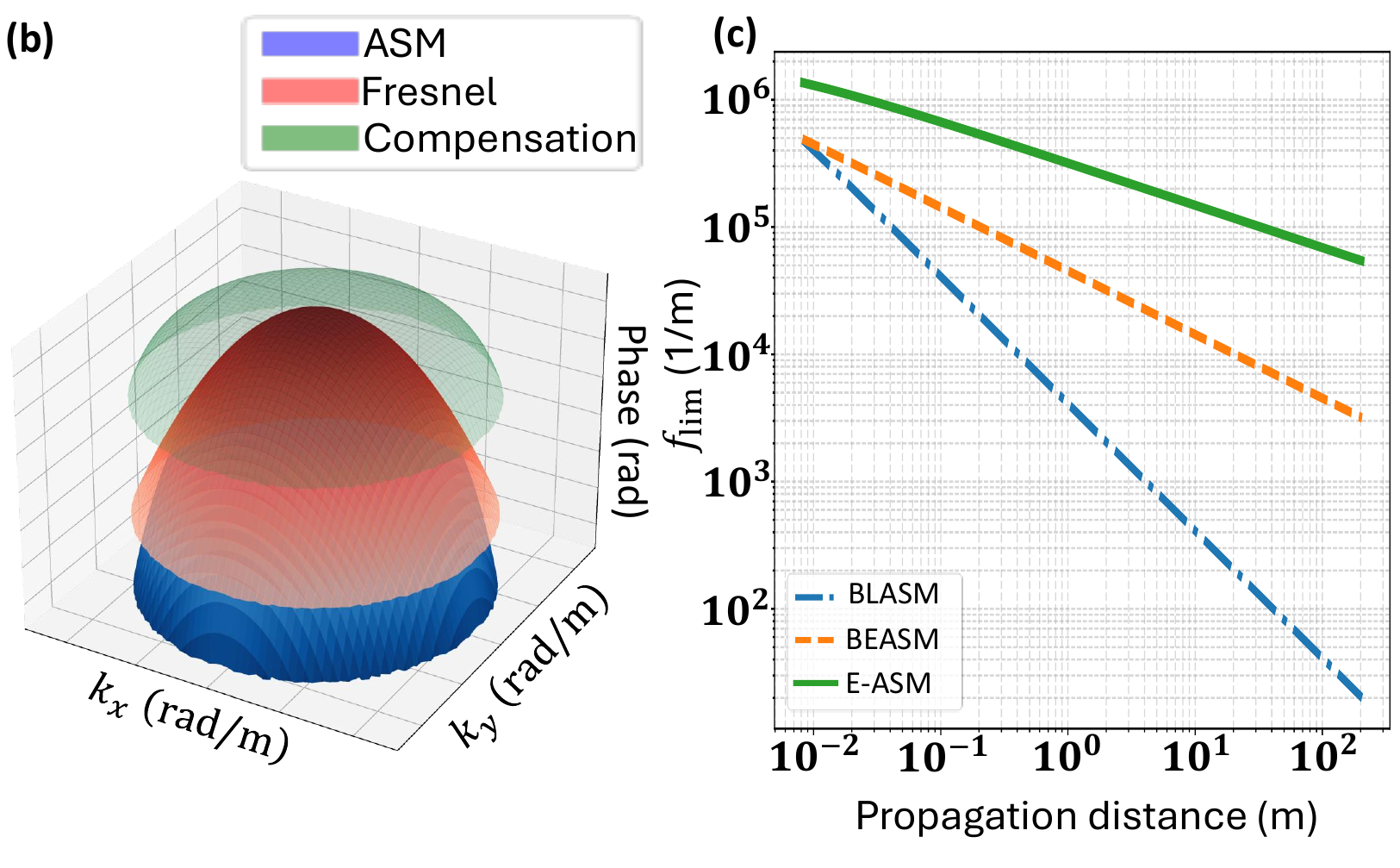}
  \caption{Core principles of E-ASM. 
\\(a) Execution flow and computational steps of E-ASM. The white dashed box indicates the size of the conventional ASM output window.
\\(b) Comparison of transfer functions for ASM, Fresnel, and compensation function of E-ASM. 
\\(c) Propagation-distance-dependent maximum spatial frequency.}
  \label{fig2}  
\end{figure}
\clearpage
\noindent Implementing angular spectrum propagation via Fresnel diffraction by pre-compensating for the difference between the ASM and Fresnel transfer functions was pioneered by the scalable angular spectrum (SAS) propagation~\cite{mansuripur1989certain, asoubar2014efficient, 
heintzmann2023scalable}. SAS linearly expands the observation window—the distinctive feature of single-step Fresnel diffraction—without the paraxial approximation, so that applying this pre-compensation before single-step Fresnel propagation captures non-paraxial behavior while 
retaining the scaling property.\\
The pre-compensating transfer function, representing the difference between the ASM and Fresnel diffraction, is given by:
\begin{flalign}
    &\phi(k_x, k_y) = \arg{(H_{\mathrm{ASM}}H_{\mathrm{Fresnel}}^*)}\notag \\
    &=\frac{2\pi}{\lambda}z\left[\sqrt{1-(\lambda f_x)^2-(\lambda f_y)^2} -\left(1-\frac{(\lambda f_x)^2}{2}-\frac{(\lambda f_y)^2}{2}\right)\right]
    \label{pre-compensation phase}
\end{flalign}
where $H_{\mathrm{Fresnel}}^*$ is the complex conjugate of the Fresnel transfer function.
Specifically, the $L \times L$ source window is pre-compensated within a padded domain of size $L_p \times L_p = 2L \times 2L$ by sequentially applying two-fold zero-padding, an FFT, the pre-compensation phase multiplier, and an IFFT. To avoid Fourier-domain aliasing, band-limiting conditions are imposed according to Eqs. \eqref{bandlimit x} and \eqref{bandlimit y}
\begin{equation}
    z\left|\frac{\lambda f_{x}}{\sqrt{1-(\lambda f_{x} )^2-(\lambda f_{y} )^2}}-\lambda f_{x}\right|\leq\frac{1}{2\Delta_{f_{x}}}
    \label{bandlimit x}
\end{equation}
\begin{equation}
    z\left|\frac{\lambda f_{y}}{\sqrt{1-(\lambda f_{x} )^2-(\lambda f_{y} )^2}}-\lambda f_{y}\right|\leq\frac{1}{2\Delta_{f_{y}}}
    \label{bandlimit y}
\end{equation}
where $\Delta_{f_{x}}$ and $\Delta_{f_{y}}$ represent the sampling frequency on the Fourier domain along the $x$ and $y$ axes. SAS subsequently propagates this pre-compensated field via single-step Fresnel diffraction:
\begin{equation}
    U(x,y;z)=Q_2(x,y)\int Q_1(x',y')U(x',y';0)\cdot\exp\left[-i\frac{2\pi}{\lambda z}(xx'+yy')\right]dx'dy'.
\end{equation}
Here, the quadratic phase factor $Q_1$ and $Q_2$ are
\begin{equation}
    Q_1(x',y')=\exp\left[i\frac{k}{2z}(x'^2+y'^2)\right]
    \label{Q1}
\end{equation}
\begin{equation}
    Q_2(x,y)=\frac{\exp(ikz)}{i\lambda z}\exp\left[i\frac{k}{2z}(x^2+y^2)\right].
    \label{Q2}
\end{equation}
After the propagation process, the padded region is cropped, and the central region of interest (RoI) is scaled by $M_{\text{SAS}} = N\lambda z/(2L^2)$, where $N$ is the pixel count.\\
Crucially, rather than implementing single-step Fresnel propagation based on a single FFT, the proposed E-ASM propagates the pre-compensated field using a real-space convolution via the matrix triple product (MTP) framework. Specifically, the discrete Fresnel diffraction on the sampling grid is formulated as:
{\small
\begin{flalign}
& U(x,y;z) = \notag\\ 
& \frac{e^{ikz}}{i\lambda z} \sum_{m,n} U(x_m, y_n;0) 
\int_{x_m - \Delta_x/2}^{x_m + \Delta_x/2} 
\int_{y_n - \Delta_y/2}^{y_n + \Delta_y/2} 
\exp\!\left[\frac{ik}{2z}\!\left((x - x_1)^2 + (y - y_1)^2\right)\right] dx_1\, dy_1 &
\label{analytic Fresnel}
\end{flalign}}\\
where $x_1, y_1$ are integration variables spanning the extent of each pixel. Because the quadratic phase is separable in $x_1$ and $y_1$, the 2D integration factorizes into a product of 1D integrals:
\begin{equation}
    U(x,y;z) = \frac{e^{ikz}}{i\lambda z} \sum_{m,n} U(x_m, y_n;0)\; I_x(x;\, x_m)\; I_y(y;\, y_n)
\end{equation}
where $I_x$ and $I_y$ are the 1D integrals:
\begin{equation}
    I_x(x;x_m)=\int_{x_m-\Delta_x/2}^{x_m+\Delta_x/2}\exp\left[\frac{ik}{2z}(x-x_1)^2\right]dx_1
    \label{kernel x}
\end{equation}
\begin{equation}
    I_y(y;y_n)=\int_{y_n-\Delta_y/2}^{y_n+\Delta_y/2}\exp\left[\frac{ik}{2z}(y-y_1)^2\right]dy_1.
    \label{kernel y}
\end{equation}
Equations \eqref{kernel x}--\eqref{kernel y} have analytic closed forms:
\begin{equation}
    I_x(x;x_m)=\frac{\sqrt{\pi}}{2\sqrt{-\alpha}}\left[\mathrm{erf}(\sqrt{-\alpha}(x-a))-\mathrm{erf}(\sqrt{-\alpha}(x-b))\right]
\end{equation}
\begin{equation}
    I_y(y;y_n)=\frac{\sqrt{\pi}}{2\sqrt{-\alpha}}\left[\mathrm{erf}(\sqrt{-\alpha}(y-a))-\mathrm{erf}(\sqrt{-\alpha}(y-b))\right]
\end{equation}
where $\alpha=\frac{ik}{2z}$, and $a$ and $b$ are the lower and upper 
integration bounds. Consequently, Eq. \eqref{analytic Fresnel} can be calculated via the MTP algorithm without encountering any spatial sampling artifacts. In practice, the analytical integration of the propagation kernel can be numerically approximated using a Riemann sum framework. Under this model, Eq. \eqref{analytic Fresnel} is discretized as:
\begin{align}
U(x,y;z)\approx &\frac{e^{ikz}}{i\lambda z} \sum_{m,n} U(x_m, y_n;0) 
\exp\!\left[\frac{ik}{2z}\!\left((x - x_m)^2 + (y - y_n)^2\right)\right]\Delta_x\Delta_y
\label{Riemann sum model}
\end{align}
which is likewise computed via the MTP framework. E-ASM thus offers two implementation modalities: a zero-order hold model with an analytical closed-form solution, and a point-source model using Riemann sum integration. Throughout this paper, the Riemann sum approach is adopted for rigorous comparison with the first RS integration ground truth and other algorithms. \\
When the point-source model is adopted and both the input and output grids are uniform, the MTP can be further accelerated by factorizing the propagation kernel. We parametrize the uniform input and 
output grids by their sample indices as
\begin{equation}
    x_m = x_{m,0} + m\,\Delta_x, \qquad 
    x_o = x_{o,0} + o\,\Delta_\text{out},
    \label{eq:grid_param}
\end{equation}
where $m \in [0, N)$ and $o \in [0, M)$ are the input and output sample 
indices, and $\Delta_x$, $\Delta_\text{out}$ are the input and output 
pixel pitches, respectively. Substituting Eq. \eqref{eq:grid_param} into the 1D propagation kernel and expanding the quadratic phase yields
\begin{equation}
    \frac{ik}{2z}(x_o - x_m)^2 
    = \phi_o(o) + \phi_m(m) + \frac{ik}{2z}\Delta_0^2 
    - \frac{ik}{z}\Delta_\text{out}\Delta_x\, o\,m,
    \label{eq:kernel_expand}
\end{equation}
where $\Delta_0 \equiv x_{o,0} - x_{m,0}$ is the offset between the grid origins, and
\begin{align}
    \phi_o(o) &= \frac{ik}{2z}\left[(o\Delta_\text{out})^2 
    + 2\Delta_0\, o\Delta_\text{out}\right], \\
    \phi_m(m) &= \frac{ik}{2z}\left[(m\Delta_x)^2 
    - 2\Delta_0\, m\Delta_x\right]
\end{align}
are phase terms depending solely on the output index $o$ and the input 
index $m$, respectively. The only term coupling the two indices is the bilinear cross term 
$-\frac{ik}{z}\Delta_\text{out}\Delta_x\, o\,m$. Although this term 
prevents a direct separation, the quadratic identity
\begin{equation}
    o\,m = \frac{o^2 + m^2 - (o - m)^2}{2}
    \label{eq:bluestein_identity}
\end{equation}
recasts it as a discrete convolution. The same algebraic identity underlies Bluestein's algorithm for the chirp z-transform~\cite{1162132}; here, however, we apply it directly to the real-space Fresnel propagation kernel rather than to accelerate a discrete Fourier transform.\\
Substituting Eq. \eqref{eq:bluestein_identity}, the $o^2$ and $m^2$ 
contributions are absorbed into the output- and input-dependent phases, which become
\begin{align}
    \tilde{\phi}_o(o) &= \phi_o(o) - \frac{ik}{2z}\Delta_\text{out}\Delta_x\,o^2
    = \frac{ik}{2z}\left[o^2\Delta_\text{out}(\Delta_\text{out}-\Delta_x) + 2\Delta_0\,o\Delta_\text{out}\right]\\
    \tilde{\phi}_m(m) &= \phi_m(m) - \frac{ik}{2z}\Delta_\text{out}\Delta_x\,m^2
    = \frac{ik}{2z}\left[m^2\Delta_x(\Delta_x-\Delta_\text{out}) - 2\Delta_0\,m\Delta_x\right],
\end{align}
leaving only the $(o-m)^2$ term as a convolution kernel. The propagated field becomes
\begin{equation}
    U(x_o) = \frac{e^{ikz}}{i\lambda z}\,\Delta_x\, 
    e^{\tilde{\phi}_o(o)}\, e^{\frac{ik}{2z}\Delta_0^2} 
    \sum_{m} \left[U(x_m;0)\, e^{\tilde{\phi}_m(m)}\right] g(o - m),
    \label{eq:bluestein_conv}
\end{equation}
with convolution kernel $g(n) = \exp\!\left[\frac{ik}{2z}\Delta_\text{out}\Delta_x\, n^2\right]$, $n=o-m$. Equation \eqref{eq:bluestein_conv} is evaluated in three steps: (i) a pre-chirp multiplication by $e^{\tilde{\phi}_m(m)}$; (ii) a linear convolution with $g(n)$ via zero-padded FFTs of length $\geq N + M - 1$ to avoid circular wrap-around; and (iii) a post-chirp multiplication by $e^{\tilde{\phi}_o(o)}$ and the constant phase $e^{\frac{ik}{2z}\Delta_0^2}$. Since each convolution costs $O(N\log N)$ per axis, the separable two-dimensional propagation drops from $O(N^3)$ to $O(N^2 \log N)$, yielding results identical to the direct MTP evaluation up to machine precision.\\
Additionally, unlike the BEASM—which crops the padded region after the propagation—the E-ASM directly computes the desired $N \times N$ output window without cropping. In the MTP formulation, this reduces the core operation to a $(N \times 2N) \cdot (2N \times 2N) \cdot (2N \times N)$ product rather than $(2N \times 2N) \cdot (2N \times 2N) \cdot (2N \times 2N)$ sequence that a cropped implementation requires. The same cropping-free advantage carries over to the kernel factorization variant: because the output is evaluated directly on the target $N \times N$ grid, this approach retains the property while further lowering the complexity from $O(N^3)$ to $O(N^2 \log N)$.\\
The exceptional accuracy of E-ASM stems from its pre-compensation phase varying much more slowly in the Fourier domain than the ASM and Fresnel transfer function phases, as illustrated in Fig. \ref{fig2}(b). The maximum spatial frequency permitted by the band-limit directly reflects this trend. For BLASM, under the Nyquist condition, it is  \cite{matsushima2009band}:
\begin{equation}
    z\left|\frac{\lambda f_{x,\mathrm{lim}}}{\sqrt{1-(\lambda f_{x,\mathrm{lim}} )^2}}\right|=N\Delta_x\rightarrow f_{x,\mathrm{lim}}=\frac{N\Delta_x}{\lambda\sqrt{z^2+(N\Delta_x)^2}}\approx \frac{N\Delta_x}{\lambda z}\propto\frac{1}{z}.
\end{equation}
For simplicity, we assume a one-dimensional (1D) case. For BEASM, the maximum spatial 
frequency scales as $1/\sqrt{z}$ rather than $1/z$ \cite{zhang2020band}:
{\small\begin{equation}
    z\left|\frac{\lambda f_{x,\mathrm{lim}}}{\sqrt{1-(\lambda f_{x,\mathrm{lim}} )^2}}\right|=\frac{N}{2f_{x, \mathrm{lim}}}\rightarrow f_{x,\mathrm{lim}}=\frac{1}{\lambda}\sqrt{\frac{-\frac{N^2\lambda^2}{4}+\frac{N\lambda}{2}\sqrt{\frac{N^2\lambda^2}{4}+4z^2}}{2z^2}}\approx \sqrt{\frac{N}{2z\lambda}}\propto\frac{1}{\sqrt{z}}.
\end{equation}}\\
For E-ASM, it scales as $1/\sqrt[3]{z}$, preserving significantly more spectral information over long propagation distances than both BLASM and BEASM:
\begin{equation}
    z\left|\frac{\lambda f_{x,\mathrm{lim}}}{\sqrt{1-(\lambda f_{x,\mathrm{lim}} )^2}}-\lambda f_{x,\mathrm{lim}}\right|=N\Delta_x\rightarrow f_{x,\mathrm{lim}}\approx\frac{1}{\lambda}\left(\frac{2N\Delta_x}{z}\right)^\frac{1}{3}\propto\frac{1}{\sqrt[3]{z}}.
\end{equation}
The maximum spatial frequency profiles for each algorithm are compared in Fig. \ref{fig2}(c).\\
By evaluating the real-space convolution for Fresnel propagation, E-ASM enables an arbitrary selection of the destination window. Nevertheless, the allowable magnification and lateral shift remain bounded because discrete Fresnel diffraction lacks evanescent decay to attenuate higher-order periodic frequency bands. This numerical zone immune to aliasing and wrap-around artifacts can be mapped out analytically (see the supplementary section~1). Because the pre-compensation step performs a circular convolution on the zero-padded array of size $2L$, the artifact-free observation window after center-cropping is bounded by $M_{\mathrm{max}} \cdot L$, where $M_{\mathrm{max}} = N \lambda z / (2 L^2)$ coincides with the natural magnification of the SAS method. For the point-source model, $M_{\mathrm{max}} > 1$ is additionally required to avoid under-sampling the quadratic phase $Q_1$ in Eq.~\eqref{Q1}, imposing the same critical minimum propagation distance as the SAS 
method~\cite{heintzmann2023scalable}. The analytical zero-order hold model removes this constraint entirely, extending the E-ASM below the critical distance threshold. However, because $M_{\mathrm{max}}$ decreases linearly with $z$, the usable field of view contracts accordingly; the selected observation window must still be restricted to the valid numerical regime. In addition to the replica bound, a vignetting constraint arises when the propagating field diverges, as in the SAS. However, at unit magnification, vignetting never occurs and no maximum propagation distance exists (see supplementary section~1). Note that this is not a universal bound like the replica condition. The implementation details of the E-ASM workflow are formalized in Algorithm \ref{alg:E-ASM}.
\clearpage
\begin{algorithm}
\caption{Implementation of the E-ASM}
\label{alg:E-ASM}
\small
\begin{algorithmic}[1]
 
\Require Source field $U(X_s, Y_s)$, wavelength $\lambda$,
         propagation distance $z$, source grid $X_s, Y_s$
         with window size $L$,source pitch $\Delta_p$ and $N \times N$ samples,
         observation grid $X_o, Y_o$
\Ensure Propagated field $U(X_o, Y_o)$
\State $M_{\max} \gets \dfrac{N \lambda z}{2L^2}$
\If{$\max|X_o| > M_{\max} \cdot \dfrac{L}{2}$
    \textbf{or} $\max|Y_o| > M_{\max} \cdot \dfrac{L}{2}$}
    \State \textbf{abort} with warning
\EndIf
\State $X_p,\, Y_p \gets 2\times$ extended grid from $X_s,\, Y_s$
\State $U_p(X_p, Y_p) \gets$ zero-pad $U(X_s, Y_s)$ onto $X_p,\, Y_p$
\State $f_X,\, f_Y \gets$ spatial frequency grid from $X_p,\, Y_p$
\State $H_\mathrm{ASM} \gets \exp\!\left(iz\sqrt{k^2 - 4\pi^2(f_X^2 + f_Y^2)}\right)$
\State $H_\mathrm{F}^* \gets \exp\!\left(i\pi\lambda z\,(f_X^2 + f_Y^2)\right)$
\State $W \gets$ band-limit window computed from $f_X,\, f_Y,\, z,\, \lambda$
\State $U_c(X_p, Y_p) \gets \mathcal{F}^{-1}\!\left[W \cdot H_\mathrm{ASM} \cdot H_\mathrm{F}^* \cdot \mathcal{F}(U_p)\right]$
\If{zero-order hold (analytic kernel)}
    \State $I_x(X_o, X_p) \gets
        \displaystyle\int_{X_p - \Delta_p/2}^{X_p + \Delta_p/2}
        \exp\!\left[\frac{ik}{2z}(X_o - x')^2\right] dx'$
    \State $I_y(Y_o, Y_p) \gets
        \displaystyle\int_{Y_p - \Delta_p/2}^{Y_p + \Delta_p/2}
        \exp\!\left[\frac{ik}{2z}(Y_o - y')^2\right] dy'$
    \State $U(X_o, Y_o) \gets
        \dfrac{e^{ikz}}{i\lambda z}
        \sum_{m,n} U_c\!\left(X_p^{(n)}, Y_p^{(m)}\right)
        I_x\!\left(X_o, X_p^{(n)}\right)
        I_y\!\left(Y_o, Y_p^{(m)}\right)$
\ElsIf{point source \textbf{and} uniform grid (Kernel factorization path)}
    \For{each axis $u \in \{x, y\}$ with input coord.\ $c_p$ (pitch $\Delta_{c_p}$), output coord.\ $c_o$ (pitch $\Delta_{c_o}$)}
        \State $\Delta_0 \gets c_o^{(0)} - c_p^{(0)}$            
        \State $\theta \gets \dfrac{k}{z}\,\Delta_{c_o}\Delta_{c_p}$    
        \State $\tilde{\phi}_p \gets \dfrac{ik}{2z}\!\left[m^2\Delta_{c_p}(\Delta_{c_p}-\Delta_{c_o}) - 2\Delta_0\, m\Delta_{c_p}\right]$
        \State $\tilde{\phi}_o \gets \dfrac{ik}{2z}\!\left[o^2\Delta_{c_o}(\Delta_{c_o}-\Delta_{c_p}) + 2\Delta_0\, o\Delta_{c_o}\right]$
        \State $g(n) \gets \exp\!\left(i\,\theta\,n^2/2\right),\quad n = o - m$
        \State $b \gets U_c \odot e^{\tilde{\phi}_p}$
        \State $c \gets \mathcal{F}^{-1}\!\left[\mathcal{F}(b)\cdot\mathcal{F}(g)\right]$
        \State $\mathcal{B}_u \gets e^{\tilde{\phi}_o} \odot e^{\frac{ik}{2z}\Delta_0^2} \odot c \cdot \Delta_{c_p}$
    \EndFor
    \State $U(X_o, Y_o) \gets
        \dfrac{e^{ikz}}{i\lambda z}\,\mathcal{B}_y\!\left[\mathcal{B}_x[U_c]\right]$
\Else
    \State $I_x(X_o, X_p) \gets
        \exp\!\left[\dfrac{ik}{2z}(X_o - X_p)^2\right] \Delta_{X_{p}}$
    \State $I_y(Y_o, Y_p) \gets
        \exp\!\left[\dfrac{ik}{2z}(Y_o - Y_p)^2\right] \Delta_{Y_{p}}$
    \State $U(X_o, Y_o) \gets
        \dfrac{e^{ikz}}{i\lambda z}
        \sum_{m,n} U_c\!\left(X_p^{(n)}, Y_p^{(m)}\right)
        I_x\!\left(X_o, X_p^{(n)}\right)
        I_y\!\left(Y_o, Y_p^{(m)}\right)$
\EndIf
\end{algorithmic}
\end{algorithm}
\clearpage
\section{Results and Discussion}
\subsection{scalar diffraction benchmarking and demonstration}
\begin{figure}[htbp]
  \centering
  \includegraphics[width=\textwidth, page=4]{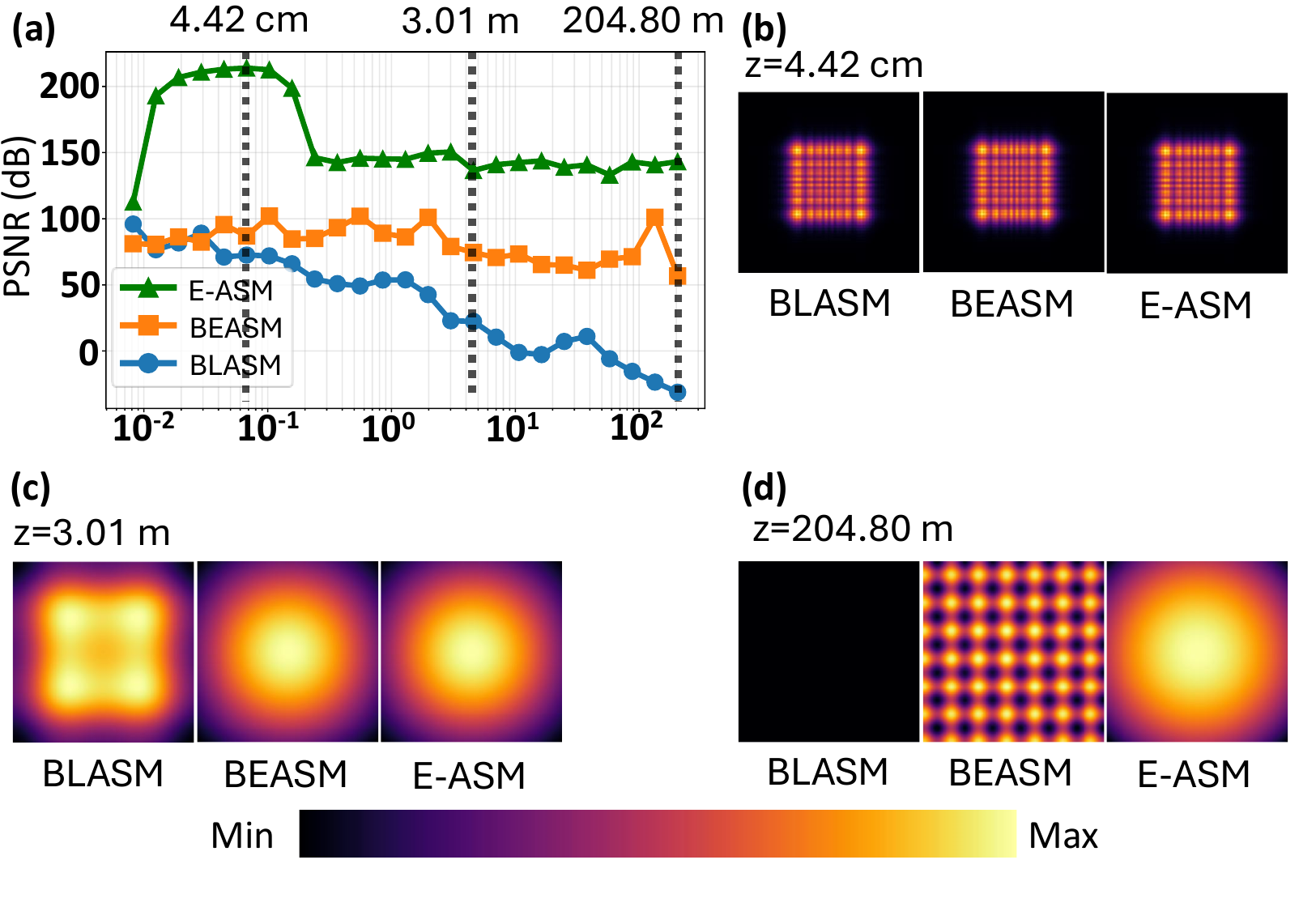}
  \caption{Scalar diffraction benchmark for a rectangular aperture. 
\\(a) Distance-dependent peak signal-to-noise ratio (PSNR)
\\(b--d) Intensity profiles for different propagation algorithms and corresponding field distributions at three selected distances.}
  \label{fig3}
\end{figure}

\begin{figure}[htbp]
  \centering
  \includegraphics[width=0.85\textwidth, page=5]{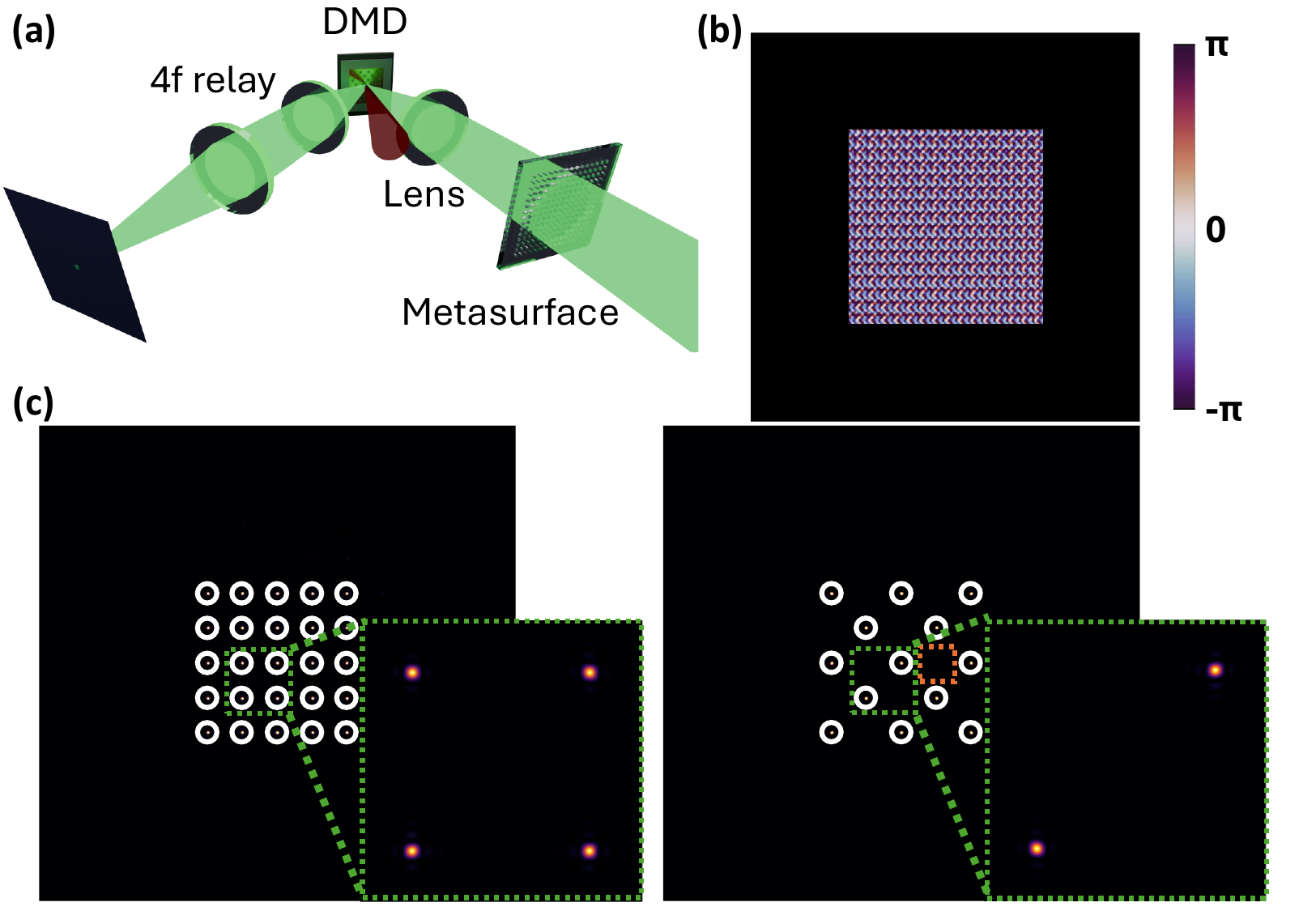}
  \includegraphics[width=0.85\textwidth, page=6]{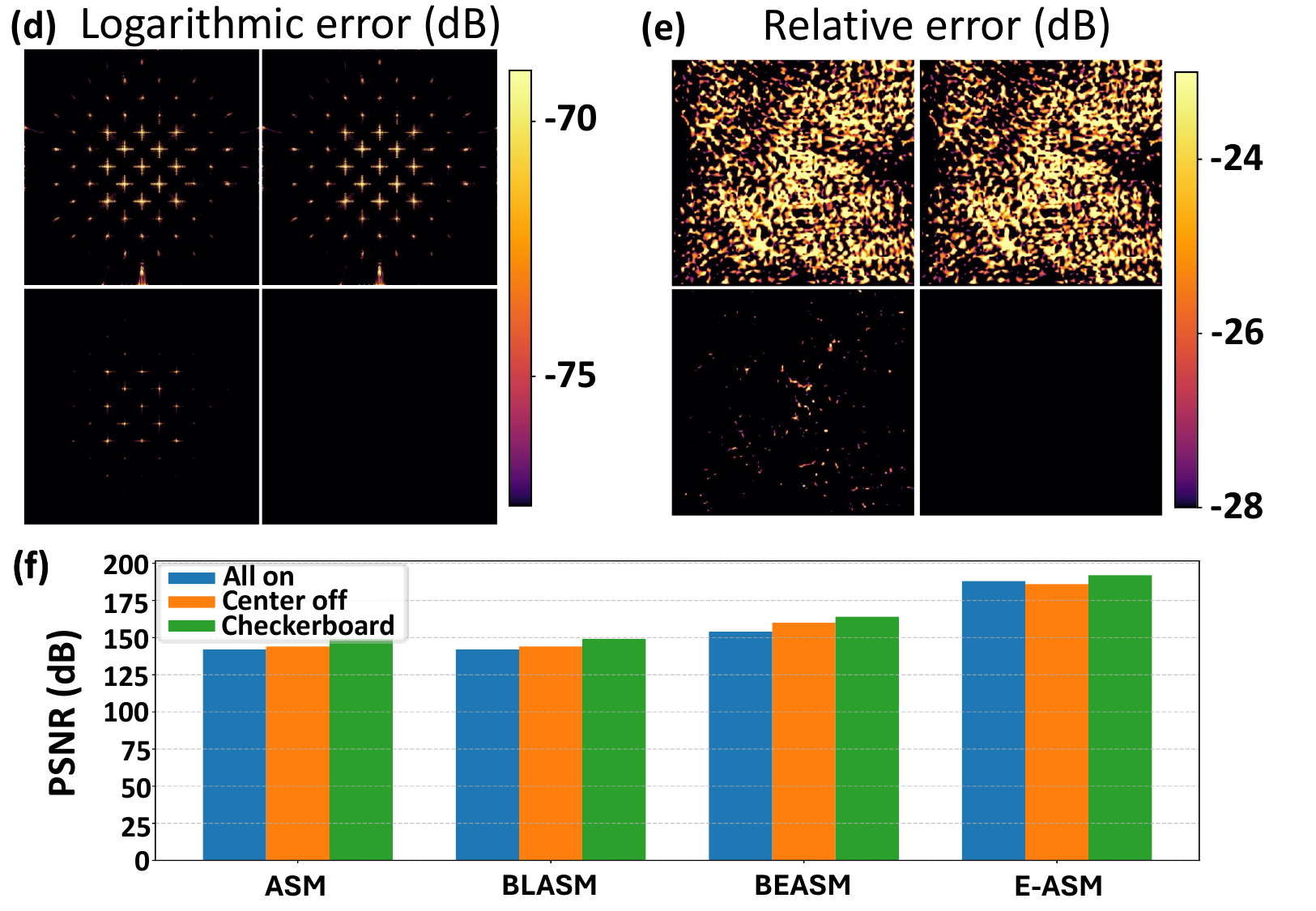}
  \caption{{\small (a) Schematic of the full-path optical calculation of the OALQG. (b) Designed phase distribution of the metasurface. (c) The intensity distribution before the DMD (left) and the observation plane with the ``checkerboard'' mode (right). The main plots are scaled with a non-linear power law ($I^{1/8}$) to enhance overall visibility, whereas the magnified insets inside the green dashed boxes present the linear intensity distribution ($I$). White circle indicates the position of each spot. (d--e) The logarithmic error on the entire observation plane (d) and the relative error inside the dashed orange box region which is one of the dark region of the checkerboard pattern (e) for ASM (upper left), BLASM (upper right), BEASM (lower left), and E-ASM (lower right), respectively. (f) The PSNR of each mode for different algorithms.}}
  \label{fig4}  
\end{figure}

To evaluate E-ASM, we benchmark scalar diffraction across various propagation distances and apply it to a real-world quantum hardware problem: optically addressed local qubit gates (OALQG). We first compute the diffraction pattern past a rectangular aperture, with the setup summarized in Table \ref{table 1}.\\
\begin{table}[h]
\vspace{-5mm}
\caption{\bf Simulation parameters for the rectangular aperture benchmark}\label{table 1}
\begin{tabular}{>{\centering\arraybackslash}p{3cm} >{\centering\arraybackslash}p{6cm}}
\toprule%
Parameters & Values \\
\midrule
Window size & $2048\, \mathrm{\mu m}\times2048\, \mathrm{\mu m}$  \\
Pixel number $(N_x, N_y)$ & $2048 \times2048$  \\
Wavelength& $500$ nm  \\
Aperture size & $1024\, \mathrm{\mu m}\times1024\, \mathrm{\mu m}$  \\
Propagation distance & 8.192 mm - 204.8 m (log spacing) \\
\botrule
\end{tabular}
\vspace{-5mm}
\end{table}\\
\noindent In Fig. \ref{fig3}(a), E-ASM retains its remarkable accuracy even at a propagation distance of 204.8 m, where BEASM and BLASM fail entirely due to severe aliasing artifacts. At the shortest distance—coinciding with the minimum critical propagation distance at a Fresnel number of 256—E-ASM performs similarly to BEASM and BLASM owing to localized numerical instability. Beyond this threshold, however, E-ASM surpasses BEASM by a large margin, reaching up to a 100 dB PSNR gap. Even after band-limiting filters out its higher spatial frequencies, E-ASM retains a 50--60 dB PSNR advantage over BEASM. It also computes faster than BEASM; execution times across sampling counts are given in Table \ref{table 2}. Specifically, E-ASM maintains an order-of-magnitude speed advantage over the two BEASM variants up to $4096^2$ sampling points, which is the maximum memory capacity handled by our RTX 3090 Ti GPU setup.\\
\clearpage
\begin{table}[h]
\footnotesize
\vspace{-10mm}
\caption{\bf Computation time (ms) comparison}\label{table 2}
\centering
\begin{tabular}{>{\centering\arraybackslash}p{2.5cm} *{5}{>{\raggedleft\arraybackslash}p{1.5cm}}}
\toprule
& \multicolumn{5}{c}{Sampling Number} \\ 
\cmidrule(lr){2-6}
Methods & $256^2$ & $512^2$ & $1024^2$ & $2048^2$ & $4096^2$\\
\midrule
ASM          & 2.88 & 5.14  & 18.91  & 65.93 & 269.39\\ 
BLASM        & 3.02 & 6.07  & 19.23  & 68.65 & 269.74\\
BEASM (MTP)  & 12.22 & 70.99 & 496.98 & 3831.33 & 30551.52\\
BEASM(NUFFT) & 51.81 & 86.35 & 281.43 & 1037.62 & 4292.98\\
E-ASM       & 7.96  & 15.01 & 42.10 & 140.27 & 599.85\\ 
\botrule
\end{tabular}
\vspace{-10mm}
\end{table}
\noindent Next, we evaluate an OALQG system using the E-ASM. The OALQG generates a 2D array of individually controllable focused spots that can be toggled on or off \cite{zhang2024scaled}, enabling parallel, site-selective gate operations on neutral atoms \cite{weitenberg2011single, xia2015randomized}, trapped ions \cite{debnath2016demonstration, crain2014individual}, and solid-state defects \cite{pfender2014single, bradley2019ten}.\\
Local optical addressing is particularly critical for trapped-ion quantum computing. Delivering independent Raman beam pairs to individual ions allows simultaneous driving of multiple motional modes to achieve all-to-all connectivity, bypassing the topological constraints of global or nearest-neighbor setups \cite{lu2019global}. Furthermore, as linear ion chains lengthen, spectral crowding complicates the isolation of motional modes, which degrades gate fidelity. Individually addressed multi-mode driving circumvents this, sustaining high-fidelity operations in long chains. This is even more essential when scaling to 2D ion crystals in Penning or surface traps \cite{guo2024site, makadia2026programmable}, where lattice complexity rules out simple row-or-column addressing. Thus, the OALQG framework is a cornerstone for scaling up fault-tolerant quantum processors.\\
Our specific OALQG configuration is schematized in Fig. \ref{fig4}(a), consisting of a metasurface, a Fourier lens, a DMD, and a $4f$ relay. All lenses feature a focal length of $f = 5$ mm and an $\mathrm{NA} = 0.1$. Since all lenses operate at $\mathrm{NA}=0.1$, the cross-polarization and longitudinal field components are negligible compared to the dominant transverse component, so a scalar treatment is adequate for this benchmark. The metasurface, with a $500\, \mathrm{nm} \times 500$ nm unit cell, projects a focal spot array onto the DMD via the Fourier lens. The metasurface spans $512\,\mathrm{\mu m} \times 512\,\mathrm{\mu m}$ within a $1024\,\mathrm{\mu m} \times 1024\,\mathrm{\mu m}$ simulation window. Its phase profile is engineered via a phase-fixed weighted Gerchberg--Saxton algorithm \cite{kim2019large} (Fig. \ref{fig4}(b)). The left side of Fig. \ref{fig4}(c) displays the intensity on the DMD, with a spot separation of $75\,\mathrm{\mu m} \times 75\,\mathrm{\mu m}$. Gating the DMD pixels enables site-selective on/off switching for individual qubit positions. The DMD pixel pitch is set to $7.5\,\mathrm{\mu m} \times 7.5\,\mathrm{\mu m}$, adhering to commercial specifications like the Texas Instruments DLP6500. Activating or deactivating a $10 \times 10$ super-cell of the DMD controls the presence of each spot.\\
We evaluated three target profiles: ``All-on'', ``Center-off'', and ``Checkerboard''. For brevity, the main text presents the logarithmic error profile of the ``Checkerboard'' mode, while the ``All-on'' and ``Center-off'' cases are detailed in Section 3 of the Supplementary Information.
\clearpage
\noindent E-ASM accurately predicts not only the full intensity distribution but also fine features within regions of extremely low intensity (orange dashed box), which is crucial for evaluating OALQG 
crosstalk. Here, because the Fourier lens and $4f$ relay collimate the beams and suppress high spatial frequencies—mitigating aliasing artifacts—the conventional ASM achieves relatively high precision compared to the free-space benchmark. This same collimation explains why BLASM remains nearly identical to the conventional ASM despite the large propagation distance relative to the wavelength. Nonetheless, E-ASM shows a clear superiority, delivering a 60 dB accuracy margin over the standard ASM and a 30 dB improvement over BEASM.

\subsection{Vectorial diffraction benchmark}
As shown in Supplementary Section 4, the Luneburg integral is an exact Maxwell solution despite its apparent inability to reconstruct $E_y$ under $x$-polarized illumination. This arises only when the scalar input is unphysical; for a true Maxwell solution, all six field components are naturally self-consistent. We utilized the Luneburg integral for the vectorial diffraction benchmark with the $\mathbf{A}$-seeding solution as an input. In the next section, we jointly applied the $\mathbf{H}$-driven PEC and $\mathbf{E}$-driven PMC models to obtain symmetric solutions instead of Luneburg integral.\\
We chose the fundamental Gaussian, $\mathrm{HG}_{21}$, and $\mathrm{LG}_{20}$ cosine beams as inputs, each with a dominant $x$-polarization and a beam waist of $w_0 = 8\,\mathrm{\mu m}$. The full vector field was generated by $\mathbf{A}$-seeding the plane-wave spectrum at the waist plane ($z = 0$), ensuring an exact free-space Maxwell solution. The $E_x$ and $E_y$ components were evaluated through analytical differentiation of the Lorenz-gauge expression, removing numerical error. Conversely, the $E_z$ component cannot be derived analytically since the $z$-derivative of the vector potential requires the spectrum away from the waist plane. Instead, $E_z$ was reconstructed from the divergence-free condition: $\tilde{E}_z = -(k_x \tilde{E}_x + k_y \tilde{E}_y)/k_z$. As a result, the $E_x$ and $E_y$ fields are free of numerical error, while the accuracy of $E_z$ is restricted only by the discrete Fourier transform discretization.\\
Given $w_0 = 8\,\mathrm{\mu m}$, the divergence angle $\theta_d = \lambda/(\pi w_0) \approx 0.02$ guarantees that the spectral energy remains confined well within the propagating region. Consequently, the $1/k_z$ factor used in evaluating $\tilde{E}_z$ maintains excellent numerical stability.\\
\begin{figure}[htbp]
  \centering
  \includegraphics[width=\textwidth, page=7]{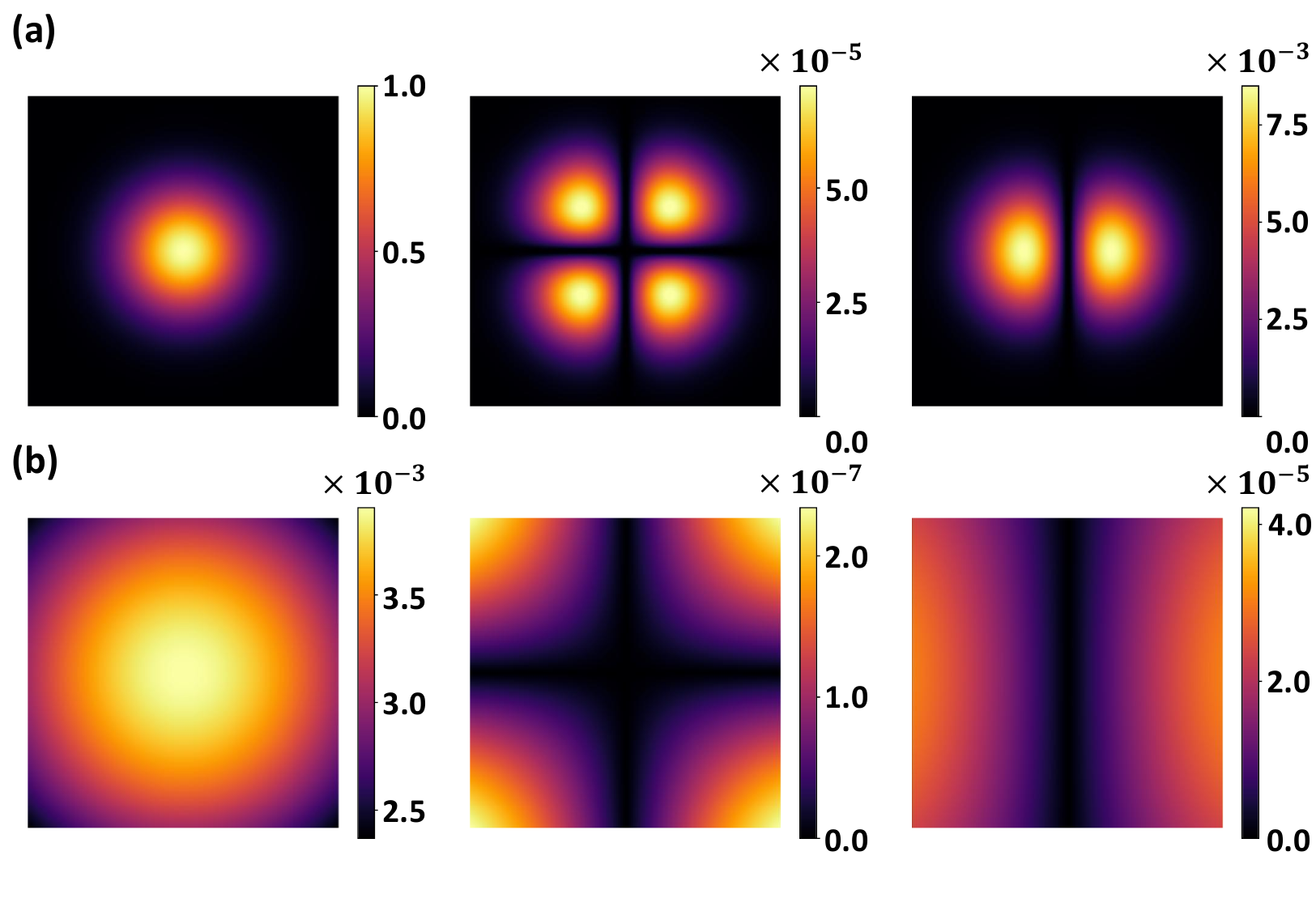}
 \caption{Input profile and propagated ground truth of the Gaussian vector beam. 
\\(a) Magnified amplitude distributions of the input electric field ($E_x, E_y, E_z$ components) for the $x$-polarized fundamental Gaussian beam, showing a $50\,\mathrm{\mu m} \times 50\,\mathrm{\mu m}$ region inside the $1024\,\mathrm{\mu m} \times 1024\,\mathrm{\mu m}$ total window. 
\\(b) Full-window ground-truth field amplitudes after a propagation distance of $z = 0.103\,\mathrm{m}$.}
  \label{fig5}  
\end{figure}\\
\noindent For the fundamental Gaussian and $\mathrm{HG}_{21}$ vector beam simulations, the window size, pixel count, and wavelength are identical to those listed in Table \ref{table 1}. In contrast, the window size for the $\mathrm{LG}_{20}$ cosine beam simulation is reduced by half to $1024\,\mathrm{\mu m} \times 1024\,\mathrm{\mu m}$ because the high spatial frequency components inherent to vortex beams necessitate a finer sampling rate. The $\mathrm{HG}_{21}$ results are omitted from the main text, as they closely follow the trends of the fundamental Gaussian and $\mathrm{LG}_{20}$ cosine beams (Supplementary Section 11).\\
Fig. \ref{fig5}(a) illustrates the amplitude distributions of the $E_x$, $E_y$, and $E_z$ field components obtained by $\mathbf{A}$-seeding the plane-wave spectrum of the fundamental Gaussian beam at its waist ($A_x$). Because the $E_y$ component is rigorously preserved and all field components are self-consistent with Maxwell's equations, no discrepancies exist among the PEC solution (the Luneburg integral), the PMC solution (the VASM), and the Stratton--Chu integral solutions.\\
\begin{figure}[htbp]
  \centering
  \includegraphics[width=0.49\textwidth, page=8]{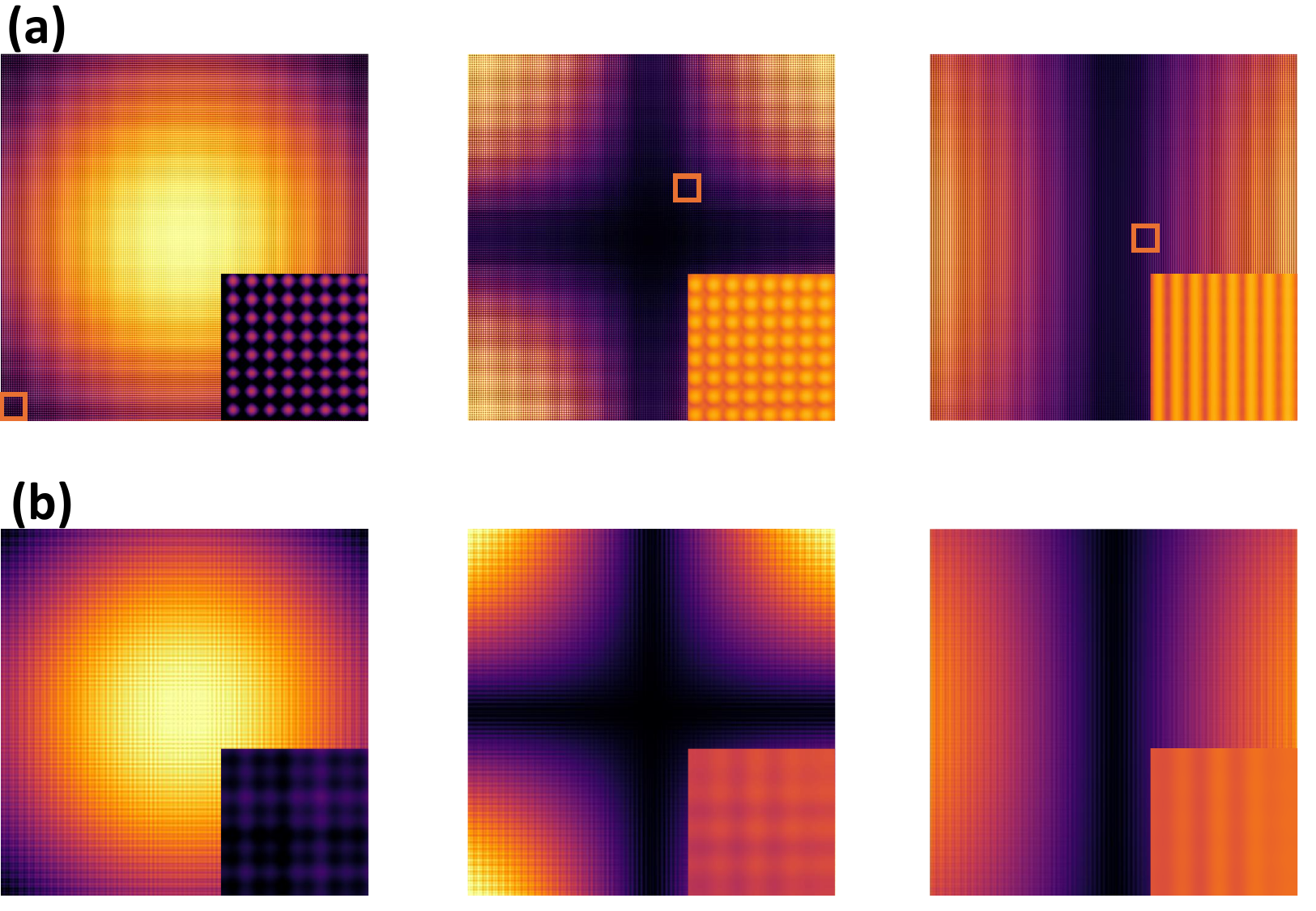}
  \includegraphics[width=0.49\textwidth, page=9]{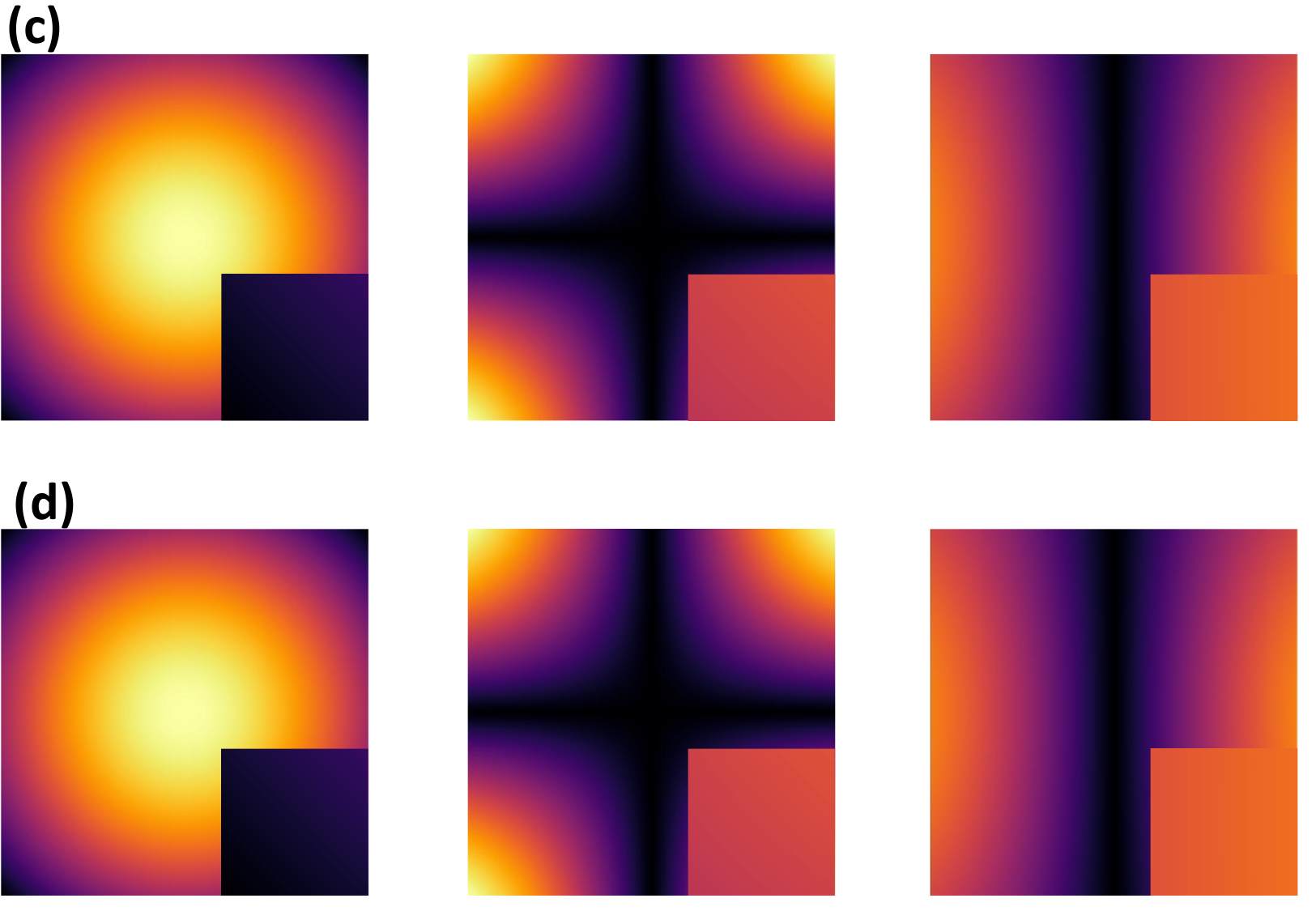}
  \includegraphics[width=0.49\textwidth, page=10, trim = 0cm 7cm 0cm 0cm, clip]{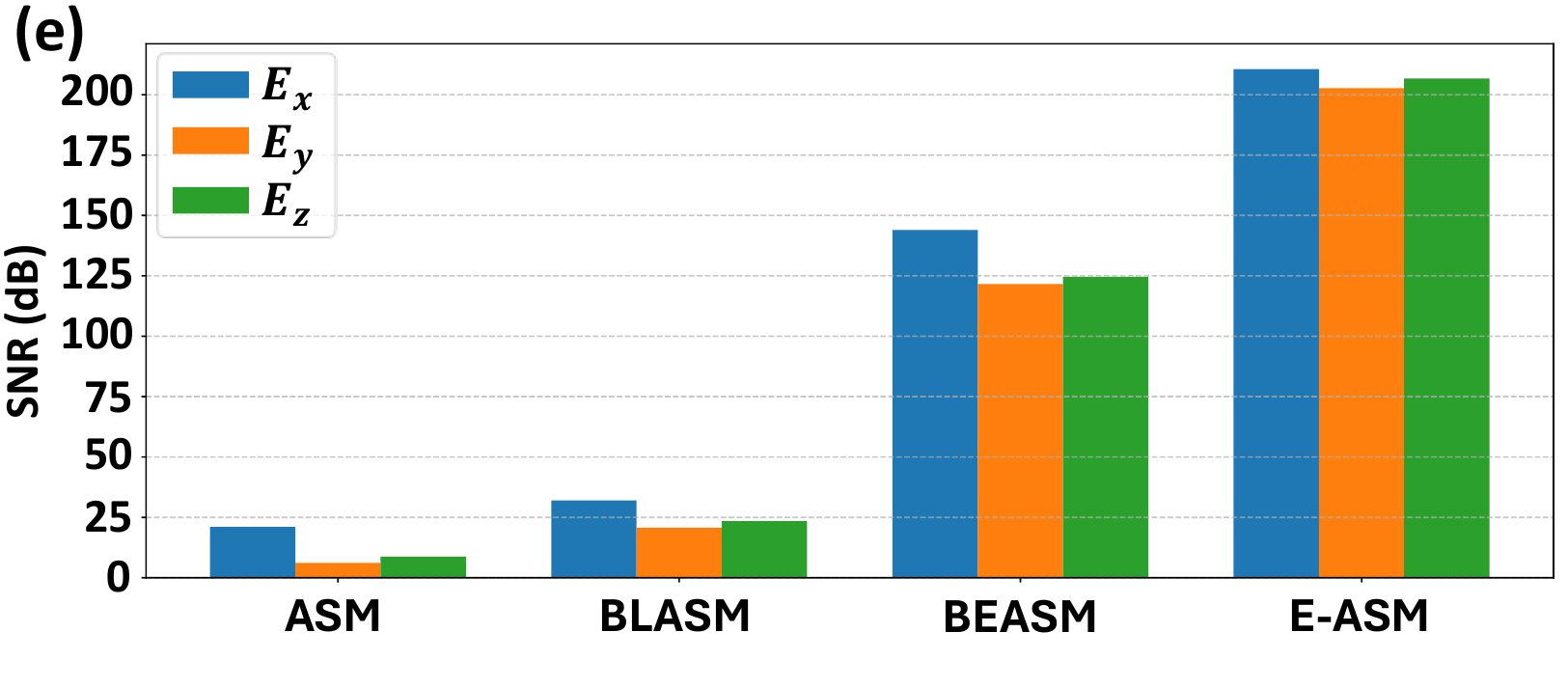}
  \includegraphics[width=0.49\textwidth, page=11, trim = 0cm 7cm 0cm 0cm, clip]{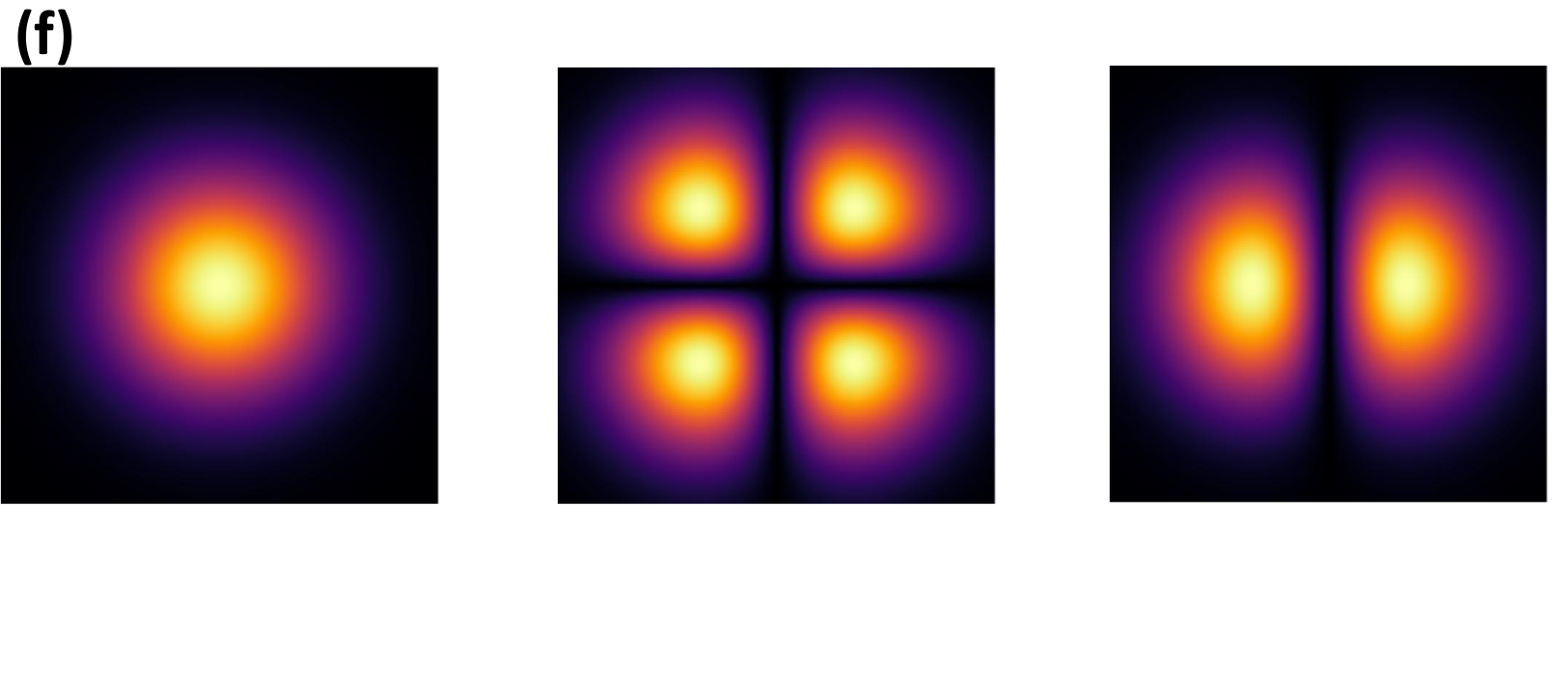}
  \caption{Comparison of propagated vectorial field and numerical accuracy. 
\\(a--d) Propagated $E_x, E_y, E_z$ field amplitudes at $z = 0.103\,\mathrm{m}$ simulated via ASM (a), BLASM (b), BEASM (c), and E-ASM (d). The inset shows the magnified orange box region of interest, plotted with a non-linear power-law scaling ($\sim\!I^{1/4}$) to enhance visibility. 
\\(e) Component-wise SNR performance for different propagation methods. 
\\(f) $4\times$ expanded observation window achieved using E-ASM.}
  \label{fig6}  
\end{figure}

\begin{figure}[htbp]
  \centering
  \includegraphics[width=0.49\textwidth, trim=0cm 0.5cm 0cm 0cm, clip, page=12]{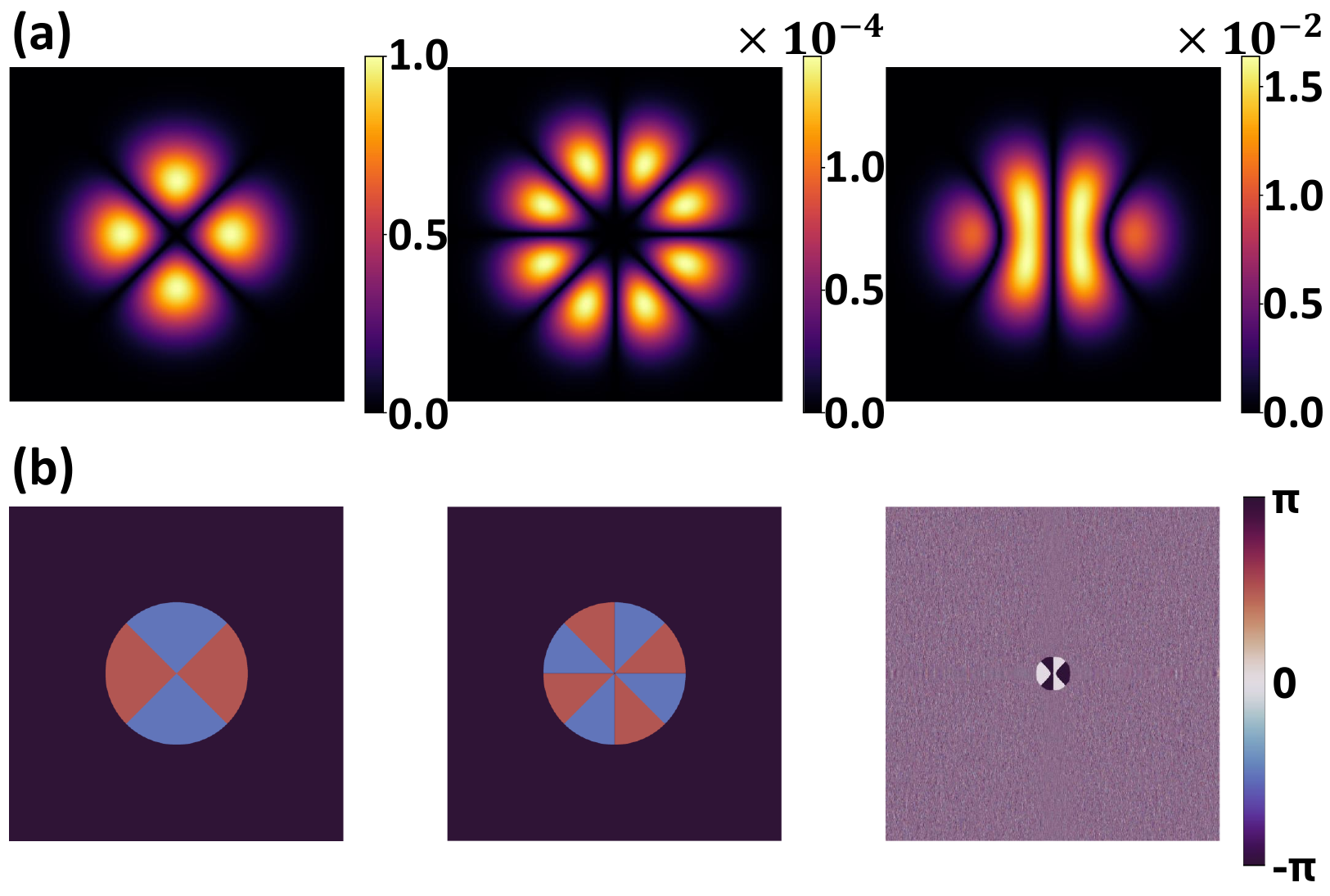}
  \includegraphics[width=0.49\textwidth, trim=0cm 0.5cm 0cm 0cm, clip, page=13]{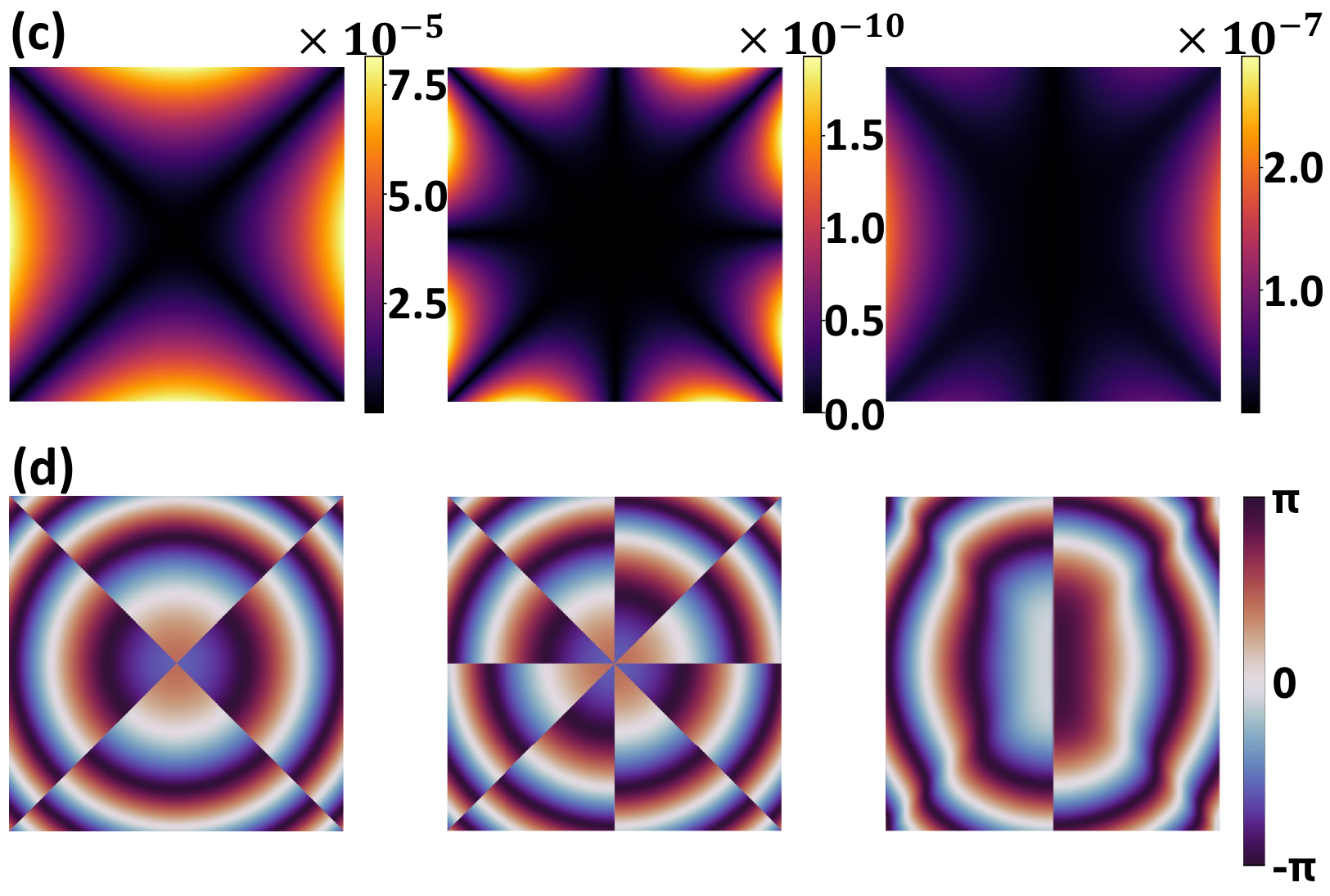}
  \caption{Input vector fields and propagated ground truth of the $\mathrm{LG}_{20}$ cosine beam. 
\\(a--b) Magnified amplitude (a) and phase (b) profiles for each vector component at the input plane, showing a $50\,\mathrm{\mu m} \times 50\,\mathrm{\mu m}$ region inside the $1024\,\mathrm{\mu m} \times 1024\,\mathrm{\mu m}$ window. 
\\(c--d) Full-window ground-truth (c) amplitude and (d) phase distributions after propagating to $z = 0.103\,\mathrm{m}$.}
  \label{fig7}  
\end{figure}

\noindent The SNR is defined as $\mathrm{SNR} = \frac{\int |U(x,y)|^2 dxdy}{\int |U(x,y) - \alpha U_{\mathrm{GT}}(x,y)|^2 dxdy}$, where the scaling factor is given by $\alpha = \frac{\int U(x,y)U_{\mathrm{GT}}^*(x,y)dxdy}{\int |U_{\mathrm{GT}}(x,y)|^2 dxdy}$ as proposed by K. Matsushima (2003) \cite{matsushima2003fast}. Unlike the PSNR, the SNR quantifies the accuracy of the complex-valued field rather than the intensity alone—essential for vector beams, which require precise phase alongside amplitude.\\
Both ASM and BLASM suffer severe artifacts from heavy spatial-frequency truncation. Although BEASM appears visually artifact-free, E-ASM surpasses it by 60 dB in SNR, and exceeds the conventional ASM and BLASM by over 150 dB. Capitalizing on the ability of E-ASM to arbitrarily select the observation window within the artifact-free region, Fig. \ref{fig6}(f) demonstrates a $4\times$ zoomed-out, wide-field image after propagation, ensuring that the overall spatial features are not truncated by a restricted computational window.\\
Our next demonstration involves the vectorial propagation of an $\mathrm{LG}_{20}$ cosine beam, synthesized via the coherent addition of two LG beams possessing orbital angular momentum (OAM) states of $+2$ and $-2$. Figs. \ref{fig7}(a) and \ref{fig7}(b) illustrate the amplitude and phase distributions at the beam waist, respectively. The amplitude profile is magnified within the central $50 \,\mathrm{\mu m}\times 50\,\mathrm{\mu m}$ region of the entire $1024\,\mathrm{\mu m} \times 1024\,\mathrm{\mu m}$ simulation window, whereas the phase plot in Fig. \ref{fig7}(b) is displayed over the full, uncropped window.\\
Interestingly, the cross-polarization component ($E_y$ field) exhibits the eight-lobed azimuthal structure of an $\mathrm{LG}_{40}$ cosine beam. This is a direct signature of spin--orbit coupling near the beam waist: the transversality projection transfers the $\pm2$ OAM content of the input into $\pm 4$ harmonics of the cross-polarized field, consistent with the geometric-phase mechanism predicted in \cite{bliokh2011spin}. Such spin–orbit-induced structure in superposed LG modes has not previously been resolved at this fidelity. Quantitative access to these exact cross-polarized and longitudinal field distributions is particularly relevant for high-fidelity qubit control in non-paraxial geometries, where such components directly perturb the intended light--atom interaction.
\begin{figure}[htbp]
  \centering
  \includegraphics[width=0.49\textwidth, page=14]{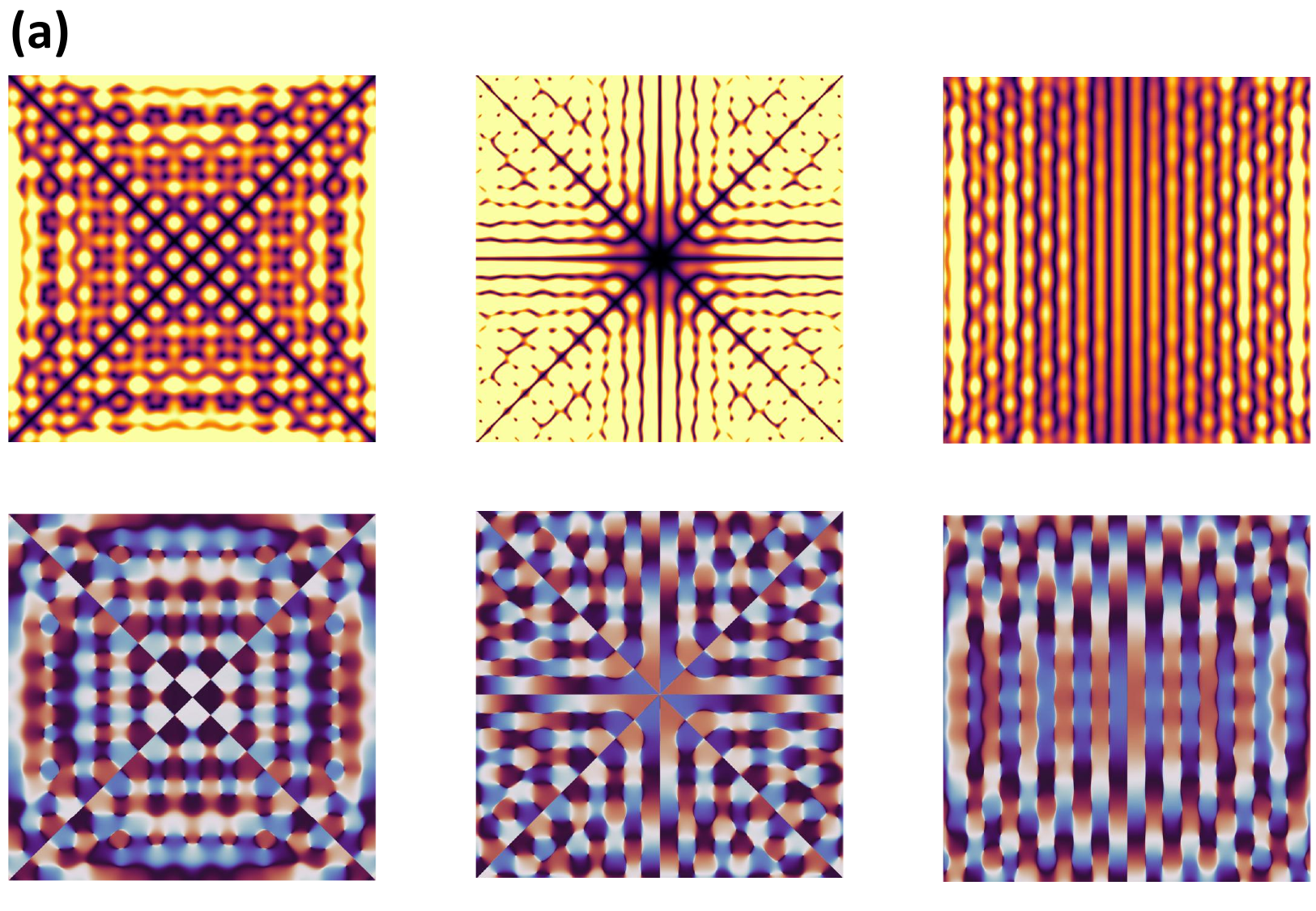}
  \includegraphics[width=0.49\textwidth, page=15]{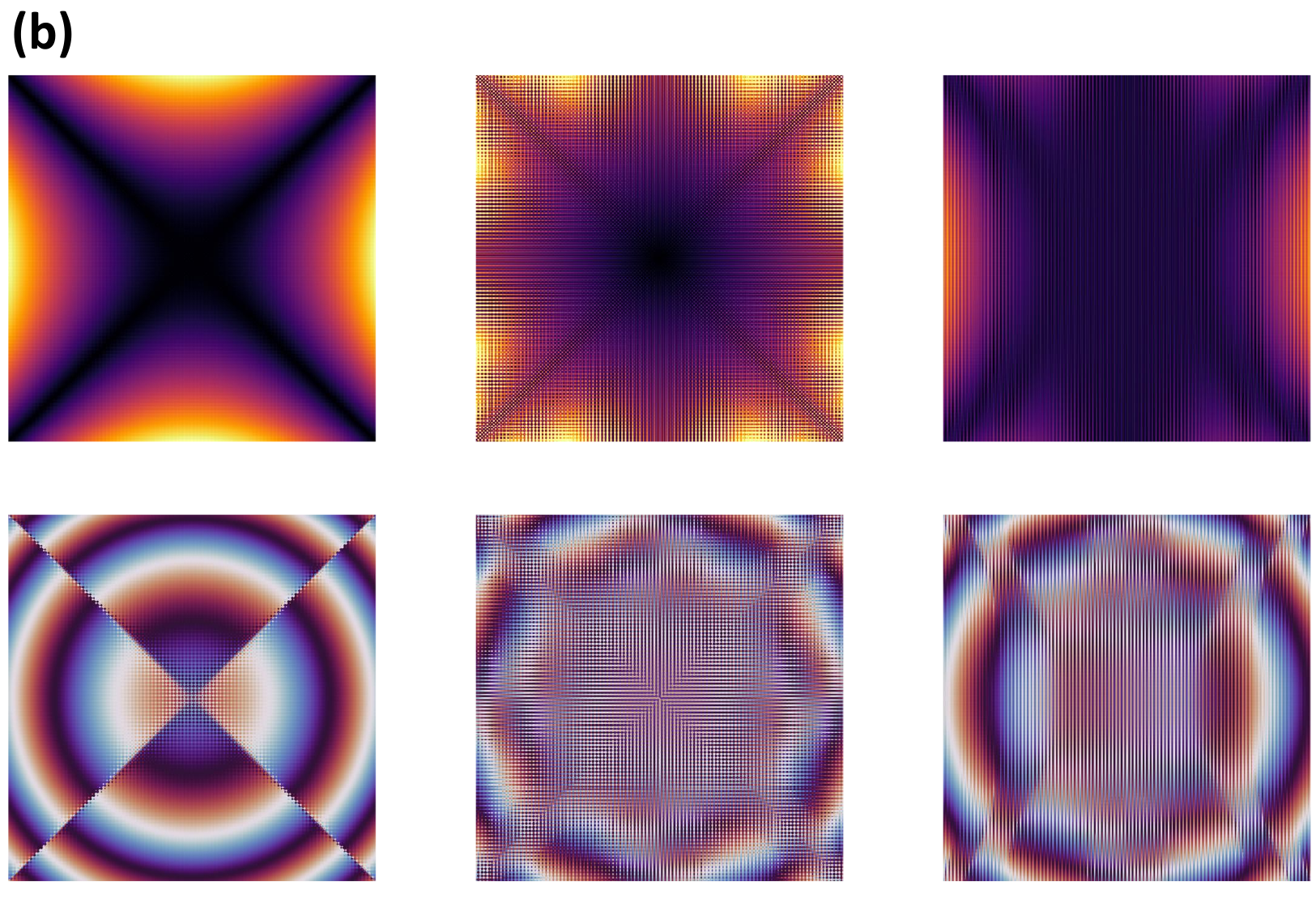}
  \includegraphics[width=0.49\textwidth, page=16]{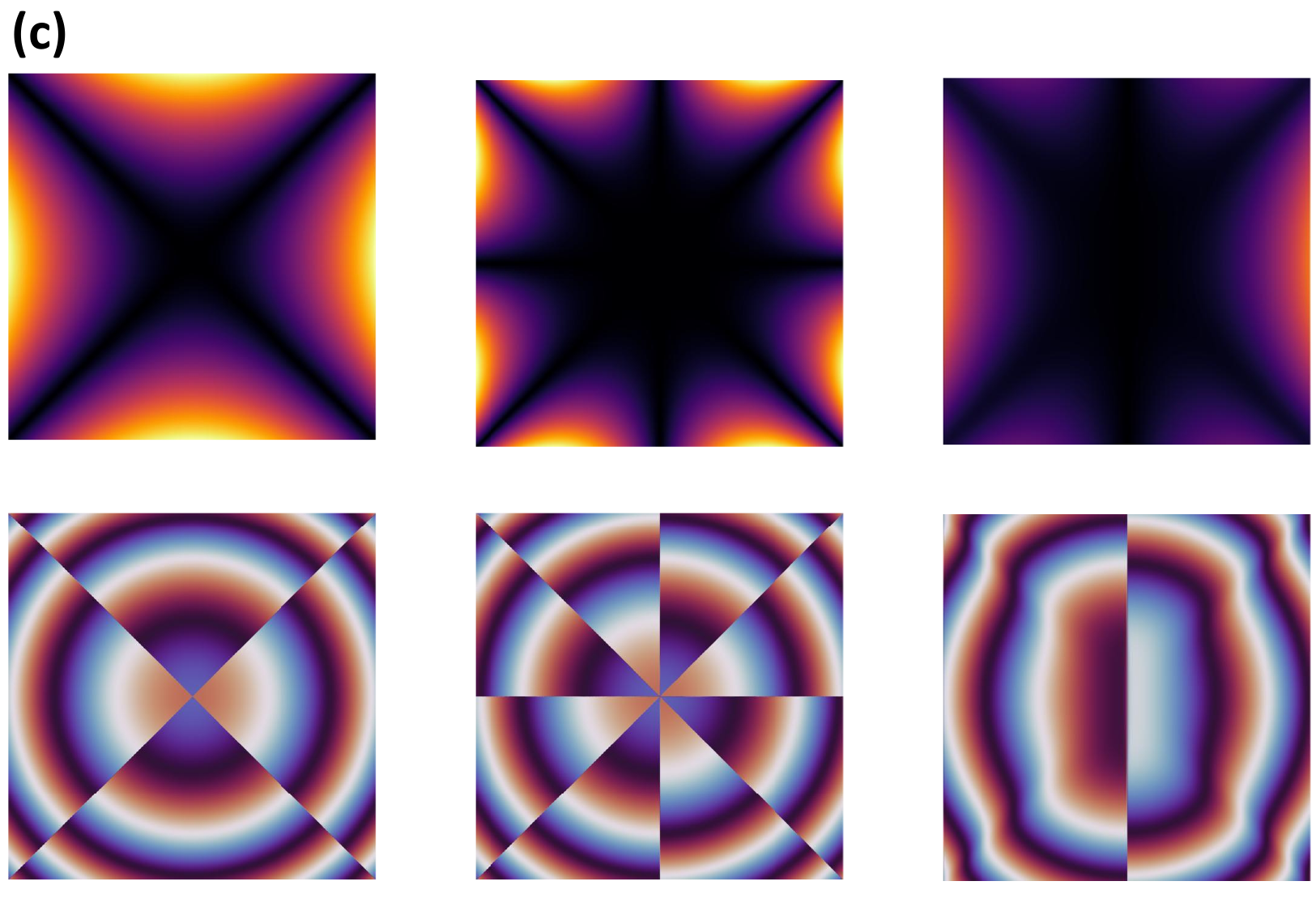}
  \includegraphics[width=0.49\textwidth, page=17]{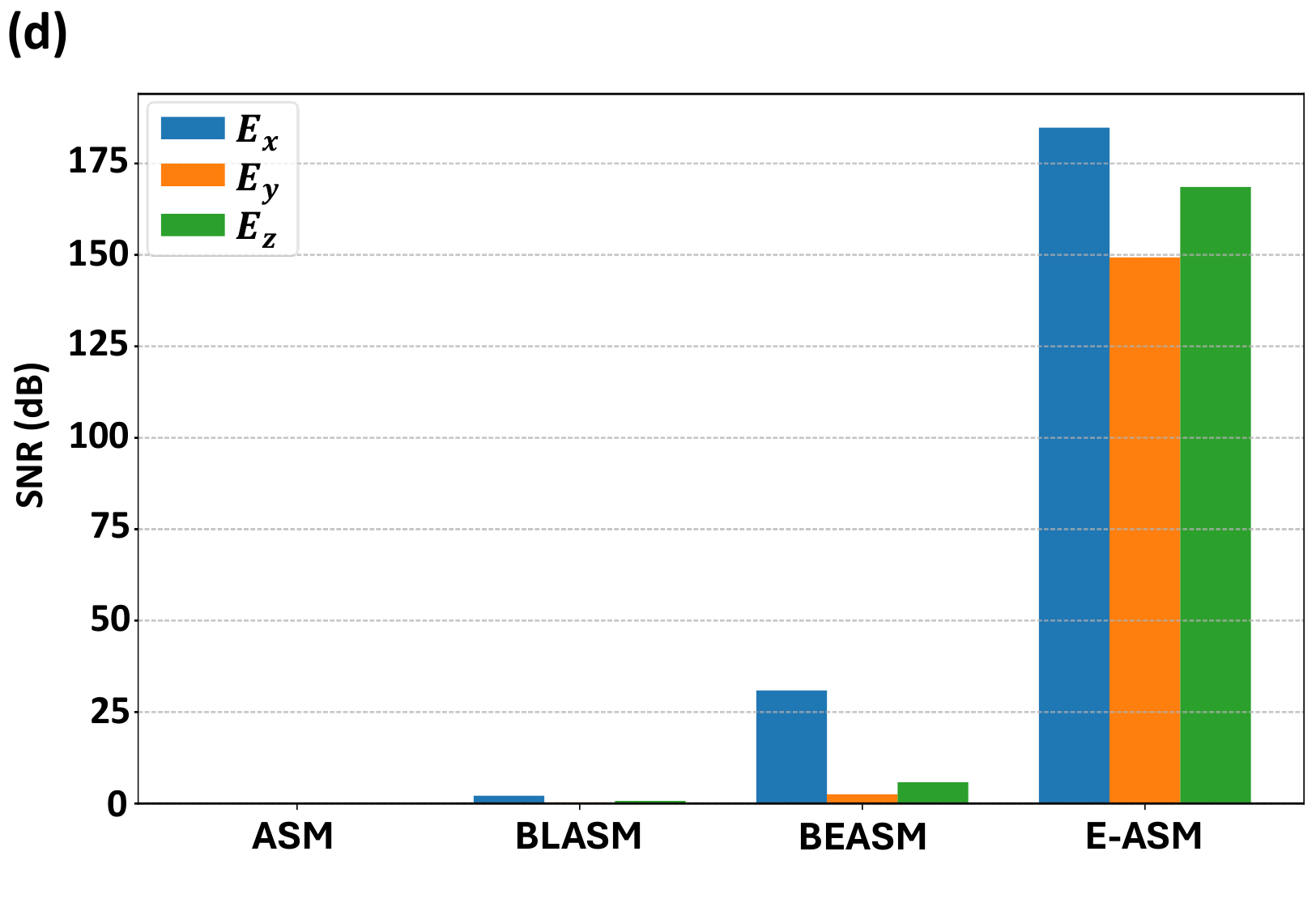} 
  \caption{Propagated vectorial field and SNR of the $\mathrm{LG}_{20}$ beam. 
\\(a--c) Propagated $E_x, E_y, E_z$ amplitudes at $z = 0.103\,\mathrm{m}$ using BLASM (a), BEASM (b), and E-ASM (c). 
\\(d) Component-wise SNR performance across different algorithms.}
  \label{fig8}  
\end{figure}\\
\noindent The conventional ASM result is omitted here because it yields no meaningful signal due to severe numerical artifacts, as indicated by the SNR plot in Fig. \ref{fig8}(d). Unlike the fundamental Gaussian and $\mathrm{HG}_{21}$ cases, BEASM also suffers severe artifacts here, primarily because the vortex beam carries substantial high-spatial-frequency content. The same trend appears between the fundamental Gaussian and $\mathrm{HG}_{21}$ beams: the richer high-frequency content of $\mathrm{HG}_{21}$ slightly lowers its SNR.\\
In this specific scenario, because the Fresnel number is 20.36 (which is much greater than unity), the state-of-the-art BEASM fails to handle the near-field propagation of the vortex beam. In contrast, only E-ASM accurately captures all three vector components ($E_x$, $E_y$, and $E_z$) of the propagated vortex beam. Since full 2D Gaussian quadrature of the vector potential seeding method 
remains computationally prohibitive, our ASM-compatible vectorial framework, powered by E-ASM, opens a promising route to direct, efficient simulation of long-range propagation for higher-order vector 
beams and complex structured light.\\
\subsection{Symmetrization and the Stratton--Chu Integral}
\begin{figure}[htbp]
  \centering
  \includegraphics[width=0.49\textwidth, page=18]{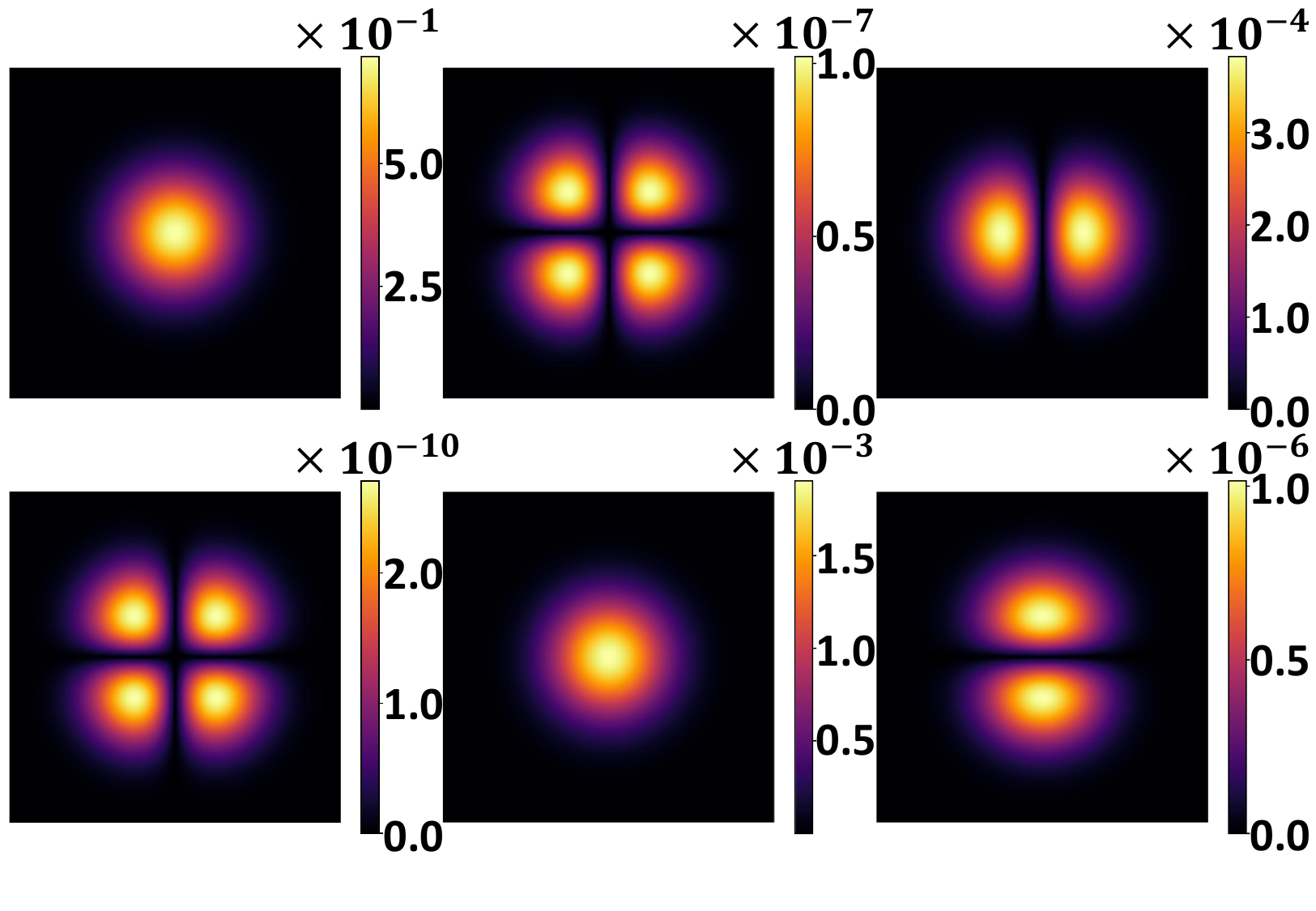}
  \includegraphics[width=0.49\textwidth, page=19]{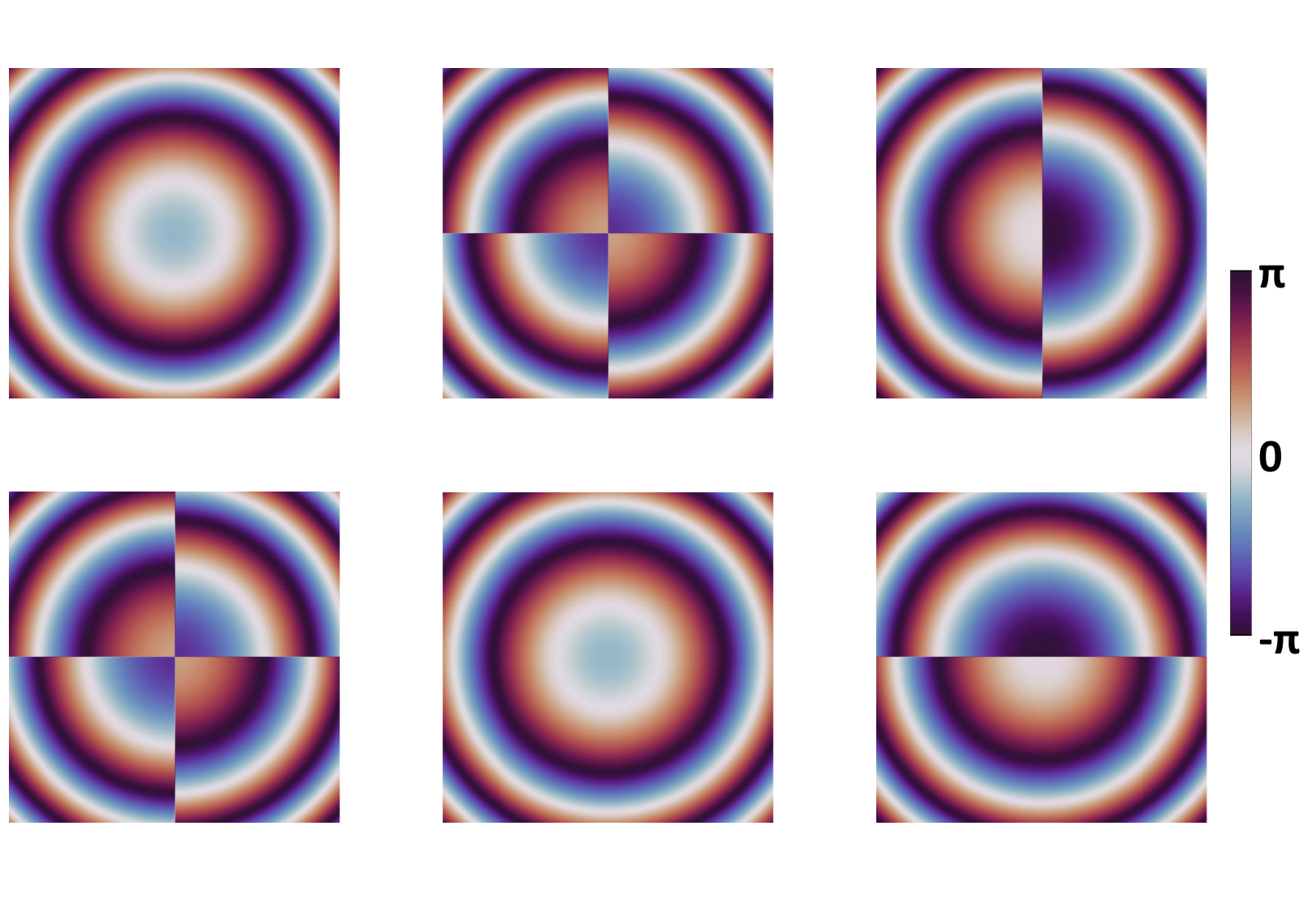}
  \caption{Propagated Gaussian vector fields via the Q-Stratton--Chu integral using E-ASM. 
    \\The six-component distribution at $z = 0.103$ m displays the electric fields ($E_x, E_y, E_z$) in the top row and the magnetic fields ($H_x, H_y, H_z$) in the bottom row, ordered left-to-right.}
  \label{fig9}  
\end{figure}

Because the standalone $\mathbf{A}$-seeding and $\mathbf{F}$-seeding formulations individually yield only five non-zero field components, averaging the two is a standard way to obtain symmetric, full six-component fields. Since the $\mathbf{A}$-seeding and $\mathbf{F}$-seeding formulations are gauge-equivalent with surface equivalence principles under the Lorenz gauge, we employ Eqs. \eqref{Matrix Schelkunoff PEC E}--\eqref{Matrix Schelkunoff PEC H} and Eqs. \eqref{Matrix Schelkunoff PMC E}--\eqref{Matrix Schelkunoff PMC H}. As detailed in Section 2.3, we simultaneously implement the $\mathbf{H}$-driven PEC and $\mathbf{E}$-driven PMC formulations: the PEC solution follows Eqs. \eqref{PEC E}--\eqref{PEC H}, and the PMC solution Eqs. \eqref{PMC E}--\eqref{PMC H}.
\begin{equation}
    \begin{pmatrix} \tilde{E}_x \\ \tilde{E}_y \\ \tilde{E}_z \end{pmatrix} 
    =Z\begin{pmatrix} 
    0 & \sigma_z & -\sigma_y \\ 
    -\sigma_z & 0 & \sigma_x \\ 
    \sigma_y & -\sigma_x & 0 
    \end{pmatrix} 
    \begin{pmatrix} \tilde{H}_x^\text{inc} \\ \tilde{H}_y^\text{inc} \\ \tilde{H}_z^\text{inc} \end{pmatrix} e^{ik_z z}
    \label{PEC E}
\end{equation}

\begin{equation}
    \begin{pmatrix} \tilde{H}_x \\ \tilde{H}_y \\ \tilde{H}_z \end{pmatrix}=
    \begin{pmatrix}
        (1-\sigma_x^2) & -\sigma_x \sigma_y & -\sigma_x \sigma_z \\
        -\sigma_x \sigma_y & (1-\sigma_y^2) & -\sigma_y \sigma_z \\
        -\sigma_x \sigma_z & -\sigma_y \sigma_z & (1-\sigma_z^2)
    \end{pmatrix}
    \begin{pmatrix} \tilde{H}_x^\text{inc} \\ \tilde{H}_y^\text{inc} \\ \tilde{H}_z^\text{inc} \end{pmatrix} e^{ik_z z}
    \label{PEC H}
\end{equation}

\begin{equation}
    \begin{pmatrix} \tilde{E}_x \\ \tilde{E}_y \\ \tilde{E}_z \end{pmatrix} = 
    \begin{pmatrix}
        (1-\sigma_x^2) & -\sigma_x \sigma_y & -\sigma_x \sigma_z \\
        -\sigma_x \sigma_y & (1-\sigma_y^2) & -\sigma_y \sigma_z \\
        -\sigma_x \sigma_z & -\sigma_y \sigma_z & (1-\sigma_z^2)
    \end{pmatrix}
    \begin{pmatrix} \tilde{E}_x^\text{inc} \\ \tilde{E}_x^\text{inc} \\ \tilde{E}_x^\text{inc} \end{pmatrix}e^{ik_z z}
    \label{PMC E}
\end{equation}

\begin{equation}
    \begin{pmatrix} \tilde{H}_x \\ \tilde{H}_y \\ \tilde{H}_z \end{pmatrix} = 
    \frac{1}{Z} \begin{pmatrix} 
        0 & -\sigma_z & \sigma_y \\ 
        \sigma_z & 0 & -\sigma_x \\ 
        -\sigma_y & \sigma_x & 0 
    \end{pmatrix} 
    \begin{pmatrix} \tilde{E}_x^\text{inc} \\ \tilde{E}_x^\text{inc} \\ \tilde{E}_x^\text{inc} \end{pmatrix}e^{ik_z z}
    \label{PMC H}
\end{equation}
\\
To verify the unified theory, the plane-wave spectrum of the vector beam is initialized at the waist plane as $(\tilde{E}_x^\text{inc}, \tilde{E}_x^\text{inc}, \tilde{E}_x^\text{inc}) = (\Psi, 0, 0)$ and $(\tilde{H}_x^\text{inc}, \tilde{H}_y^\text{inc}, \tilde{H}_z^\text{inc}) = (0, \Psi/Z, 0)$, leveraging gauge equivalence to bypass traditional $\mathbf{A}$- and $\mathbf{F}$-seeding schemes. The six field components of the fundamental Gaussian beam at $z = 0.103$ m plane are calculated by averaging the Eqs.~\eqref{PEC E}, \eqref{PMC E} and \eqref{PEC H}, \eqref{PMC H}. The symmetric solutions at the beam waist for fundamental Gaussian, $\text{HG}_{21}$, and $\text{LG}_{20}$ beams are provided in Section 12 of the Supplementary Information. Since this averaging operation mirrors the reconstruction of the Stratton--Chu integral from the combination of PEC and PMC models, we term this method the Quasi-Stratton--Chu (Q-Stratton--Chu) integral. It differs slightly from the rigorous Stratton--Chu integral due to the unphysical input field; a physical input would eliminate any discrepancy between the PEC and PMC solutions. Nevertheless, each combination remains an exact Maxwell solution because it stems directly from the surface equivalence principle. As verified in Tables~S2--S9 within Section 10 of the Supplementary Information, all individual solutions and their combinations maintain perfect Maxwell consistency.\\
The setup features a $1024 \,\mathrm{\mu m}\times 1024\,\mathrm{\mu m}$ window with $2048 \times 2048$ pixels, a beam waist of $w_0 = 128\,\mathrm{\mu m}$, and a wavelength of $\lambda = 500$ nm. Fig. \ref{fig9} displays the six-component distribution of the fundamental Gaussian at $z = 0.103$ m. The free-space transfer function, characterized by the $e^{ik_z z}$ factor, is computed using E-ASM. As concluded in the preceding section, the results exhibit no observable aliasing or numerical artifacts, avoiding the massive computational overhead of conventional Gaussian quadrature numerical integration.\\

\section{conclusion}\label{sec12}
The results above establish two things. First, the major vectorial diffraction formalisms are not distinct physical theories but alternative representations of the same boundary-value problem, each of which can be evaluated exactly by a small number of scalar angular-spectrum operations. Second, E-ASM performs those operations at an accuracy and speed unattainable with existing scalar solvers. The value of this combination extends beyond the present work. Because the vectorial framework operates entirely through scalar angular-spectrum operations, any future improvement in scalar propagation will raise the fidelity of every vectorial formalism without requiring changes to the vectorial layer.\\
The accuracy of E-ASM comes from the pre-compensating function, which varies far more slowly than the ASM or Fresnel transfer functions and therefore allows a maximum spatial frequency orders of magnitude higher than that of BEASM. Because the Fresnel step is evaluated as a real-space convolution, the observation window can be placed, sized, and sampled freely, and the fields are computed only where they are needed. This convolution can be evaluated directly through the MTP at $O(N^3)$ cost, or accelerated to $O(N^2 \log N)$ by kernel factorization in the point-source mode on uniform grids. In free-space benchmarks, E-ASM exceeds BEASM by more than 100 dB where the band-limit is inactive and stays close to 60 dB once band-limiting removes the highest spatial frequencies, while running an order of magnitude faster.\\
The two implementation modes of E-ASM have different ranges of validity. The point-source model—the faster of the two, thanks to kernel factorization—requires the propagation distance to exceed a minimum critical value to avoid undersampling the kernel; in our benchmark, this threshold corresponds to a Fresnel number above 100, deep in the near field. The zero-order-hold model, with its analytic closed-form kernel, has no such constraint and is valid at all distances. In both modes, the observation window must stay within the artifact-free region of width $M_{\text{max}} \cdot L$. E-ASM is also fully compatible with existing ASM variants, so FFT-based methods such as BLASM can be deployed at ultra-short distances when speed matters most.\\
Together, the unified theory and E-ASM extend rigorous vectorial field computation into regimes that were previously out of reach. Exact Maxwell vector beams of high order are propagated over macroscopic distances, where quadrature-based methods are prohibitively slow, with their cross-polarized and longitudinal components resolved at high fidelity. E-ASM can also serve as the missing link between simulation domains that have so far been treated separately. Ray-based methods handle bulk optics and full-wave solvers handle metasurface scattering, but the macroscopic vectorial propagation between them has lacked a solver that is simultaneously rigorous and fast enough for practical multi-scale use. Coupling through Gaussian beamlet decomposition \cite{harvey2015modeling, ashcraft2020open} or field tracing \cite{wyrowski2011introduction} would close that gap. Because the entire computation is differentiable and runs on GPU, such hybrid pipelines can be optimized end to end, whether for systems combining bulk optics with metasurfaces or for the simultaneous inverse design of multiple metasurface layers.\\
In conclusion, we have unified the fragmented landscape of vectorial diffraction theory, showing that all of its formalisms ultimately reduce to scalar angular-spectrum computation. E-ASM makes that foundation accurate and fast enough to be used without much compromise. Vectorial field computation that once required slow, case-by-case treatment now proceeds as a standard differentiable operation, available to the broader optics community.

\section{Materials and methods}
\subsection{Computational environment}
All the calculations are done on a high-performance desktop with an Intel i5-14600K, 32 GB RAM, and NVIDIA RTX 3090 Ti. The operating system is Ubuntu 22.04 of the Windows subsystem for Linux (WSL 2) with Windows 11. The code is written in both CuPy and PyTorch for GPU computation.

\section{Acknowledgments}
This work is supported by National Research Foundation of Korea (NRF) grants funded by the Korea government (MSIT) (RS-2024-00414119, RS-2025-02217649).

\section{Author contribution}
D.K. conceived the idea, implemented the code, and performed the numerical simulations; J.S. supervised the project. D.K. and J.S. conducted the theoretical analyses and prepared the manuscript.

\section{Data availability statement}
A Python+PyTorch implementation of the E-ASM and the vectorial diffraction theory can be found on our \href{https://codeberg.org/KDH136/E-ASM_and_vectorial_diffraction_theory}{Codeberg repository}.

\section{Conflict of interest}
The authors declare that they have no conflict of interest.
\clearpage

\textbf{supplementary text}

\setcounter{equation}{0}    
\setcounter{figure}{0}      
\setcounter{table}{0}       
\setcounter{section}{0}
\renewcommand{\theequation}{S\arabic{equation}}
\renewcommand{\thefigure}{S\arabic{figure}}
\renewcommand{\thetable}{S\arabic{table}}
\renewcommand{\thesection}{S\arabic{section}}

\section{Maximum magnification and minimum critical propagation distance of the E-ASM and the limits of the observation window}\label{secS1}
\begin{figure}[htbp]
  \centering
  \includegraphics[width=0.9\textwidth, page=20]{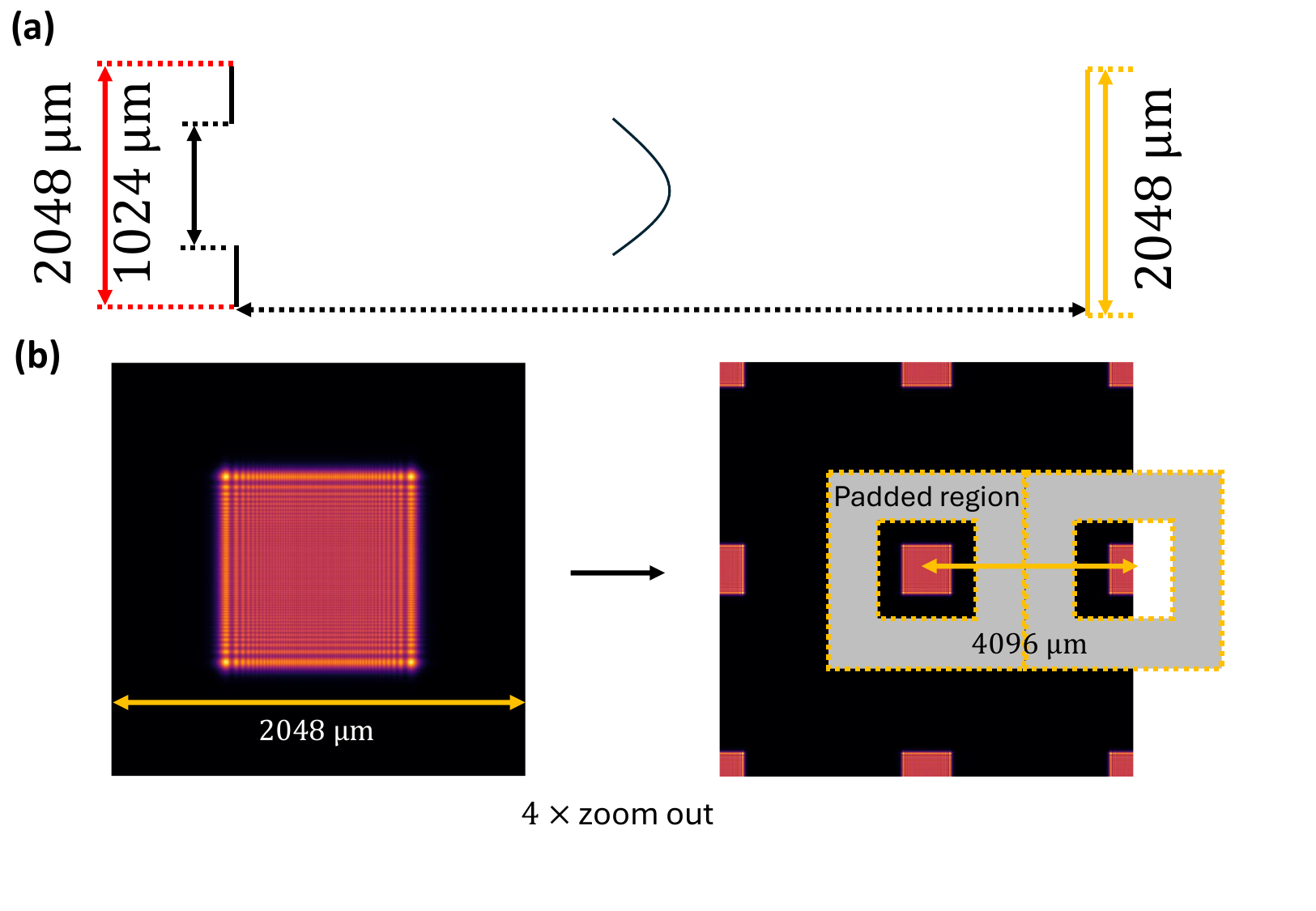}
  \includegraphics[width=0.9\textwidth, page=21]{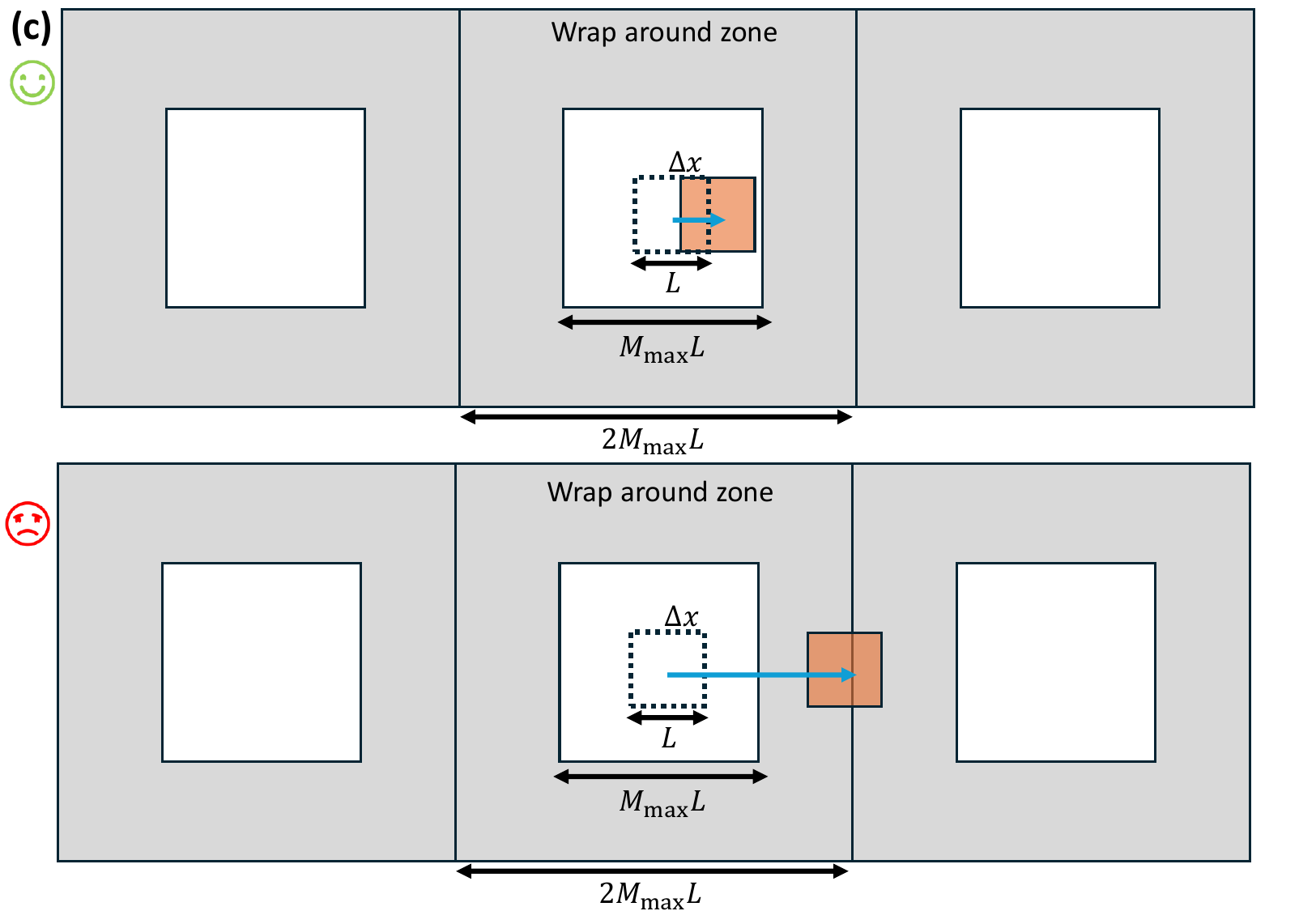}
  \caption{Observation window selection rule and the replicas. (a) The simulation setup at the minimum critical propagation distance. (b) The result with unity magnification, and the lower-right panel shows the result after $4\times$ zoom-out. (c) The observation window (orange box) selection rule. Only the white region inside the gray-shaded zone is free from aliasing and wrap-around artifacts.}
  \label{figS1}
\end{figure}

The discrete input signal is periodic in the Fourier domain with period $1/\Delta x$. Unlike the angular spectrum method, whose evanescent cutoff suppresses aliased components beyond the free-space equi-frequency contour, the Fresnel transfer function carries no such decay; every replica propagates without attenuation.
For an input field sampled on $N$ points with pitch $\Delta x$ and zero-padded to $L_p = 2L$ ($L = N\Delta x$), the first-order replica center is mapped to \begin{equation}\label{eq:replica_center}
    x_{\text{replica}} = \frac{\lambda z}{\Delta x}
\end{equation}
in the observation plane. Fig.~\ref{figS1}(b) shows the replica pattern after a 4$\times$ zoom-out at the propagation distance where the maximum magnification is unity. Because the zero-padded array doubles the effective sampling density in the Fourier domain, the replica spacing is $\lambda z / \Delta x = 2\,M_{\text{max}}\,L$, where
\begin{equation}\label{eq:Mmax}
    M_{\text{max}} \;=\; \frac{\lambda z}{2\,L\,\Delta x}
                   \;=\; \frac{N\,\lambda\,z}{2\,L^2}\,,
\end{equation}
which coincides with the magnification of the scalable angular spectrum (SAS) method~\cite{heintzmann2023scalable}. Replica overlap is therefore avoided as long as the observation window stays within $2\,M_{\text{max}}\,L$.\\
However, the pre-compensation step of the E-ASM is performed via FFT on the zero-padded array of length $L_p = 2L$ and therefore computes a circular convolution. For an input whose support is confined to the central $L$, the circular convolution coincides with the linear convolution only within the central $L$ of the output; the remaining margin of width $L/2$ on each side may contain wrap-around artifacts. Consequently, the artifact-free region after pre-compensation is at most $L$, and after the subsequent Fresnel propagation with magnification $M$ it becomes $M\cdot L$. The usable observation window is therefore \begin{equation}\label{eq:obs_window}
    W_{\text{obs}} \;\le\; M_{\text{max}}\cdot L\,,
\end{equation}
which is half the replica-free range.\\
In the E-ASM, this center crop is realized not by discarding array elements, but by restricting the pointwise evaluation of the Fresnel propagation matrices to the observation window, which inherently excludes the contaminated margin. Note that the crop must be applied after Fresnel propagation, not before: the pre-compensation shifts each spatial-frequency component $f_x$ laterally by $\delta_x = z(\tan\theta - \sin\theta)$ (with $\sin\theta = 
\lambda f_x$), pushing energy from the source region 
into the zero-padded margin. The subsequent Fresnel propagation returns this energy to the observation window by an opposing shift of $z\sin\theta$. Cropping before propagation would remove these displaced components, degrading accuracy within the RoI.\\
The Fresnel propagation kernel in the E-ASM involves a quadratic phase factor $\exp\!\bigl(i\pi x^2/(\lambda z)\bigr)$, whose local spatial frequency increases linearly with $|x|$. In the point-model implementation, this kernel is sampled directly on the computational 
grid with pitch $\Delta x$. At the edge of the zero-padded array ($|x|=L$), the phase increment between adjacent samples is
\begin{equation}\label{eq:phase_inc}
    \Delta\phi \;=\; \frac{2\pi\,L\,\Delta x}{\lambda z}
               \;=\; \frac{\pi}{M_{\text{max}}}\,.
\end{equation}
The Nyquist condition $\Delta\phi < \pi$ requires 
$M_{\text{max}}>1$, or equivalently
\begin{equation}\label{eq:zmin}
    z \;>\; z_{\min} \;=\; \frac{2\,L\,\Delta x}{\lambda}
                     \;=\; 2RL\,,
\end{equation}
where $R=\Delta x/\lambda$. This is identical to the minimum critical propagation distance of the scalable angular spectrum method~\cite{heintzmann2023scalable} and arises from the same underlying constraint: under-sampling of the propagation kernel's quadratic phase. For $z < z_{\min}$, the kernel is aliased and the point-model result is contaminated by artifacts. The zero-order hold (ZoH) model evaluates the Fresnel propagation integral in analytic closed form by treating each input pixel as a uniform rectangular source. Because no point-sampled quadratic phase appears in the computation, the kernel under-sampling problem of Eq.~\eqref{eq:phase_inc} is eliminated entirely: the ZoH model produces accurate results even for $z < z_{\min}$.
The maximum-magnification constraint of Eq.~\eqref{eq:obs_window}, however, remains in force. It originates from the periodicity of the discrete Fourier transform used in the pre-compensation step and is independent of the propagation model. At short propagation distances, $M_{\text{max}}$ falls below unity and the usable observation window shrinks below the original aperture size $L$, limiting the practical utility of near-field evaluation. In summary, the ZoH model extends the E-ASM to propagation distances below the critical distance $z_{\min}$, but the observation window contracts as $M_{\text{max}}\cdot L = N\lambda z/(2L)$, so that very short distances yield progressively smaller fields of view. Figure~\ref{figS1}(c) summarizes the permissible observation region: only positions within the white area, bounded by the replica-free range and the center-cropped margin, yield artifact-free results. The surrounding gray region is contaminated by either replica overlap or wrap-around from the circular convolution in the pre-compensation step.\\
Next, we derive the vignetting-free condition for the E-ASM. Unlike the SAS, the E-ASM can freely choose the observation window, so there is no maximum propagation distance. However, for naturally divergent signals, vignetting can occur: spatial frequencies that should contribute to the destination RoI may be suppressed by the pre-compensation bandlimit. Below we derive the condition under which this does not happen.\\
The pre-compensation phase $\phi_{2\mathrm{D}} = \arg(H_{\mathrm{AS}}H_{\mathrm{Fr}}^{*})$ is applied on the zero-padded grid of size $L_p = 2L$.
Its local signal frequency must not exceed the sampling rate; the resulting Nyquist condition on each axis is~\cite{heintzmann2023scalable}
\begin{equation}
  \left|\frac{s_x}{\sqrt{1 - s_x^2 - s_y^2}} - s_x\right| \leq \frac{L}{z}, \qquad
  \left|\frac{s_y}{\sqrt{1 - s_x^2 - s_y^2}} - s_y\right| \leq \frac{L}{z},
  \label{eq:nyquist}
\end{equation}
where $s_x = \lambda f_x$ and $s_y = \lambda f_y$ are direction sines.
The left-hand side is the difference between the exact lateral shift per unit distance ($\tan\theta$) and the Fresnel approximation ($\sin\theta$), which is precisely the quantity corrected by the pre-compensation.\\
The source field occupies $[-L/2,\; L/2]^2$ (before padding), and the destination RoI is $[-ML/2,\; ML/2]^2$.
Since the overall cascade is equivalent to AS propagation, a plane wave with direction sine $s_x$ undergoes a lateral shift $\delta_x = z\,s_x/\sqrt{1-s_x^2-s_y^2}$.
The most extreme frequency needed by the RoI corresponds to a ray emitted from one edge of the source reaching the opposite edge of the destination:
\begin{equation}
  |\delta_x| = \frac{(M+1)L}{2}, \qquad |\delta_y| = \frac{(M+1)L}{2}.
  \label{eq:corner_shift}
\end{equation}
Because the two Nyquist conditions in Eq.~\eqref{eq:nyquist} are checked independently per axis, the most restrictive point is the RoI corner ($\delta_x = \delta_y$): at fixed $\delta_x$, a nonzero $s_y$ reduces $\sqrt{1-s_x^2-s_y^2}$, steepening the pre-compensation phase gradient along $x$.\\
At the corner, symmetry gives $s_x = s_y \equiv s$, so Eq.~\eqref{eq:corner_shift} becomes
\begin{equation}
  \frac{s}{\sqrt{1 - 2s^2}} = \frac{(M+1)L}{2z},
  \label{eq:corner_tangent}
\end{equation}
with solution
\begin{equation}
  s = \frac{(M+1)L}{\sqrt{4z^2 + 2(M+1)^2 L^2}}.
  \label{eq:s_corner}
\end{equation}
Substituting into the $x$-component of Eq.~\eqref{eq:nyquist} and using Eq.~\eqref{eq:corner_tangent} to replace $s/\sqrt{1-2s^2}$ by $(M+1)L/(2z)$:
\begin{equation}
  \frac{(M+1)L}{2z} - \frac{(M+1)L}{\sqrt{4z^2 + 2(M+1)^2 L^2}} \leq \frac{L}{z}.
\end{equation}
Dividing both sides by $L/z$ yields the vignetting-free condition:
\begin{equation}
  \frac{M+1}{2}\!\left(1 - \frac{2z}{\sqrt{4z^2 + 2(M+1)^2 L^2}}\right) \leq 1.
  \label{eq:vignetting_condition}
\end{equation}
Given $z$ and $L$, the maximum vignetting-free magnification $M$ is determined by the equality case of Eq.~\eqref{eq:vignetting_condition}.\\
For unit magnification ($M=1$), the left-hand side reduces to $1 - 2z/\sqrt{4z^2+8L^2} < 1$ for all $z>0$, so equal magnification is always vignetting-free.
For large magnification ($M \gg 1$), expanding to leading order gives
\begin{equation}
  M \;\lesssim\; 2 + \frac{\sqrt{2}\,z}{L}\,,
\end{equation}
so the maximum allowed magnification grows linearly with propagation distance, following the same trend as the replica-free magnification of Eq.~\eqref{eq:Mmax}.\\
We note that for unit magnification, Eq.~\eqref{eq:vignetting_condition} is satisfied for all $z>0$, so the E-ASM operates without vignetting at any propagation distance—in contrast to the SAS, whose magnification is locked to the propagation distance and therefore inevitably encounters vignetting beyond the limit set by Eq.~\eqref{eq:vignetting_condition}. More generally, Eq.~\eqref{eq:vignetting_condition} is relevant only when the signal diverges upon propagation (e.g., diffraction from a finite aperture); for confined or converging fields whose angular content remains well within the pre-compensation bandlimit, the maximum magnification is only limited by the replicas, not vignetting.

\clearpage

\section{The two results of the scalar diffraction benchmark}\label{secS2}

\begin{figure}[htbp]
  \centering
  \includegraphics[width=0.8\textwidth, page=22, trim = 0cm 1cm 0cm 0cm, clip]{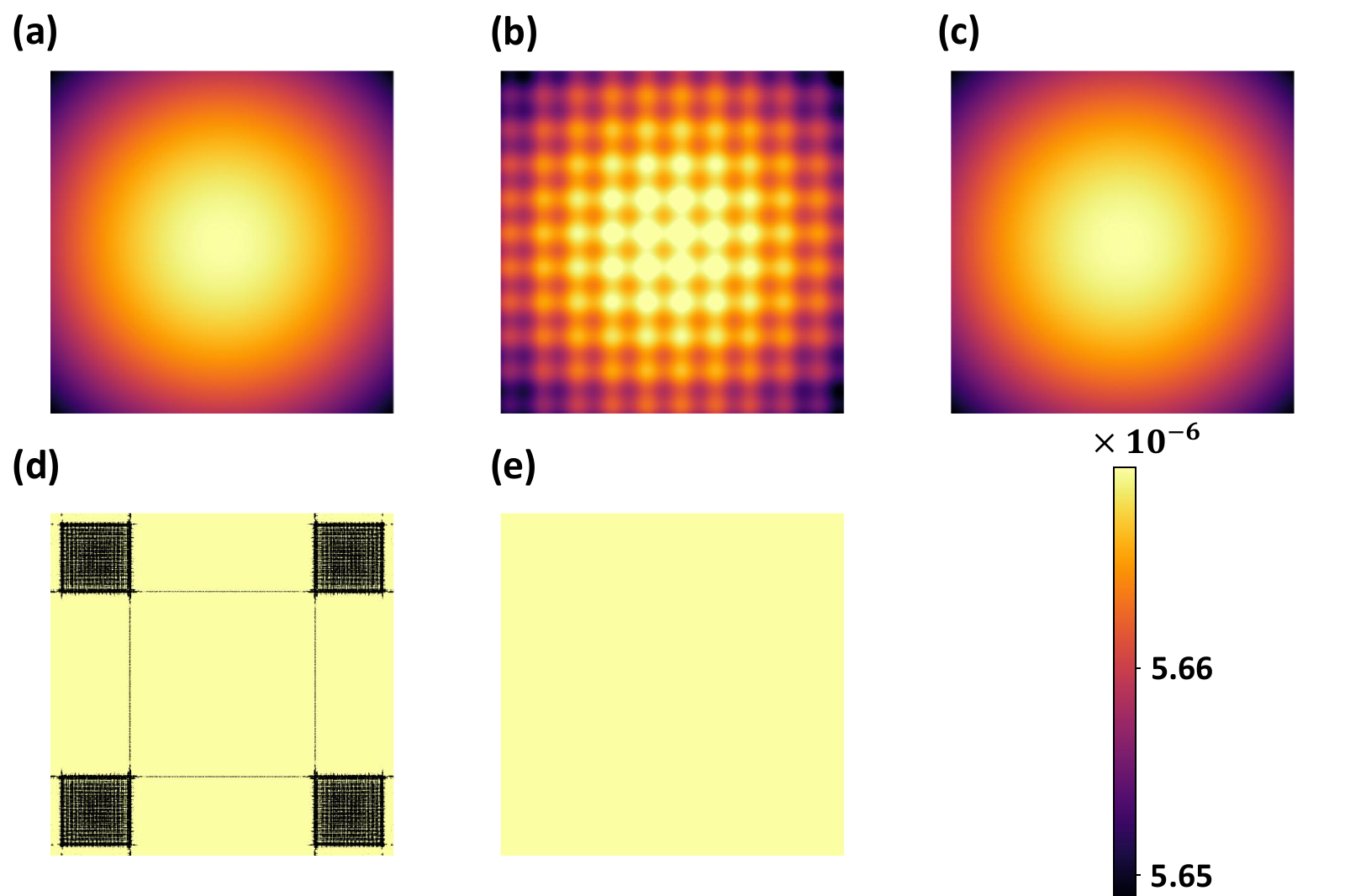}  
  \includegraphics[width=0.8\textwidth, page=23, trim = 0cm 1cm 0cm 0cm, clip]{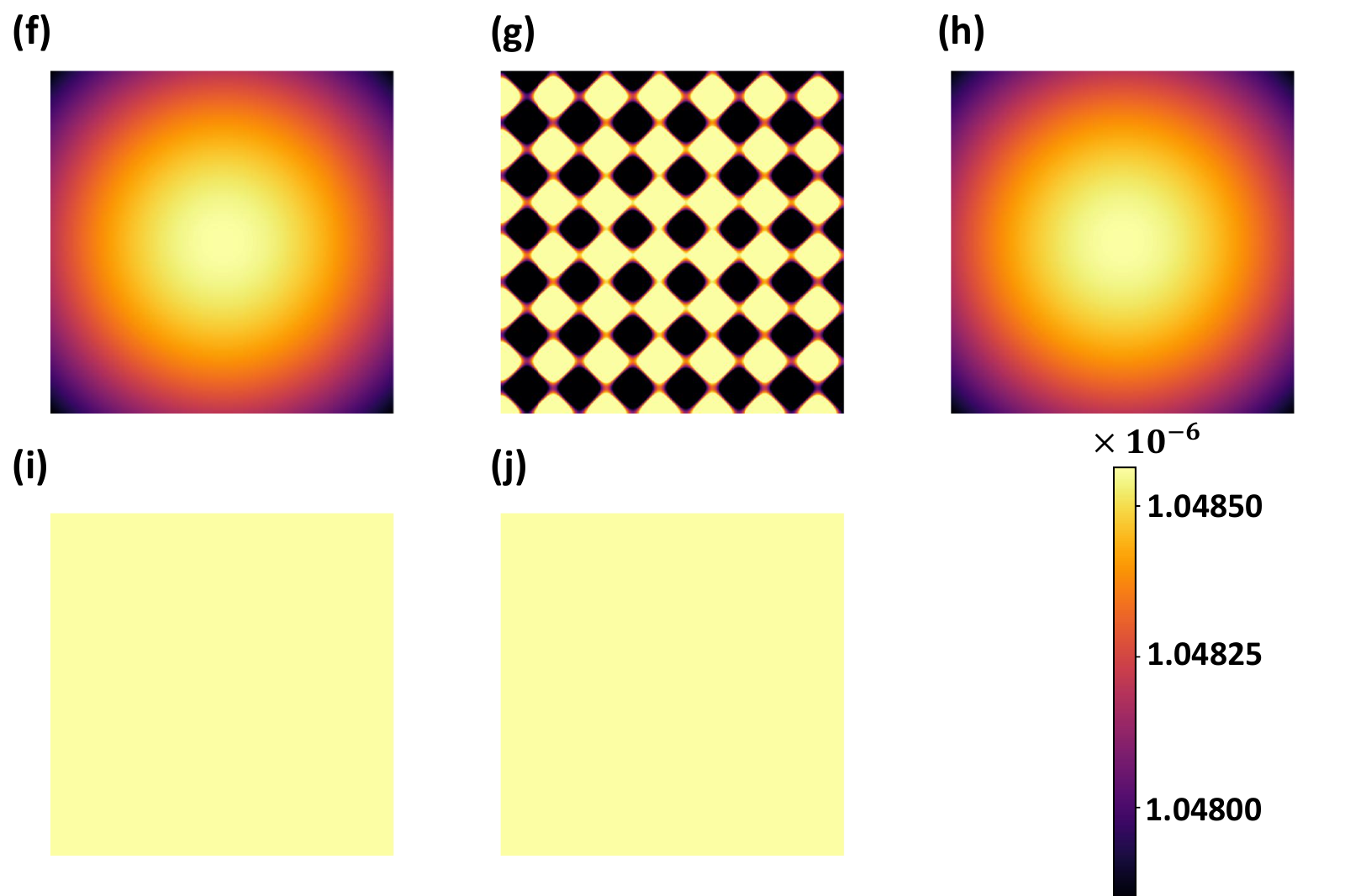}  
  \caption{Benchmark results of the scalar diffraction theory at different propagation distances. 
  (a--e) Numerical field distributions sampled at $z = 88.07$ m calculated by GT (a), BEASM (b), E-ASM (c), ASM (d), and BLASM (e). 
  (f--j) Corresponding benchmark distributions sampled at the far-field distance of $z = 204.8$ m via GT (f), BEASM (g), E-ASM (h), ASM (i), and BLASM (j).}
  \label{figS2}
\end{figure}
\clearpage

\section{Simulation results of optically addressed local qubit gates under ``All-on'' and ``Center-off'' operation modes}\label{secS3}
\begin{figure}[htbp]
  \centering
  \includegraphics[width=1.0\textwidth, page=24]{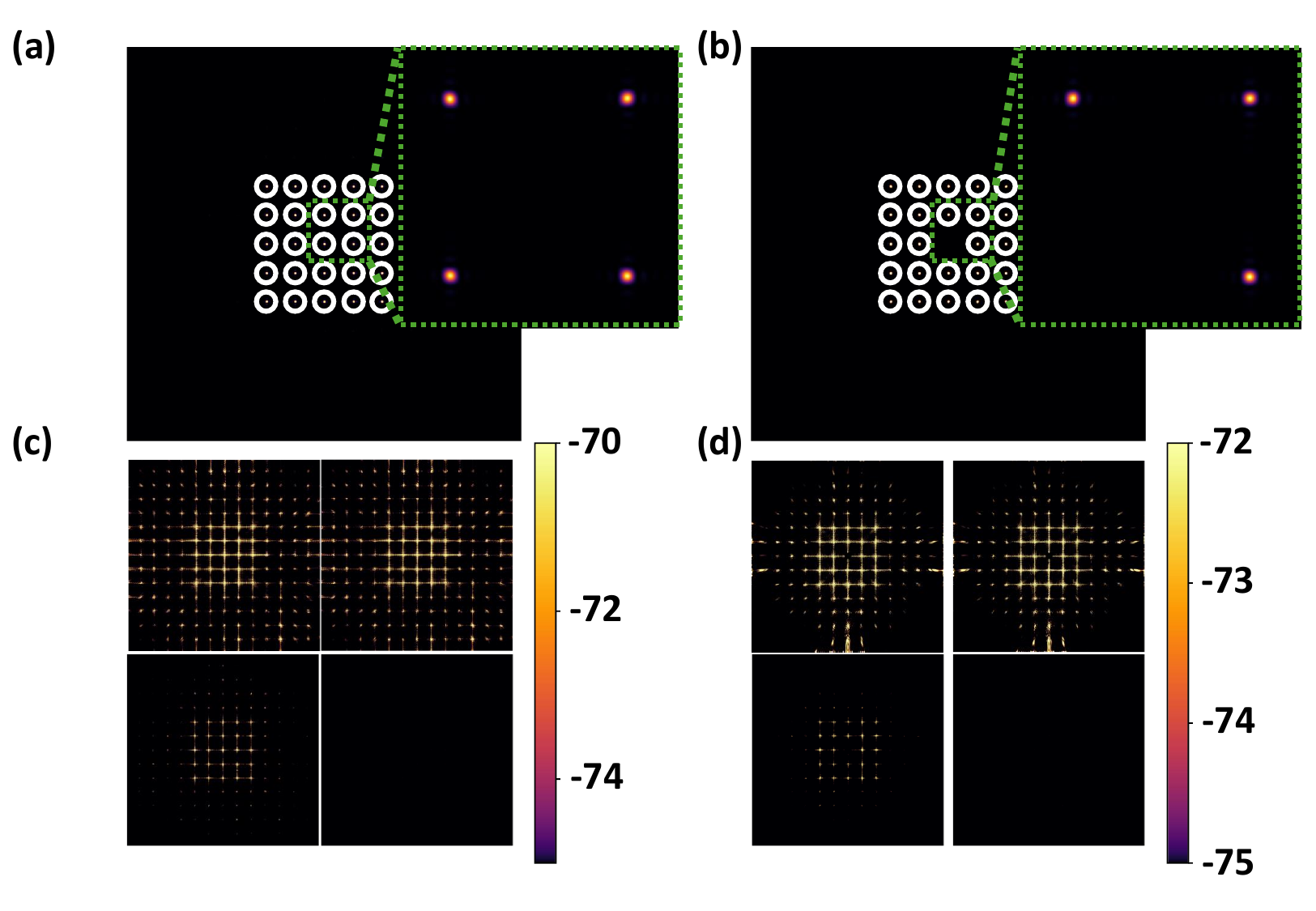}  
  \caption{Numerical simulation results of optically addressed local qubit gates under different operation modes. 
  \\(a--b) Reference intensity profiles for the ``All-on'' (a) and ``Center-off'' (b) modes. The main plots are scaled with a non-linear power law ($I^{1/8}$) to enhance overall visibility, whereas the magnified insets inside the green dashed boxes present the linear intensity distribution ($I$).
  \\(c--d) Corresponding logarithmic reconstruction errors on a dB scale for the ``All-on'' (c) and ``Center-off'' (d) modes.
  \\For each sub figure template, the four quadrants represent the results calculated by the conventional ASM (top-left), BLASM (top-right), BEASM (bottom-left), and the proposed E-ASM (bottom-right), respectively.}
  \label{figS3}
\end{figure}
\clearpage

\section{Equivalence of Stratton-Chu integral and Luneburg integral}\label{secS4}

This section establishes the mathematical equivalence among Love’s and Schelkunoff’s equivalence principles, as well as the Luneburg and Stratton-Chu integrals. The key variables employed throughout the unification of the vectorial diffraction theory are summarized below.

\begin{tabular}{ll}    
    $\sigma_x, \sigma_y, \sigma_z$ & Direction cosine of $\mathbf{k}$ vector\\
    $Z$ & Impedance ($\Omega$) \\
    $\omega$ & Angular frequency\\
    $\mu$ & Permeability of the medium \\
    $\varepsilon$ & Permittivity of the medium \\
    $\mathbf{r}$ & Position vector of the observation point \\
    $\mathbf{r}'$ & Position vector of the source point \\
    $R$ & Distance between source and observation points, $R = |\mathbf{r} - \mathbf{r}'|$ \\
    $\Sigma$ & Aperture region \\
    $G$ & Scalar Green's function \\
    $\hat{\mathbf{n}}$ & Unit normal vector to the surface \\
    $\hat{\mathbf{z}}$ & Unit normal vector to the XY plane \\
    $\lambda$ & Wavelength of the incident light \\
    $\mathbf{E}^\text{inc}, \mathbf{H}^\text{inc}$ & Incident electric and magnetic fields at the source plane\\
    $E_x^\text{inc},E_y^\text{inc}, E_z^\text{inc}, H_x^\text{inc}, H_y^\text{inc}, H_z^\text{inc}$ & Components of incident electric and magnetic fields\\
    $\tilde{E}_x^\text{inc},\tilde{E}_x^\text{inc}, \tilde{E}_x^\text{inc}, \tilde{H}_x^\text{inc}, \tilde{H}_y^\text{inc}, \tilde{H}_z^\text{inc}$ & The Fourier spectrum of each component\\
    $\mathbf{J}, \mathbf{M}$ & Electric current and magnetic current\\
    $\rho_e, \rho_m$ & Electric charge and magnetic charge density\\
    $\mathbf{F}, \mathbf{A}$ & Electric and magnetic vector potential\\
    $\Phi_e, \Phi_m$ & Electric and magnetic scalar potential
    \end{tabular}\\
\\
The unification of diverse vectorial diffraction theories is predicated upon the symmetric Maxwell equations, the Lorenz gauge, and the surface equivalence principle. A central premise of this framework is the observation that the average of the first and second Rayleigh-Sommerfeld integrals ($RS_1$ and $RS_2$) is mathematically identical to the Kirchhoff-Helmholtz integral \cite{Goodman}. This equivalence arises from the mutual cancellation of image sources—a structural symmetry that is similarly manifested in vectorial diffraction theory. Furthermore, while the Stratton-Chu integral predates the Franz formula, the former can be readily derived from the latter. The Franz formula itself is established by introducing hypothetical magnetic monopoles and applying the Lorenz gauge. Adopting the $\exp(-i\omega t)$ time-harmonic convention, the symmetric Maxwell equations \cite{Jackson, Balanis} are formulated as follows:
\begin{alignat}{4}
    \nabla \cdot \mathbf{D}_e &= \rho_e, \quad & \nabla \times \mathbf{E}_e &= i\omega\mathbf{B}_e, \quad & \nabla \cdot \mathbf{B}_e &= 0, \quad & \nabla \times \mathbf{H}_e &= -i\omega\mathbf{D}_e + \mathbf{J}
    \label{symmetric maxwell}
\end{alignat}
\begin{alignat}{4}   
    \nabla \cdot \mathbf{D}_m &= 0, \quad & \nabla \times \mathbf{E}_m &= i\omega\mathbf{B}_m - \mathbf{M}, \quad & \nabla \cdot \mathbf{B}_m &= \rho_m, \quad & \nabla \times \mathbf{H}_m &= -i\omega\mathbf{D}_m.
    \label{symmetric maxwell 2}
\end{alignat}\\
In this formulation, the total electric and magnetic fields are expressed as the superposition of the components originating from the electric and magnetic source equations, respectively.
\begin{align}
\mathbf{E}=\mathbf{E}_e+\mathbf{E}_m, \quad & \mathbf{H}=\mathbf{H}_e+\mathbf{H}_m
\end{align}
Under the Lorenz gauge, the vector and scalar potentials are related by
\begin{align}
\nabla \cdot \mathbf{A}=i\omega \mu \varepsilon \Phi_e \quad & \nabla \cdot \mathbf{F}=i\omega \mu \varepsilon \Phi_m.
\end{align}
Since the electric and magnetic vector potentials satisfy the Helmholtz equation,
\begin{align}
\left(\nabla^2+k^2\right) \mathbf{A}=-\mu\mathbf{J}, \quad & \left(\nabla^2+k^2\right) \mathbf{F}=-\varepsilon\mathbf{M}.
\end{align}
the electric and magnetic fields derived from the respective source equations in Eqs. \eqref{symmetric maxwell} and \eqref{symmetric maxwell 2} are given by \cite{Jackson, Balanis}

\begin{align}
\mathbf{B}_e=\nabla\times\mathbf{A}, \quad & \mathbf{E}_e=i\omega \mathbf{A}+\frac{i}{\omega\varepsilon\mu}\nabla\left(\nabla\cdot\mathbf{A}\right)\\
\mathbf{D}_m=-\nabla\times\mathbf{F}, \quad & \mathbf{H}_m=i\omega \mathbf{F}+\frac{i}{\omega\varepsilon\mu}\nabla\left(\nabla\cdot\mathbf{F}\right).
\end{align}\\
Consequently, the total electric and magnetic fields are obtained as the superposition of these components \cite{Jackson, Balanis}:

\begin{align}
\mathbf{E}=i\omega\mathbf{A}+\frac{i}{\omega\varepsilon\mu}\nabla\left(\nabla\cdot\mathbf{A}\right)-\frac{1}{\varepsilon}\nabla\times\mathbf{F}
\label{Lorenz E previous}
\end{align}\\
\begin{align}
\mathbf{H}=i\omega\mathbf{F}+\frac{i}{\omega\varepsilon\mu}\nabla\left(\nabla\cdot\mathbf{F}\right)+\frac{1}{\mu}\nabla\times\mathbf{A}.
\label{Lorenz H previous}
\end{align}\\
Given that $\mathbf{A}$ and $\mathbf{F}$ satisfy the Helmholtz equation, these potentials can be determined directly using the scalar Green’s function:
\begin{align}
\mathbf{A}=\mu\oiint_{s'}\mathbf{J}\left(\mathbf{r'}\right)\frac{\exp(ik|\mathbf{r}-\mathbf{r'}|)}{4\pi|\mathbf{r}-\mathbf{r'}|}ds'
\label{Freespace A}
\end{align}\\
\begin{align}
\mathbf{F}=\varepsilon\oiint_{s'}\mathbf{M}\left(\mathbf{r'}\right)\frac{\exp(ik|\mathbf{r}-\mathbf{r'}|)}{4\pi|\mathbf{r}-\mathbf{r'}|}ds'.
\label{Freespace F}
\end{align}\\
In a source-free region, the vector potentials $\mathbf{A}$ and $\mathbf{F}$ satisfy the homogeneous Helmholtz equation, which leads to \cite{Jackson, Balanis}
\begin{align}
    \mathbf{E} = \frac{i}{\omega\mu\varepsilon} \left[ \nabla \times \nabla \times \mathbf{A} \right] - \frac{1}{\varepsilon} \nabla \times \mathbf{F} \label{Lorenz E}\\
    \mathbf{H} = \frac{i}{\omega\mu\varepsilon} \left[ \nabla \times \nabla \times \mathbf{F} \right] + \frac{1}{\mu} \nabla \times \mathbf{A} \label{Lorenz H}.
\end{align}\\
The electric and magnetic current densities can be determined via three distinct equivalence principles: Love’s equivalence, Schelkunoff’s PEC equivalence, and Schelkunoff’s PMC equivalence. Love’s equivalence principle, often characterized as a zero-field equivalence, defines the electric and magnetic surface currents as $\mathbf{J} = \hat{\mathbf{n}}' \times \mathbf{H}^\text{inc}$ and $\mathbf{M} = \mathbf{E}^\text{inc} \times \hat{\mathbf{n}}'$, respectively. By substituting these current densities into Eqs. \eqref{Freespace A} and \eqref{Freespace F}, and subsequently inserting the resulting potentials into Eqs. \eqref{Lorenz E} and \eqref{Lorenz H}, the Franz formulas (Eqs. \eqref{Franz E} and \eqref{Franz H}) are recovered \cite{franz1948formulierung, tai1972kirchhoff}.\\
\begin{align}
    \mathbf{E} &= \nabla \times \oiint_S \left[ \mathbf{\hat{n}}' \times \mathbf{E}^\text{inc} \left( \mathbf{r}' \right) \right] G \, ds' - \frac{1}{i\omega\varepsilon} \nabla \times \nabla \times \oiint_S \left[ \mathbf{\hat{n}}' \times \mathbf{H}^\text{inc} \left( \mathbf{r}' \right) \right] G \, ds' \label{Franz E} \\
    \mathbf{H} &= \nabla \times \oiint_S \left[ \mathbf{\hat{n}}' \times \mathbf{H}^\text{inc} \left( \mathbf{r}' \right) \right] G \, ds' + \frac{1}{i\omega\mu} \nabla \times \nabla \times \oiint_S \left[ \mathbf{\hat{n}}' \times \mathbf{E}^\text{inc} \left( \mathbf{r}' \right) \right] G \, ds' \label{Franz H}
\end{align}\\
It should be noted that the Green’s function employed here is the free-space Green’s function. In the first term, since the curl operator involves derivatives with respect to the $\mathbf{r}'$ coordinates, the derivative of the electric field vanishes. Furthermore, by applying the identity $\nabla G = -\nabla' G$ inside the integral, the first term can be rewritten as
\begin{equation}
\oiint_S \nabla\times\left[\left[\hat{\mathbf{n}}'\times\mathbf{E}^\text{inc}(\mathbf{r}')\right]G\right] ds' = \oiint_S \left[\hat{\mathbf{n}}'\times\mathbf{E}^\text{inc}(\mathbf{r}')\right]\times\nabla'G \, ds'
\label{vector identity 1}
\end{equation}\\
Similarly, the second term can be simplified by
\begin{equation}
\oiint_S \nabla\times\nabla\times\left[\left[\hat{\mathbf{n}}'\times\mathbf{H}^\text{inc}(\mathbf{r}')\right]G\right] ds' = \oiint_S \left( k^2\left[\hat{\mathbf{n}}'\times\mathbf{H}^\text{inc}\right]G - i\omega\varepsilon\left[\hat{\mathbf{n}'}\cdot\mathbf{E}^\text{inc}\right]\nabla'G \right) ds'
\label{vector identity 2}
\end{equation}\\
Substituting Eqs. \eqref{vector identity 1} and \eqref{vector identity 2} into Eq. \eqref{Franz E} yields the Stratton-Chu formula \cite{tai1972kirchhoff, stratton1939diffraction}.
\begin{equation}
    \mathbf{E} = \oiint_S \left\{ i \omega \mu \left[ \mathbf{\hat{n}}' \times \mathbf{H}^\text{inc} \left( \mathbf{r}' \right) \right] G + \left[ \mathbf{\hat{n}}' \times \mathbf{E}^\text{inc} \left( \mathbf{r}' \right) \right] \times \nabla' G + \left[ \mathbf{\hat{n}}' \cdot \mathbf{E}^\text{inc} \right] \nabla' G \right\} ds'
    \label{Stratton-Chu E}
\end{equation}
Similarly, the $\mathbf{H}$ field is recovered as \cite{tai1972kirchhoff, stratton1939diffraction}
\begin{equation}
    \mathbf{H} = \oiint_S \left\{ -i \omega \varepsilon \left[ \mathbf{\hat{n}}' \times \mathbf{E}^\text{inc} \left( \mathbf{r}' \right) \right] G + \left[ \mathbf{\hat{n}}' \times \mathbf{H}^\text{inc} \left( \mathbf{r}' \right) \right] \times \nabla' G + \left[ \mathbf{\hat{n}}' \cdot \mathbf{H}^\text{inc} \right] \nabla' G \right\} ds'.
    \label{Stratton-Chu H}
\end{equation}\\
Given our focus on planar geometry, Eqs. \eqref{Stratton-Chu E} and \eqref{Stratton-Chu H} can be reformulated as Eqs. \eqref{Stratton-Chu E planar-2} and \eqref{Stratton-Chu H planar} under the Kirchhoff boundary condition \cite{kim2018calculation}.
\begin{equation}
    \mathbf{E} = \iint_{\Sigma} \left\{ i \omega \mu \left[ \mathbf{\hat{z}}' \times \mathbf{H}^\text{inc} \left( \mathbf{r}' \right) \right] G + \left[ \mathbf{\hat{z}}' \times \mathbf{E}^\text{inc} \left( \mathbf{r}' \right) \right] \times \nabla' G + \left[ \mathbf{\hat{z}}' \cdot \mathbf{E}^\text{inc} \right] \nabla' G \right\} ds'
    \label{Stratton-Chu E planar-2}
\end{equation}    
\begin{equation}
    \mathbf{H} = \iint_{\Sigma} \left\{ -i \omega \varepsilon \left[ \mathbf{\hat{z}}' \times \mathbf{E}^\text{inc} \left( \mathbf{r}' \right) \right] G + \left[ \mathbf{\hat{z}}' \times \mathbf{H}^\text{inc} \left( \mathbf{r}' \right) \right] \times \nabla' G + \left[ \mathbf{\hat{z}}' \cdot \mathbf{H}^\text{inc} \right] \nabla' G \right\} ds'.
    \label{Stratton-Chu H planar}
\end{equation}\\
Conversely, Schelkunoff’s PEC equivalence defines the electric and magnetic surface currents as $\mathbf{J}=0$ and $\mathbf{M}=2\mathbf{E}^\text{inc} \times \hat{\mathbf{z}}'$ for planar geometries. This formulation can be understood through the image theory associated with an infinite planar PEC boundary. It should be noted, however, that this approach cannot be generalized to arbitrary geometries, as analytical expressions for image currents do not exist for arbitrary shapes. Under Schelkunoff’s PEC equivalence, the magnetic vector potential $\mathbf{A}$ vanishes \cite{PhysRev.56.308}.
\begin{equation}
    \mathbf{E} = -\frac{1}{\varepsilon}\nabla\times\mathbf{F}
    \label{PEC E-S}
\end{equation}
\begin{equation}
    \mathbf{H} = \frac{i}{\omega\mu\varepsilon}\nabla\times\nabla\times\mathbf{F}
    \label{PEC H-S}
\end{equation}\\
By substituting Eq. \eqref{Freespace F} into Eq. \eqref{PEC E-S}, electric field becomes
\begin{equation}
    \mathbf{E} = -\nabla\times\iint_{\Sigma}2\left(\mathbf{E}^\text{inc}\times\hat{\mathbf{z}}'\right)Gds'=-2\iint_{\Sigma}\left(\hat{\mathbf{z}}'\times\mathbf{E}^\text{inc}\right)\times\nabla Gds'.
    \label{Luneburg E-S}
\end{equation}\\
Equation \eqref{Luneburg E-S} is, in fact, equivalent to the Luneburg integral. When expressed component-wise, it becomes evident that Eq. \eqref{Luneburg E-S} is mathematically identical to the Luneburg formulation \cite{baker2003mathematical}.
\begin{align}
        E_{x,y} = -2 \iint_{\Sigma} E^\text{inc}_{x,y} \left( \mathbf{r}' \right) \frac{\partial G}{\partial z} \, ds' \qquad & E_z = 2 \iint_{\Sigma} \left[ E_x^\text{inc} (\mathbf{r}') \frac{\partial G}{\partial x} + E_y^\text{inc} (\mathbf{r}') \frac{\partial G}{\partial y} \right] \, ds'
    \label{Luneburg RS-S}
\end{align}\\
The $x$ and $y$ components in Eq. \eqref{Luneburg RS-S} are identical to those of the first Rayleigh--Sommerfeld ($RS_1$) solution, while the $z$ component is recovered via the divergence-free condition. Although several studies have suggested that the Luneburg integral is inaccurate due to its perceived inability to predict polarization coupling effects, this is not the case provided the input field is a physically consistent Maxwellian solution.\\
Under the same assumptions, Schelkunoff’s PMC equivalence yields current densities of $\mathbf{J}=2\hat{\mathbf{z}}' \times \mathbf{H}^\text{inc}$ and $\mathbf{M}=0$. In a manner analogous to the PEC case, a magnetic version of the Luneburg integral can be derived under the PMC equivalence \cite{PhysRev.56.308}.
\begin{equation}
    \mathbf{H} = -2\iint_{\Sigma}\left(\hat{\mathbf{z}}'\times\mathbf{H}^\text{inc}\right)\times\nabla Gds'
\end{equation}
\noindent The electric field under the PMC equivalence framework is obtained via the curl relation:
\begin{equation}
    \mathbf{E}=-\frac{2i}{\omega\varepsilon}\nabla \times \iint_{\Sigma}\left(\hat{\mathbf{z}}'\times\mathbf{H}^\text{inc}\right)\times\nabla Gds'=\frac{2i}{\omega\mu\varepsilon} \nabla\times\nabla\times\mathbf{A}.
    \label{PMC E-S}
\end{equation}\\
Following the procedure applied to Eqs. \eqref{vector identity 1} and \eqref{vector identity 2}, Eq. \eqref{PMC E-S} can be further simplified as
\begin{equation}
    \mathbf{E}=\frac{2i}{\omega\varepsilon}\iint_{\Sigma}k^2\left[\hat{\mathbf{z}}'\times \mathbf{H}^\text{inc}\right]G-i\omega\varepsilon\left[\hat{\mathbf{z}}'\cdot\mathbf{E}^\text{inc}\right]\nabla'Gds'.
\end{equation}\\
Notably, the arithmetic average of the PEC solution (Eq. \eqref{Luneburg E-S}) and the PMC solution (Eq.\eqref{PMC E-S}) is mathematically identical to the Stratton-Chu formula. This correspondence extends to the magnetic field $\mathbf{H}$ as well. The average of the $\mathbf{H}$ field from the PEC and PMC solution is also identical to the Stratton-Chu formula. This unification stems from a mechanism identical to that found in scalar diffraction theory: the image currents from the PEC and PMC boundaries undergo mutual cancellation, thereby eliminating any non-physical image contributions.\\
This relationship underscores the fundamental physical nature of these formulations. Specifically, the obliquity factor of the Kirchhoff-Helmholtz integral identifies it as a Huygens source. Similarly, Love’s equivalence principle serves as its electromagnetic analogue, acting as a Huygens dipole source. Just as averaging the first and second Rayleigh-Sommerfeld solutions yields a Huygens source through the cancellation of image sources, averaging the PEC and PMC formulations reconstructs the electromagnetic Huygens source. According to the electromagnetic uniqueness theorem, provided that the incident field is a physically consistent Maxwellian solution, the fields propagated via the PEC, PMC, and Love equivalence principles are all mathematically identical.

\section{Equivalence of generic vector angular spectrum method and Schelkunoff's PMC solution}\label{secS5}
Song et al. (2025) \cite{song2025generic} proposed the generic vector angular spectrum method (VASM), which accurately captures polarization coupling effects while maintaining the computational scalability of the standard ASM algorithm. The VASM performs a $p/s$ polarization projection in the Fourier domain prior to the propagation step, ensuring that each electric field component inherently satisfies the transversal condition of the electric field to the $\mathbf{k}$ vector as well as the divergence-free condition. Their projection matrix is inspired by the Richards-Wolf integral \cite{richards1959electromagnetic}, which applies $p/s$ projection in accordance with the sine condition of an aplanatic lens.\\
Although the authors contend that VASM is fundamentally distinct from the Luneburg integral, which is often cited as failing to capture polarization mixing due to the independent propagation of $E_x$ and $E_y$, we demonstrate that VASM is, in fact, equivalent to the PMC equivalence solution. This solution can be interpreted as the magnetic counterpart to the Luneburg integral. To establish this equivalence, we assume that the input field $\left( E_x^\text{inc}, E_y^\text{inc}, E_z^\text{inc} \right), \left( H_x^\text{inc}, H_y^\text{inc}, H_z^\text{inc} \right)$ is physical.
In the Fourier domain, the curl-equation relationship between the electric and magnetic fields simplifies to Eqs. \eqref{curl E} and \eqref{curl H}.
\begin{equation}
    \left( \tilde{E}_x^\text{inc}, \tilde{E}_x^\text{inc}, \tilde{E}_x^\text{inc} \right) = Z \left( \sigma_z \tilde{H}_y^\text{inc} - \sigma_y \tilde{H}_z^\text{inc}, \sigma_x \tilde{H}_z^\text{inc} - \sigma_z \tilde{H}_x^\text{inc}, \sigma_y \tilde{H}_x^\text{inc} - \sigma_x \tilde{H}_y^\text{inc} \right)\label{curl E}
\end{equation}
\begin{equation}
    \left( \tilde{H}_x^\text{inc}, \tilde{H}_y^\text{inc}, \tilde{H}_z^\text{inc} \right) = \frac{1}{Z} \left( \sigma_y \tilde{E}_x^\text{inc} - \sigma_z \tilde{E}_x^\text{inc}, \sigma_z \tilde{E}_x^\text{inc} - \sigma_x \tilde{E}_x^\text{inc}, \sigma_x \tilde{E}_x^\text{inc} - \sigma_y \tilde{E}_x^\text{inc} \right)\label{curl H}
\end{equation}
By substituting Eqs. \eqref{curl E} and \eqref{curl H} into the PMC boundary solution, the magnetic field $\mathbf{H}$ can be expressed as:
\begin{equation}
    \begin{pmatrix} \tilde{H}_x \\ \tilde{H}_y \\ \tilde{H}_z \end{pmatrix} = 
\begin{pmatrix} 
1 & 0 & 0 \\ 
0 & 1 & 0 \\ 
-\frac{k_x}{k_z} & -\frac{k_y}{k_z} & 0 
\end{pmatrix} 
\begin{pmatrix} \tilde{H}_x^\text{inc} \\ \tilde{H}_y^\text{inc} \\ \tilde{H}_z^\text{inc} \end{pmatrix} e^{ik_z z} = 
\frac{1}{Z} 
\begin{pmatrix} 
0 & -\sigma_z & \sigma_y \\ 
\sigma_z & 0 & -\sigma_x \\ 
-\sigma_y & \sigma_x & 0 
\end{pmatrix} 
\begin{pmatrix} \tilde{E}_x^\text{inc} \\ \tilde{E}_x^\text{inc} \\ \tilde{E}_x^\text{inc} \end{pmatrix} e^{ik_z z}.
\label{VASM PMC H-S}
\end{equation}\\
Notably, the $H_x$ and $H_y$ components propagate independently via the ASM, while $H_z$ is recovered through the divergence-free condition, which is the magnetic counterpart of the Luneburg integral. The electric field can then be derived directly from either the curl equation or the PMC solution. In Eqs. \eqref{VASM PMC Ex}–\eqref{VASM PMC Ez}, the common phase factor $\exp(ik_z z)$ has been omitted for brevity.
\begin{align}
    \tilde{E}_x &= Z \left( \sigma_z \tilde{H}_y^\text{inc} + \frac{\sigma_y k_x \tilde{H}_x^\text{inc} + \sigma_y k_y \tilde{H}_y^\text{inc}}{k_z} \right) \notag \\
    &= \left( \sigma_z^2 \tilde{E}_x^\text{inc} - \sigma_z \sigma_x \tilde{E}_x^\text{inc} + \frac{\sigma_y k_x \left( \sigma_y \tilde{E}_x^\text{inc} - \sigma_z \tilde{E}_x^\text{inc} \right) + \sigma_y k_y \left( \sigma_z \tilde{E}_x^\text{inc} - \sigma_x \tilde{E}_x^\text{inc} \right)}{k_z} \right) \label{VASM PMC Ex}\\[10pt]
    \tilde{E}_y &= Z \left( -\sigma_z \tilde{H}_x^\text{inc} + \frac{-\sigma_x k_x \tilde{H}_x^\text{inc} - \sigma_x k_y \tilde{H}_y^\text{inc}}{k_z} \right) \notag \\
    &= \left( \sigma_z^2 \tilde{E}_x^\text{inc} - \sigma_z \sigma_y \tilde{E}_x^\text{inc} + \frac{-\sigma_x k_x \left( \sigma_y \tilde{E}_x^\text{inc} - \sigma_z \tilde{E}_x^\text{inc} \right) - \sigma_x k_y \left( \sigma_z \tilde{E}_x^\text{inc} - \sigma_x \tilde{E}_x^\text{inc} \right)}{k_z} \right) \label{VASM PMC Ey}\\[10pt]
    \tilde{E}_z &= Z \left( \sigma_y \tilde{H}_x^\text{inc} - \sigma_x \tilde{H}_y^\text{inc} \right) = \sigma_y \left( \sigma_y \tilde{E}_x^\text{inc} - \sigma_z \tilde{E}_x^\text{inc} \right) - \sigma_x \left( \sigma_z \tilde{E}_x^\text{inc} - \sigma_x \tilde{E}_x^\text{inc} \right)\label{VASM PMC Ez}
\end{align}
Consequently, the Fourier-domain representation of the electric field is identical to the VASM formulation \cite{song2025generic}.
\begin{equation}
    \begin{pmatrix}
    \tilde{E}_x \\
    \tilde{E}_y \\
    \tilde{E}_z
\end{pmatrix} = 
\begin{pmatrix}
    \left( 1 - \sigma_x^2 \right) & -\sigma_x \sigma_y & -\sigma_x \sigma_z \\
    -\sigma_x \sigma_y & \left( 1 - \sigma_y^2 \right) & -\sigma_y \sigma_z \\
    -\sigma_x \sigma_z & -\sigma_y \sigma_z & \left( 1 - \sigma_z^2 \right)
\end{pmatrix}
\begin{pmatrix}
    \tilde{E}_x^\text{inc} \\
    \tilde{E}_x^\text{inc} \\
    \tilde{E}_x^\text{inc}
\end{pmatrix} e^{ik_z z}
\label{VASM PMC E}
\end{equation}\\
This equivalence arises from the fact that the $p/s$ polarization projection is fundamentally achieved by solving the curl equation for a given magnetic field $\mathbf{H}$. Even if the input electric field is unphysical, the $p/s$ polarization projection induced by the curl equation naturally redistributes the energy such that the resulting electric field satisfies the transversality condition. In a similar manner, the PEC-based solution can also be derived.
\begin{equation}
    \begin{pmatrix} \tilde{E}_x \\ \tilde{E}_y \\ \tilde{E}_z \end{pmatrix} = 
    \begin{pmatrix} 
    1 & 0 & 0 \\ 
    0 & 1 & 0 \\ 
    -\frac{k_x}{k_z} & -\frac{k_y}{k_z} & 0 
    \end{pmatrix} 
    \begin{pmatrix} \tilde{E}_x^\text{inc} \\ \tilde{E}_x^\text{inc} \\ \tilde{E}_x^\text{inc} \end{pmatrix} e^{ik_z z} = 
    Z
    \begin{pmatrix} 
    0 & \sigma_z & -\sigma_y \\ 
    -\sigma_z & 0 & \sigma_x \\ 
    \sigma_y & -\sigma_x & 0 
    \end{pmatrix} 
    \begin{pmatrix} \tilde{H}_x^\text{inc} \\ \tilde{H}_y^\text{inc} \\ \tilde{H}_z^\text{inc} \end{pmatrix} e^{ik_z z}
    \label{VASM PEC E}
\end{equation}
\begin{align}
    \tilde{H}_x &= \frac{1}{Z} \left( -\sigma_z \tilde{E}_x^\text{inc} + \frac{-\sigma_y k_x \tilde{E}_x^\text{inc} - \sigma_y k_y \tilde{E}_x^\text{inc}}{k_z} \right) \notag \\
    &= \left( \sigma_z^2 \tilde{H}_x^\text{inc} - \sigma_z \sigma_x \tilde{H}_z^\text{inc} + \frac{\sigma_y k_x \left( \sigma_y \tilde{H}_z^\text{inc} - \sigma_z \tilde{H}_y^\text{inc} \right) + \sigma_y k_y \left( \sigma_z \tilde{H}_x^\text{inc} - \sigma_x \tilde{H}_z^\text{inc} \right)}{k_z} \right) \label{eq:VASM_PEC_Hx} \\[10pt]
    \tilde{H}_y &= \frac{1}{Z} \left( \sigma_z \tilde{E}_x^\text{inc} + \frac{\sigma_x k_x \tilde{E}_x^\text{inc} + \sigma_x k_y \tilde{E}_x^\text{inc}}{k_z} \right) \notag \\
    &= \left( \sigma_z^2 \tilde{H}_y^\text{inc} - \sigma_z \sigma_y \tilde{H}_z^\text{inc} + \frac{-\sigma_x k_x \left( \sigma_y \tilde{H}_z^\text{inc} - \sigma_z \tilde{H}_y^\text{inc} \right) - \sigma_x k_y \left( \sigma_z \tilde{H}_x^\text{inc} - \sigma_x \tilde{H}_z^\text{inc} \right)}{k_z} \right) \label{eq:VASM_PEC_Hy} \\[10pt]
    \tilde{H}_z &= \frac{1}{Z} \left( -\sigma_y \tilde{E}_x^\text{inc} + \sigma_x \tilde{E}_x^\text{inc} \right) \notag \\
    &= \sigma_y \left( \sigma_y \tilde{H}_z^\text{inc} - \sigma_z \tilde{H}_y^\text{inc} \right) - \sigma_x \left( \sigma_z \tilde{H}_x^\text{inc} - \sigma_x \tilde{H}_z^\text{inc} \right) \label{eq:VASM_PEC_Hz} 
\end{align}
\begin{equation}
    \begin{pmatrix}
    \tilde{H}_x \\
    \tilde{H}_y \\
    \tilde{H}_z
    \end{pmatrix} = 
    \begin{pmatrix}
    \left( 1 - \sigma_x^2 \right) & -\sigma_x \sigma_y & -\sigma_x \sigma_z \\
    -\sigma_x \sigma_y & \left( 1 - \sigma_y^2 \right) & -\sigma_y \sigma_z \\
    -\sigma_x \sigma_z & -\sigma_y \sigma_z & \left( 1 - \sigma_z^2 \right)
    \end{pmatrix}
    \begin{pmatrix}
    \tilde{H}_x^\text{inc} \\
    \tilde{H}_y^\text{inc} \\
    \tilde{H}_z^\text{inc}
\end{pmatrix} e^{ik_z z}
\label{eq:VASM_PEC_H}
\end{equation}\\
In Eqs.\eqref{eq:VASM_PEC_Hx}–\eqref{eq:VASM_PEC_Hz}, the phase factor $\exp(ik_z z)$ has been omitted for simplicity. The PEC-based formulation corresponds to the magnetic analogue of the VASM.

\section{Stratton-Chu formula implemented by the VASM}\label{secS6}
Building upon the conclusion in Section S4 that the average of the PEC and PMC solutions is equivalent to Love’s equivalence, the average of Eqs. \eqref{VASM PMC E} and \eqref{VASM PEC E} yields the electric field of the Stratton-Chu formula.
\begin{equation}
    \begin{pmatrix}
    \tilde{E}_x \\
    \tilde{E}_y \\
    \tilde{E}_z
    \end{pmatrix} = 
    \begin{pmatrix}
    1 - \frac{\sigma_x^2}{2} & -\frac{\sigma_x \sigma_y}{2} & -\frac{\sigma_x \sigma_z}{2} \\
    -\frac{\sigma_x \sigma_y}{2} & 1 - \frac{\sigma_y^2}{2} & -\frac{\sigma_y \sigma_z}{2} \\
    -\frac{1}{2} \left( \sigma_x \sigma_z + \frac{\sigma_x}{\sigma_z} \right) & -\frac{1}{2} \left( \sigma_y \sigma_z + \frac{\sigma_y}{\sigma_z} \right) & \frac{1-\sigma_z^2}{2}
    \end{pmatrix}
    \begin{pmatrix}
    \tilde{E}_x^\text{inc} \\
    \tilde{E}_x^\text{inc} \\
    \tilde{E}_x^\text{inc}
\end{pmatrix}e^{ik_z z}
\label{Stratton-Chu VASM E}
\end{equation}
Similarly, the magnetic field can be obtained by averaging Eqs. \eqref{VASM PMC H-S} and \eqref{eq:VASM_PEC_H}.
\begin{equation}
    \begin{pmatrix}
    \tilde{H}_x \\
    \tilde{H}_y \\
    \tilde{H}_z
    \end{pmatrix} = 
    \begin{pmatrix}
    1 - \frac{\sigma_x^2}{2} & -\frac{\sigma_x \sigma_y}{2} & -\frac{\sigma_x \sigma_z}{2} \\
    -\frac{\sigma_x \sigma_y}{2} & 1 - \frac{\sigma_y^2}{2} & -\frac{\sigma_y \sigma_z}{2} \\
    -\frac{1}{2} \left( \sigma_x \sigma_z + \frac{\sigma_x}{\sigma_z} \right) & -\frac{1}{2} \left( \sigma_y \sigma_z + \frac{\sigma_y}{\sigma_z} \right) & \frac{1-\sigma_z^2}{2}
    \end{pmatrix}
    \begin{pmatrix}
    \tilde{H}_x^\text{inc} \\
    \tilde{H}_y^\text{inc} \\
    \tilde{H}_z^\text{inc}
\end{pmatrix}e^{ik_z z}
\label{Stratton-Chu VASM H}
\end{equation}\\
Since this representation is compatible with acceleration via E-ASM, BLASM, and BEASM, Eqs. \eqref{Stratton-Chu VASM E} and \eqref{Stratton-Chu VASM H} offer a novel approach to accelerating the Stratton-Chu formula for planar geometries. However, as demonstrated in the subsequent section, it is numerically more robust to compute the independently propagated H-driven PEC and E-driven PMC fields and average them in the spatial domain, rather than applying the combined transfer matrices in Eqs. \eqref{Stratton-Chu VASM E} and \eqref{Stratton-Chu VASM H} directly (see section~\ref{secS9}). This approach helps mitigate the numerical instabilities caused by the singular nature of the $\sigma_x/\sigma_z$ and $\sigma_y/\sigma_z$ terms near the evanescent boundary.

\section{Derivation of spherical vectorial diffraction from the VASM representation of the Stratton-Chu formula}\label{secS7}

Recently, Liu et al. (2026) \cite{liu2026vectorial} introduced the spherical vectorial diffraction (SVD) method in an attempt to unify vectorial diffraction theory. However, we demonstrate that the SVD algorithm is essentially a paraxial approximation of the formulations presented in Eqs. \eqref{Stratton-Chu VASM E} and \eqref{Stratton-Chu VASM H}. By extracting the $(3,1)$ and $(3,2)$ components of the matrix in Eq. \eqref{Stratton-Chu VASM E}, we obtain:
\begin{align}
-\frac{1}{2}\left(\sigma_x\sigma_z+\frac{\sigma_x}{\sigma_z}\right)\approx-\frac{\sigma_x}{2}\left(1-\frac{\sigma_t^2}{2} + 1 + \frac{\sigma_t^2}{2}\right)=-\sigma_x \\
-\frac{1}{2}\left(\sigma_y\sigma_z+\frac{\sigma_y}{\sigma_z}\right)\approx-\frac{\sigma_y}{2}\left(1-\frac{\sigma_t^2}{2} + 1 + \frac{\sigma_t^2}{2}\right)=-\sigma_y
\end{align}
where $\sigma_t^2=\sigma_x^2+\sigma_y^2$. Following the assumption of a transversely polarized incident field ($\tilde{E}_{z}^\text{inc}=0$), as adopted by Liu (2026), the third column of the matrix vanishes. Finally, by applying the relations $\sigma_x=\lambda f_x, \sigma_y = \lambda f_y$, the SVD formula is obtained.
\begin{equation}
    \begin{pmatrix}
    \tilde{E}_x \\
    \tilde{E}_y \\
    \tilde{E}_z
    \end{pmatrix} = 
    \begin{pmatrix}
    1 - \frac{\lambda^2 f_x^2}{2} & -\frac{\lambda^2 f_x f_y}{2} & 0 \\
    -\frac{\lambda^2 f_x f_y}{2} & 1 - \frac{\lambda^2 f_y^2}{2} & 0 \\
    -\lambda f_x & -\lambda f_y & 0
    \end{pmatrix}
    \begin{pmatrix}
    \tilde{E}_x^\text{inc} \\
    \tilde{E}_x^\text{inc} \\
    \tilde{E}_x^\text{inc}
    \end{pmatrix}e^{ik_z z}
\end{equation}

\section{Equivalence of vector potential seeding and PEC/PMC equivalence}\label{secS8}
The vector potential seeding method has been established as a robust mathematical framework for analyzing complex vectorial beams, such as vectorial Gaussian and Hermite-Gaussian beams \cite{levy2019mathematics, agrawal1979gaussian, davis1979theory, levy2016weakly}. The core principle of this approach involves 'seeding' a paraxial scalar solution into the individual components of the vector potentials. For instance, to describe an $x$-polarized fundamental Gaussian beam, a scalar field $\Psi$ is embedded into the $x$-component of the magnetic vector potential as $\mathbf{A} = \frac{1}{i\omega}(\Psi, 0, 0)$.\\
This method exploits the fact that in a homogeneous medium, each Cartesian component of the vector potentials $\mathbf{A}$ and $\mathbf{F}$ must satisfy the Helmholtz equation. Consequently, any exact or paraxial scalar solution to the Helmholtz equation can serve as a legitimate 'seed' for constructing a full vectorial field. In practical applications, since standard solutions—including Gaussian, Hermite-Gaussian, Ince-Gaussian, and Laguerre-Gaussian modes—are typically derived under the paraxial approximation, the Fourier spectrum of the profile at the beam waist is seeded rather than the spatial function itself. This ensures that the propagated fields, calculated via the Angular Spectrum Method (ASM) by applying the $\exp(ik_z z)$ propagator, strictly satisfy the 3D Helmholtz equation and, by extension, Maxwell’s equations.
When the input field is defined by its components $\left( E_x^\text{inc}, E_y^\text{inc}, E_z^\text{inc} \right)$, the vector potential $\mathbf{A}$ is seeded such that
\begin{equation}
    \mathbf{A}=\frac{1}{i\omega}\left(E_x^\text{inc}, E_y^\text{inc}, E_z^\text{inc} \right).
\end{equation}
It should be emphasized that the initial components $(E_x^\text{inc}, E_y^\text{inc}, E_z^\text{inc})$ are not required to be exact solutions to Maxwell’s equations. Because each Cartesian component of the vector potential independently satisfies the Helmholtz equation, it is sufficient for the 'seed' fields to be solutions to the Helmholtz equation (including paraxial ones). Consequently, the resulting electric and magnetic fields derived from these potentials under the Lorenz gauge are guaranteed to be physically consistent Maxwellian solutions. From Eqs. \eqref{Lorenz E previous} and \eqref{Lorenz H previous}, we obtain
\begin{flalign}
    &\begin{aligned}
        \tilde{E}_x &= i\omega \left[ \tilde{A}_x - \frac{k_x^2}{k^2}\tilde{A}_x - \frac{k_x k_y}{k^2}\tilde{A}_y - \frac{k_x k_z}{k^2}\tilde{A}_z \right] \\
        \tilde{E}_y &= i\omega \left[ \tilde{A}_y - \frac{k_x k_y}{k^2}\tilde{A}_x - \frac{k_y^2}{k^2}\tilde{A}_y - \frac{k_y k_z}{k^2}\tilde{A}_z \right] \\
        \tilde{E}_z &= i\omega \left[ \tilde{A}_z - \frac{k_x k_z}{k^2}\tilde{A}_x - \frac{k_y k_z}{k^2}\tilde{A}_y - \frac{k_z^2}{k^2}\tilde{A}_z \right]
    \end{aligned} \notag \\
    &\rightarrow \begin{pmatrix} \tilde{E}_x \\ \tilde{E}_y \\ \tilde{E}_z \end{pmatrix} = 
    \begin{pmatrix}
        (1-\sigma_x^2) & -\sigma_x \sigma_y & -\sigma_x \sigma_z \\
        -\sigma_x \sigma_y & (1-\sigma_y^2) & -\sigma_y \sigma_z \\
        -\sigma_x \sigma_z & -\sigma_y \sigma_z & (1-\sigma_z^2)
    \end{pmatrix}
    \begin{pmatrix} \tilde{E}_x^\text{inc} \\ \tilde{E}_y^\text{inc} \\ \tilde{E}_z^\text{inc} \end{pmatrix}
    \label{A-seeding E}
\end{flalign}
\begin{align}
    \begin{aligned}
        \tilde{H}_x &= \frac{1}{\omega\mu} \left[ k_y \tilde{E}_z^\text{inc} - k_z \tilde{E}_y^\text{inc} \right] \\
        \tilde{H}_y &= \frac{1}{\omega\mu} \left[ k_z \tilde{E}_x^\text{inc} - k_x \tilde{E}_z^\text{inc} \right] \\
        \tilde{H}_z &= \frac{1}{\omega\mu} \left[ k_x \tilde{E}_y^\text{inc} - k_y \tilde{E}_x^\text{inc} \right]
    \end{aligned}
    \rightarrow
    \begin{pmatrix} \tilde{H}_x \\ \tilde{H}_y \\ \tilde{H}_z \end{pmatrix} = 
    \frac{1}{Z}
    \begin{pmatrix}
        0 & -\sigma_z & \sigma_y \\
        \sigma_z & 0 & -\sigma_x \\
        -\sigma_y & \sigma_x & 0
    \end{pmatrix}
    \begin{pmatrix} \tilde{E}_x^\text{inc} \\ \tilde{E}_y^\text{inc} \\ \tilde{E}_z^\text{inc} \end{pmatrix}.
    \label{A-seeding H}
\end{align}\\
For brevity, the common propagator $\exp(ik_z z)$ has been omitted 
from Eqs.~\eqref{A-seeding E} and \eqref{A-seeding H}. It becomes evident that the resulting expressions are identical to the PMC equivalence solutions in Eqs. \eqref{VASM PMC E} and \eqref{VASM PMC H-S}. Consequently, the magnetic vector potential ($\mathbf{A}$) seeding method is gauge equivalent to the VASM representation of the PMC solution under the Lorenz gauge.

\section{Symmetrization, Quasi-Stratton-Chu integral}\label{secS9}
When the $x$-polarized scalar field is seeded solely into the magnetic vector potential as $\mathbf{A}=\frac{1}{i\omega}(\Psi, 0, 0)$, the resulting magnetic field lacks the $H_y$ component. To resolve this asymmetry, a symmetrization process has been proposed \cite{cullen1979complex, levy2019mathematics, barton1989fifth, erikson1994polarization}, which involves seeding the electric vector potential $\mathbf{F}$ as $\mathbf{F}=\frac{1}{i\omega}\left(0, \frac{\Psi}{Z}, 0\right)$, where $Z$ denotes the intrinsic impedance of the medium.\\
Subsequently, the corresponding electric and magnetic fields are derived using Eqs. \eqref{Lorenz E previous} and \eqref{Lorenz H previous}. While the $\mathbf{A}$-seeding approach results in a field where certain components are absent, this $\mathbf{F}$-seeding solution conversely lacks the $E_y$ component. By averaging the results from both $\mathbf{A}$-seeding and $\mathbf{F}$-seeding, one can obtain a fully symmetrized solution encompassing all six electromagnetic field components. Following the procedure detailed in the previous section, we seed $\mathbf{F}$ as follows:
\begin{equation}
    \mathbf{F}=\frac{1}{i\omega}\left(H_x^\text{inc}, H_y^\text{inc}, H_z^\text{inc} \right).
    \label{F seeding}
\end{equation}\\
The electric and magnetic fields are obtained via Eqs. \eqref{PEC E-S} and \eqref{PEC H-S}.
By substituting Eq. \eqref{F seeding} into Eqs. \eqref{PEC E-S} and \eqref{PEC H-S}, we obtain
\begin{flalign}
    &\begin{aligned}
        \tilde{H}_x &= i\omega \left[ \tilde{F}_x - \frac{k_x^2}{k^2}\tilde{F}_x - \frac{k_x k_y}{k^2}\tilde{F}_y - \frac{k_x k_z}{k^2}\tilde{F}_z \right] \\
        \tilde{H}_y &= i\omega \left[ \tilde{F}_y - \frac{k_x k_y}{k^2}\tilde{F}_x - \frac{k_y^2}{k^2}\tilde{F}_y - \frac{k_y k_z}{k^2}\tilde{F}_z \right] \\
        \tilde{H}_z &= i\omega \left[ \tilde{F}_z - \frac{k_x k_z}{k^2}\tilde{F}_x - \frac{k_y k_z}{k^2}\tilde{F}_y - \frac{k_z^2}{k^2}\tilde{F}_z \right]
    \end{aligned}
    \notag \\
    & \rightarrow \begin{pmatrix} \tilde{H}_x \\ \tilde{H}_y \\ \tilde{H}_z \end{pmatrix} = 
    \begin{pmatrix}
        (1-\sigma_x^2) & -\sigma_x \sigma_y & -\sigma_x \sigma_z \\
        -\sigma_x \sigma_y & (1-\sigma_y^2) & -\sigma_y \sigma_z \\
        -\sigma_x \sigma_z & -\sigma_y \sigma_z & (1-\sigma_z^2)
    \end{pmatrix}
    \begin{pmatrix} \tilde{H}_x^\text{inc} \\ \tilde{H}_y^\text{inc} \\ \tilde{H}_z^\text{inc} \end{pmatrix}
\end{flalign}
\begin{align}
    \begin{aligned}
        \tilde{E}_x &= \frac{1}{\omega\varepsilon} \left[ -k_y \tilde{H}_z^\text{inc} + k_z \tilde{H}_y^\text{inc} \right] \\
        \tilde{E}_y &= \frac{1}{\omega\varepsilon} \left[ -k_z \tilde{H}_x^\text{inc} + k_x \tilde{H}_z^\text{inc} \right] \\
        \tilde{E}_z &= \frac{1}{\omega\varepsilon} \left[ -k_x \tilde{H}_y^\text{inc} + k_y \tilde{H}_x^\text{inc} \right]
    \end{aligned}
    \rightarrow
    \begin{pmatrix} \tilde{E}_x \\ \tilde{E}_y \\ \tilde{E}_z \end{pmatrix} = 
    Z
    \begin{pmatrix}
        0 & \sigma_z & -\sigma_y \\
        -\sigma_z & 0 & \sigma_x \\
        \sigma_y & -\sigma_x & 0
    \end{pmatrix}
    \begin{pmatrix} \tilde{H}_x^\text{inc} \\ \tilde{H}_y^\text{inc} \\ \tilde{H}_z^\text{inc} \end{pmatrix}.
\end{align}\\

\noindent As before, the common propagator $\exp(ik_z z)$ has been omitted 
from all transfer matrices in this section for brevity; restoring 
it recovers full agreement with the PEC and PMC solutions in 
Sections~\ref{secS5} and \ref{secS6}. The $\mathbf{F}$-seeding solution is also gauge equivalent to the PEC solution. It should be noted, however, that $\mathbf{A}$- and $\mathbf{F}$-seeding utilize vector potentials defined differently from those in the PEC and PMC equivalence frameworks. While PEC and PMC equivalence principles construct electric and magnetic surface currents solely from tangential $\mathbf{E}$ and $\mathbf{H}$ components—thereby lacking a $z$-component—the $\mathbf{A}$- and $\mathbf{F}$-seeding methods can incorporate a longitudinal ($z$) component. Despite these distinct boundary representations, they are mathematically guaranteed to be gauge equivalent under the Lorenz gauge.\\
Consequently, as the symmetrization process is achieved by averaging the $\mathbf{A}$- and $\mathbf{F}$-seeding solutions, it can be fundamentally interpreted as a numerical implementation of the Stratton-Chu integral, analogous to the averaging of the PEC and PMC equivalence formulations.\\
Notably, even if the input field is unphysical, the resulting symmetrized field remains a mathematically exact Maxwellian solution, as do the individual PEC, PMC, or $\mathbf{A}$/$\mathbf{F}$-seeding results. It is important to observe, however, that this symmetrized solution may mathematically deviate from the field defined by Love’s equivalence principle. This discrepancy arises because the $\mathbf{A}$- and $\mathbf{F}$-seeding components represent distinct physical field distributions when the input is not a self-consistent Maxwellian solution. Nevertheless, in practical applications, such numerical differences are typically negligible \cite{levy2019mathematics}. This suggests that while all these methods enforce physical consistency, the choice of seeding (or equivalence principle) acts as a mathematical filter that slightly alters the reconstructed field when the source information is incomplete or inconsistent. The scalar-wave symmetrization process can therefore be regarded as a quasi-Stratton-Chu integral.\\
Regarding the symmetrization process or the Quasi-Stratton-Chu integral, there are four possible configurations for the input field. If the input field is physically consistent, no discrepancies arise among these choices. This equivalence stems from the fact that the $s/p$ polarization projection matrix effectively acts as an identity operator when applied to a physical input field. By examining the longitudinal component $\tilde{E}_z$ in Eq.~\eqref{VASM PMC E}, we can rewrite the expression as
\begin{equation}
    \tilde{E}_z=\frac{k_z^2}{k^2}\left(-\frac{k_x\tilde{E}_x+k_y\tilde{E}_y}{k_z}\right)+\frac{k_t^2}{k^2}\tilde{E}_z.
    \label{div decompoesed Ez}
\end{equation}
Recognizing that $-\frac{k_x\tilde{E}_x+k_y\tilde{E}_y}{k_z}$ represents the reconstruction of the $E_z$ component via the divergence-free condition, which is the foundational principle of the Luneburg integral, we define
\begin{equation}
\tilde{E}_z^{\mathrm{div-free}}=-\frac{k_x\tilde{E}_x+k_y\tilde{E}_y}{k_z}.
\end{equation}
With this substitution, Eq. \eqref{div decompoesed Ez} can be rewritten as
\begin{equation}
    \tilde{E}_z=\frac{k_z^2}{k^2}\tilde{E}_z^{\mathrm{div-free}}+\frac{k_t^2}{k^2}\tilde{E}_z.
\end{equation}
For a physically consistent, divergence-free input field, the projection reduces to an identity operation as follows:
\begin{equation}
    \tilde{E}_z=\frac{k_z^2}{k^2}\tilde{E}_z+\frac{k_t^2}{k^2}\tilde{E}_z=\tilde{E}_z.
\end{equation}
Similarly, the transverse components $E_x$ and $E_y$ undergo a trivial identity projection, leaving the original field distribution unchanged.
Conversely, if the input field is non-Maxwellian, the four possible input configurations yield distinct results. Based on Eqs.\eqref{A-seeding E}–-\eqref{A-seeding H}, the $\mathbf{A}$-seeding (or PMC) solution can be expressed as
\begin{equation}
    \begin{pmatrix} \tilde{H}_x \\ \tilde{H}_y \\ \tilde{H}_z \end{pmatrix} = 
    \begin{pmatrix} 
        1 & 0 & 0 \\ 
        0 & 1 & 0 \\ 
        -\frac{k_x}{k_z} & -\frac{k_y}{k_z} & 0 
    \end{pmatrix} 
    \begin{pmatrix} \tilde{H}_x^\text{inc} \\ \tilde{H}_y^\text{inc} \\ \tilde{H}_z^\text{inc} \end{pmatrix} = 
    \frac{1}{Z} \begin{pmatrix} 
        0 & -\sigma_z & \sigma_y \\ 
        \sigma_z & 0 & -\sigma_x \\ 
        -\sigma_y & \sigma_x & 0 
    \end{pmatrix} 
    \begin{pmatrix} \tilde{E}_x^\text{inc} \\ \tilde{E}_y^\text{inc} \\ \tilde{E}_z^\text{inc} \end{pmatrix}
    \label{A-seeding H full}
\end{equation}
\begin{flalign}
    &\begin{pmatrix} \tilde{E}_x \\ \tilde{E}_y \\ \tilde{E}_z \end{pmatrix} = 
    Z \begin{pmatrix} 
        \frac{\sigma_y \sigma_x}{\sigma_z} & \sigma_z + \frac{\sigma_y^2}{\sigma_z} & 0 \\ 
        -\sigma_z - \frac{\sigma_x^2}{\sigma_z} & -\frac{\sigma_y \sigma_x}{\sigma_z} & 0 \\ 
        \sigma_y & -\sigma_x & 0 
    \end{pmatrix} 
    \begin{pmatrix} \tilde{H}_x^\text{inc} \\ \tilde{H}_y^\text{inc} \\ \tilde{H}_z^\text{inc} \end{pmatrix} \notag \\
    &= 
    \begin{pmatrix}
        (1-\sigma_x^2) & -\sigma_x \sigma_y & -\sigma_x \sigma_z \\
        -\sigma_x \sigma_y & (1-\sigma_y^2) & -\sigma_y \sigma_z \\
        -\sigma_x \sigma_z & -\sigma_y \sigma_z & (1-\sigma_z^2)
    \end{pmatrix}
    \begin{pmatrix} \tilde{E}_x^\text{inc} \\ \tilde{E}_y^\text{inc} \\ \tilde{E}_z^\text{inc} \end{pmatrix}.
    \label{A-seeding E full}
\end{flalign}
\\
Similarly, the PEC-based formulation (or $\mathbf{F}$-seeding approach) can be expressed as
\begin{equation}
    \begin{pmatrix} \tilde{E}_x \\ \tilde{E}_y \\ \tilde{E}_z \end{pmatrix} = 
    \begin{pmatrix} 
        1 & 0 & 0 \\ 
        0 & 1 & 0 \\ 
        -\frac{k_x}{k_z} & -\frac{k_y}{k_z} & 0 
    \end{pmatrix} 
    \begin{pmatrix} \tilde{E}_x^\text{inc} \\ \tilde{E}_y^\text{inc} \\ \tilde{E}_z^\text{inc} \end{pmatrix} = 
    Z \begin{pmatrix} 
        0 & \sigma_z & -\sigma_y \\ 
        -\sigma_z & 0 & \sigma_x \\ 
        \sigma_y & -\sigma_x & 0 
    \end{pmatrix} 
    \begin{pmatrix} \tilde{H}_x^\text{inc} \\ \tilde{H}_y^\text{inc} \\ \tilde{H}_z^\text{inc} \end{pmatrix}
    \label{F-seeding E full}
\end{equation}

\begin{flalign}
    &\begin{pmatrix} \tilde{H}_x \\ \tilde{H}_y \\ \tilde{H}_z \end{pmatrix} = 
    \frac{1}{Z} \begin{pmatrix} 
        -\frac{\sigma_y \sigma_x}{\sigma_z} & -\sigma_z - \frac{\sigma_y^2}{\sigma_z} & 0 \\ 
        \sigma_z + \frac{\sigma_x^2}{\sigma_z} & \frac{\sigma_y \sigma_x}{\sigma_z} & 0 \\ 
        -\sigma_y & \sigma_x & 0 
    \end{pmatrix} 
    \begin{pmatrix} \tilde{E}_x^\text{inc} \\ \tilde{E}_y^\text{inc} \\ \tilde{E}_z^\text{inc} \end{pmatrix} \notag \\
    &=\begin{pmatrix}
        (1-\sigma_x^2) & -\sigma_x \sigma_y & -\sigma_x \sigma_z \\
        -\sigma_x \sigma_y & (1-\sigma_y^2) & -\sigma_y \sigma_z \\
        -\sigma_x \sigma_z & -\sigma_y \sigma_z & (1-\sigma_z^2)
    \end{pmatrix}
    \begin{pmatrix} \tilde{H}_x^\text{inc} \\ \tilde{H}_y^\text{inc} \\ \tilde{H}_z^\text{inc} \end{pmatrix}.
    \label{F-seeding H full}
\end{flalign}
\\
Consequently, four distinct input configurations can be considered for the symmetrization process:\\\\
1. Purely electric-based ($E^{PMC}$ and $E^{PEC}$),\\\\
2. Purely magnetic-based ($H^{PMC}$ and $H^{PEC}$),\\\\
3. H-driven PMC / E-driven PEC, and\\\\
4. E-driven PMC / H-driven PEC.\\\\
In practice, the 'E-driven PMC / H-driven PEC' combination is the most advantageous, as it eliminates the $1/\sigma_z$ factor that otherwise leads to numerical singularities in the Fourier domain near the evanescent cutoff. Notably, this particular configuration aligns perfectly with the protocols established in prior vector potential seeding research. In the following section, we will analyze the numerical discrepancies among these four combinations using a fundamental Gaussian beam.
\section{Discrepancy of the PEC and PMC solution}\label{secS10}
In this section, we analyze the discrepancy between all input configurations. Here, we used the fundamental Gaussian mode seeded by $\mathbf{A}=\frac{1}{i\omega}\left(\Psi,0,0\right), \mathbf{F}=\frac{1}{i\omega} \left(0, \frac{\Psi}{Z}, 0\right)$. Then, the discrepancy of the $E_x$ field can be obtained from Eqs.\eqref{A-seeding H full}--\eqref{F-seeding H full}.
\begin{equation}
    \Delta E_{x, \mathrm{E-E}} = \sigma_x^2\tilde{E}_x^\text{inc}
    \label{E E discrepancy}
\end{equation}
\begin{equation}
    \Delta E_{x, \mathrm{H-H}} =\frac{\sigma_y^2}{\sigma_z}\tilde{E}_x^\text{inc}\approx\sigma_y^2\left(1+\frac{\sigma_x^2+\sigma_y^2}{2}\right)\tilde{E}_x^\text{inc}
    \label{H H discrepancy}
\end{equation}
\begin{equation}
    \Delta E_{x, \mathrm{H-E}} =\left(1-\left(\sigma_z+\frac{\sigma_y^2}{\sigma_z}\right)\right)\tilde{E}_x^\text{inc}\approx \frac{\sigma_x^2-\sigma_y^2-\sigma_x^2\sigma_y^2-\sigma_y^4}{2}\tilde{E}_x^\text{inc}
    \label{H E discrepancy}
\end{equation}
\begin{equation}
    \Delta E_{x, \mathrm{E-H}} =\left(\sigma_z-\left(1-\sigma_x^2\right)\right)\tilde{E}_x^\text{inc}\approx \frac{\sigma_x^2-\sigma_y^2}{2}\tilde{E}_x^\text{inc}
    \label{E H discrepancy}
\end{equation}
Eqs. \eqref{E E discrepancy}-\eqref{E H discrepancy} correspond to the discrepancy of $E_x$ component of `Purely electric-based', `Purely magnetic-based', `H-driven PMC / E-driven PEC', and `E-driven PMC / H-driven PEC' respectively. Because the average of the $\sigma_x^2, \sigma_y^2, \sigma_x^2\sigma_y^2, \sigma_y^4$ is $\frac{\theta_d^2}{4}, \frac{\theta_d^2}{4}, \frac{\theta_d^4}{16}, \frac{3\theta_d^4}{16}$, the average of the factor in front of the $\Delta E_x$ follows:
\begin{equation}
    \langle\mathrm{Prefactor}_{\mathrm{E-E}}\rangle = \frac{\theta_d^2}{4}
\end{equation}
\begin{equation}
    \langle\mathrm{Prefactor}_{\mathrm{H-H}}\rangle = \frac{2\theta_d^2+\theta_d^4}{8}
\end{equation}
\begin{equation}
    \langle\mathrm{Prefactor}_{\mathrm{H-E}}\rangle= -\frac{\theta_d^4}{8}
\end{equation}
\begin{equation}
    \langle\mathrm{Prefactor}_{\mathrm{E-H}}\rangle =0
\end{equation}
The staggered input, i.e., 'E-driven PMC / H-driven PEC' and 'H-driven PMC / E-driven PEC' show minimal discrepancy. This is because the input fields undergo symmetric matrix operations. Table~\ref{Discrepancy} shows the discrepancy of the fundamental Gaussian beam whose waist is $128\,\mathrm{\mu m}$ with a wavelength of $500\,\mathrm{nm}$ sampled by a $1024\,\mathrm{\mu m} \times 1024\,\mathrm{\mu m}$ window with $500\,\mathrm{nm}$ pixel pitch.

\begin{table}[h]
\caption{\bf The discrepancy of $E_x$ under the four different combinations}\label{Discrepancy}
\begin{tabular}{>{\centering\arraybackslash}p{3cm} >{\centering\arraybackslash}p{6cm}}
\toprule%
Input combination (PMC/PEC) &Discrepancy \\
\midrule
E-E & $1.120423\times10^{-13}$  \\
H-H & $1.120426\times10^{-13}$  \\
H-E & $3.734751\times10^{-14}$  \\
E-H & $3.734745\times10^{-14}$  \\
\botrule
\end{tabular}
\end{table}

The discrepancy is defined by
\begin{equation}
    \Delta=\frac{|E_{x, PEC}-E_{x,PMC}|^2}{|E_{x,PEC}+E_{x, PMC}|^2}.
\end{equation}
However, if the waist of the seeded fundamental Gaussian beam is much smaller than the wavelength such as high-NA focusing configuration, a significant amount of energy is spilled over the evanescent boundary. If the sampling pitch is beyond the Nyquist frequency so that the small beam waist can be sufficiently sampled, `Purely electric-based', `Purely magnetic-based', and `H-driven PMC / E-driven PEC' combinations fail to recover $E_z$ components due to the singularity in the Fourier domain. Only the `E-driven PMC / H-driven PEC' combination can successfully recover all the six components. Even though there is no significant discrepancy between all four input combinations for a narrow-bandwidth Gaussian beam, it is recommended to use `E-driven PMC / H-driven PEC' input to avoid singularity for the case of large-bandwidth signal.\\
Next, we confirm that the PEC / PMC solutions (or $\mathbf{A}$-seeding and $\mathbf{F}$-seeding) and Quasi-Stratton-Chu solution (or symmetrization) are Maxwellian solutions. We formulate the Maxwell residual power defined by Levy et al. (2019) \cite{levy2019mathematics},
\begin{equation}
    M_1=\frac{\iint|\nabla\cdot \mathbf{E}|^2dxdy}{\mathscr{P}\left(|E_x|^2\right)}
\end{equation}
\begin{equation}
    M_2=\frac{\iint|\nabla\cdot \mathbf{H}|^2dxdy}{\mathscr{P}\left(|H_y|^2\right)}
\end{equation}
\begin{equation}
    M_{3, 4, 5}=\frac{\iint|\nabla\times \mathbf{E}-i\omega\mu \mathbf{H}|_{x,y,z}^2dxdy}{\mathscr{P}\left(|H_y|^2\right)}
\end{equation}
\begin{equation}
    M_{6, 7, 8}=\frac{\iint|\nabla\times \mathbf{H}+i\omega\varepsilon \mathbf{E}|_{x,y,z}^2dxdy}{\mathscr{P}\left(|E_x|^2\right)}
\end{equation}
where $\mathscr{P}\left(|E_x|^2\right)$ and $\mathscr{P}\left(|H_y|^2\right)$ are the powers of the dominant electric and magnetic field components. Here, unlike Levy's definition, each residual is normalized by the power of the corresponding dominant field component, electric or magnetic, to ensure dimensional consistency. As shown, all combinations exhibit extremely low Maxwell residual power because each is an exact Maxwell solution. The residual is limited only by the numerical precision of the FFT.

\begin{table}[h]
\caption{\bf Maxwell residual power of the E-driven PMC solution}\label{Maxwell residual power of E-driven PMC}
\begin{tabular}{>{\centering\arraybackslash}p{3cm} >{\centering\arraybackslash}p{6cm}}
\toprule%
$M_i$ &Maxwell residual power \\
\midrule
$M_1$ & $2.69\times10^{-33}$  \\
$M_2$ & $2.30\times10^{-33}$  \\
$M_3$& $8.94\times10^{-40}$  \\
$M_4$ & $1.74\times10^{-20}$  \\
$M_5$ & $6.74\times10^{-27}$ \\
$M_6$ & $1.74\times10^{-20}$ \\
$M_7$ & $2.60\times10^{-33}$ \\
$M_8$ & $6.74\times10^{-27}$ \\
\botrule
\end{tabular}
\end{table}

\begin{table}[h]
\caption{\bf Maxwell residual power of the H-driven PMC solution}\label{Maxwell residual power of H-driven PMC}
\begin{tabular}{>{\centering\arraybackslash}p{3cm} >{\centering\arraybackslash}p{6cm}}
\toprule%
$M_i$ &Maxwell residual power \\
\midrule
$M_1$ & $2.60\times10^{-33}$  \\
$M_2$ & $2.19\times10^{-33}$  \\
$M_3$& $9.43\times10^{-40}$  \\
$M_4$ & $1.74\times10^{-20}$  \\
$M_5$ & $6.74\times10^{-27}$ \\
$M_6$ & $1.74\times10^{-20}$ \\
$M_7$ & $2.60\times10^{-33}$ \\
$M_8$ & $6.74\times10^{-27}$ \\
\botrule
\end{tabular}
\end{table}

\begin{table}[h]
\caption{\bf Maxwell residual power of the E-driven PEC solution}\label{Maxwell residual power of E-driven PEC}
\begin{tabular}{>{\centering\arraybackslash}p{3cm} >{\centering\arraybackslash}p{6cm}}
\toprule%
$M_i$ &Maxwell residual power \\
\midrule
$M_1$ & $2.54\times10^{-33}$  \\
$M_2$ & $2.25\times10^{-33}$  \\
$M_3$& $2.60\times10^{-33}$  \\
$M_4$ & $1.74\times10^{-20}$  \\
$M_5$ & $6.74\times10^{-27}$ \\
$M_6$ & $1.74\times10^{-20}$ \\
$M_7$ & $1.05\times10^{-39}$ \\
$M_8$ & $6.74\times10^{-27}$ \\
\botrule
\end{tabular}
\end{table}

\begin{table}[h]
\caption{\bf Maxwell residual power of the H-driven PEC solution}\label{Maxwell residual power of H-driven PEC}
\begin{tabular}{>{\centering\arraybackslash}p{3cm} >{\centering\arraybackslash}p{6cm}}
\toprule%
$M_i$ &Maxwell residual power \\
\midrule
$M_1$ & $2.55\times10^{-33}$  \\
$M_2$ & $2.27\times10^{-33}$  \\
$M_3$& $2.60\times10^{-33}$  \\
$M_4$ & $1.74\times10^{-20}$  \\
$M_5$ & $6.74\times10^{-27}$ \\
$M_6$ & $1.74\times10^{-20}$ \\
$M_7$ & $1.05\times10^{-39}$ \\
$M_8$ & $6.74\times10^{-27}$ \\
\botrule
\end{tabular}
\end{table}

\begin{table}[h]
\caption{\bf Maxwell residual power of the symmetrized solution of the purely electric-based combination}\label{Maxwell residual power of E-E}
\begin{tabular}{>{\centering\arraybackslash}p{3cm} >{\centering\arraybackslash}p{6cm}}
\toprule%
$M_i$ &Maxwell residual power \\
\midrule
$M_1$ & $2.39\times10^{-33}$  \\
$M_2$ & $1.83\times10^{-33}$  \\
$M_3$& $6.51\times10^{-34}$  \\
$M_4$ & $1.75\times10^{-20}$  \\
$M_5$ & $6.74\times10^{-27}$ \\
$M_6$ & $1.74\times10^{-20}$ \\
$M_7$ & $6.51\times10^{-34}$ \\
$M_8$ & $6.74\times10^{-27}$ \\
\botrule
\end{tabular}
\end{table}

\begin{table}[h]
\caption{\bf Maxwell residual power of the symmetrized solution of the purely magnetic-based combination}\label{Maxwell residual power of H-H}
\begin{tabular}{>{\centering\arraybackslash}p{3cm} >{\centering\arraybackslash}p{6cm}}
\toprule%
$M_i$ &Maxwell residual power \\
\midrule
$M_1$ & $2.39\times10^{-33}$  \\
$M_2$ & $1.74\times10^{-33}$  \\
$M_3$& $6.51\times10^{-34}$  \\
$M_4$ & $1.75\times10^{-20}$  \\
$M_5$ & $6.74\times10^{-27}$ \\
$M_6$ & $1.74\times10^{-20}$ \\
$M_7$ & $6.51\times10^{-34}$ \\
$M_8$ & $6.74\times10^{-27}$ \\
\botrule
\end{tabular}
\end{table}

\begin{table}[h]
\caption{\bf Maxwell residual power of the symmetrized solution of the H-driven PMC / E-driven PEC combination}\label{Maxwell residual power of H-E}
\begin{tabular}{>{\centering\arraybackslash}p{3cm} >{\centering\arraybackslash}p{6cm}}
\toprule%
$M_i$ &Maxwell residual power \\
\midrule
$M_1$ & $2.35\times10^{-33}$  \\
$M_2$ & $1.66\times10^{-33}$  \\
$M_3$& $6.51\times10^{-34}$  \\
$M_4$ & $1.74\times10^{-20}$  \\
$M_5$ & $6.74\times10^{-27}$ \\
$M_6$ & $1.74\times10^{-20}$ \\
$M_7$ & $6.51\times10^{-34}$ \\
$M_8$ & $6.74\times10^{-27}$ \\
\botrule
\end{tabular}
\end{table}

\begin{table}[!ht]
\caption{\bf Maxwell residual power of the symmetrized solution of the E-driven PMC / H-driven PEC combination}\label{Maxwell residual power of E-H}
\begin{tabular}{>{\centering\arraybackslash}p{3cm} >{\centering\arraybackslash}p{6cm}}
\toprule%
$M_i$ &Maxwell residual power \\
\midrule
$M_1$ & $2.43\times10^{-33}$  \\
$M_2$ & $1.74\times10^{-33}$  \\
$M_3$& $6.51\times10^{-34}$  \\
$M_4$ & $1.74\times10^{-20}$  \\
$M_5$ & $6.74\times10^{-27}$ \\
$M_6$ & $1.74\times10^{-20}$ \\
$M_7$ & $6.51\times10^{-34}$ \\
$M_8$ & $6.74\times10^{-27}$ \\
\botrule
\end{tabular}
\end{table}
\mbox{}
\clearpage 

\section{Results of the vectorial diffraction benchmark of the Hermite-Gaussian beam}\label{secS11}
This section shows the simulation results of the vectorial diffraction benchmark of the $\text{HG}_{21}$ mode. The simulation setup is identical to that used in the Gaussian beam benchmark. 
\begin{figure}[htbp]
  \centering
  \includegraphics[width=\textwidth, page=25]{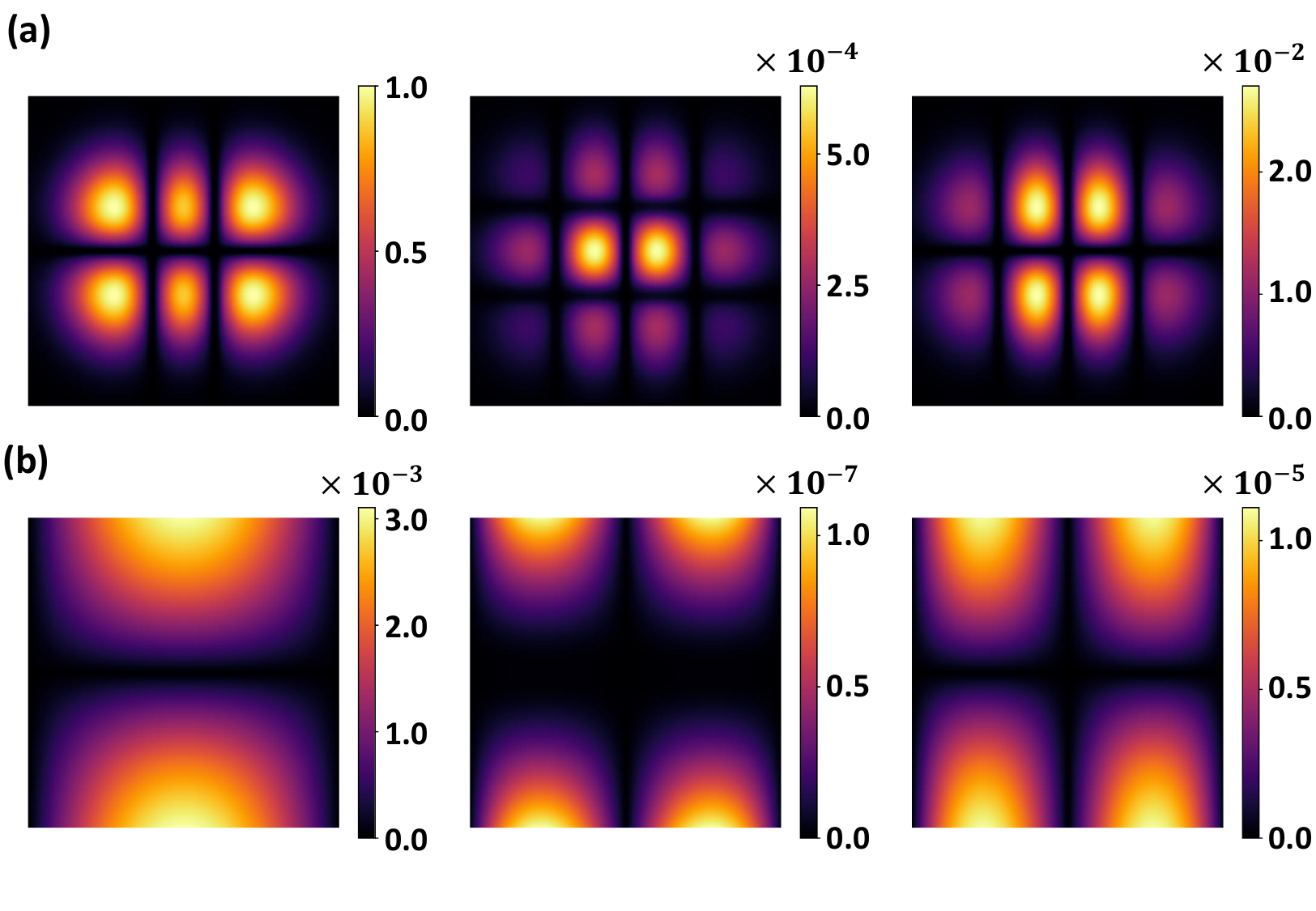}
  \caption{(a) The zoomed view of the $E_x, E_y, E_z$ components at the beam waist within a $50\,\mathrm{\mu m}$ region at the center of the source plane. (b) The ground truth of the $E_x, E_y, E_z$ components on the destination plane at $z = 0.103\,\mathrm{m}$.}
  \label{figS4}
\end{figure}
\clearpage

\begin{figure}[htbp]
  \centering
  \includegraphics[width=0.49\textwidth, page=26]{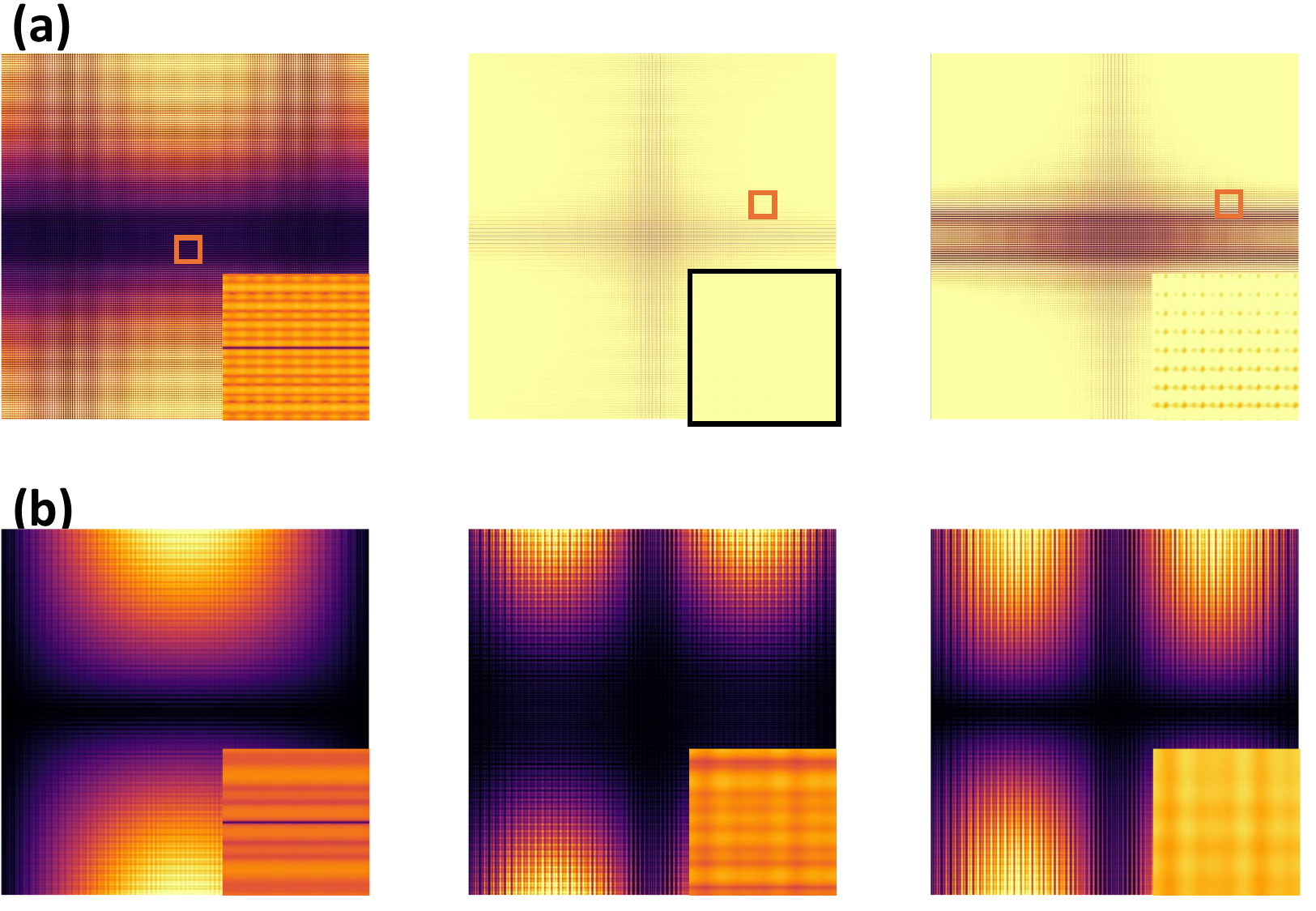}
  \includegraphics[width=0.49\textwidth, page=27]{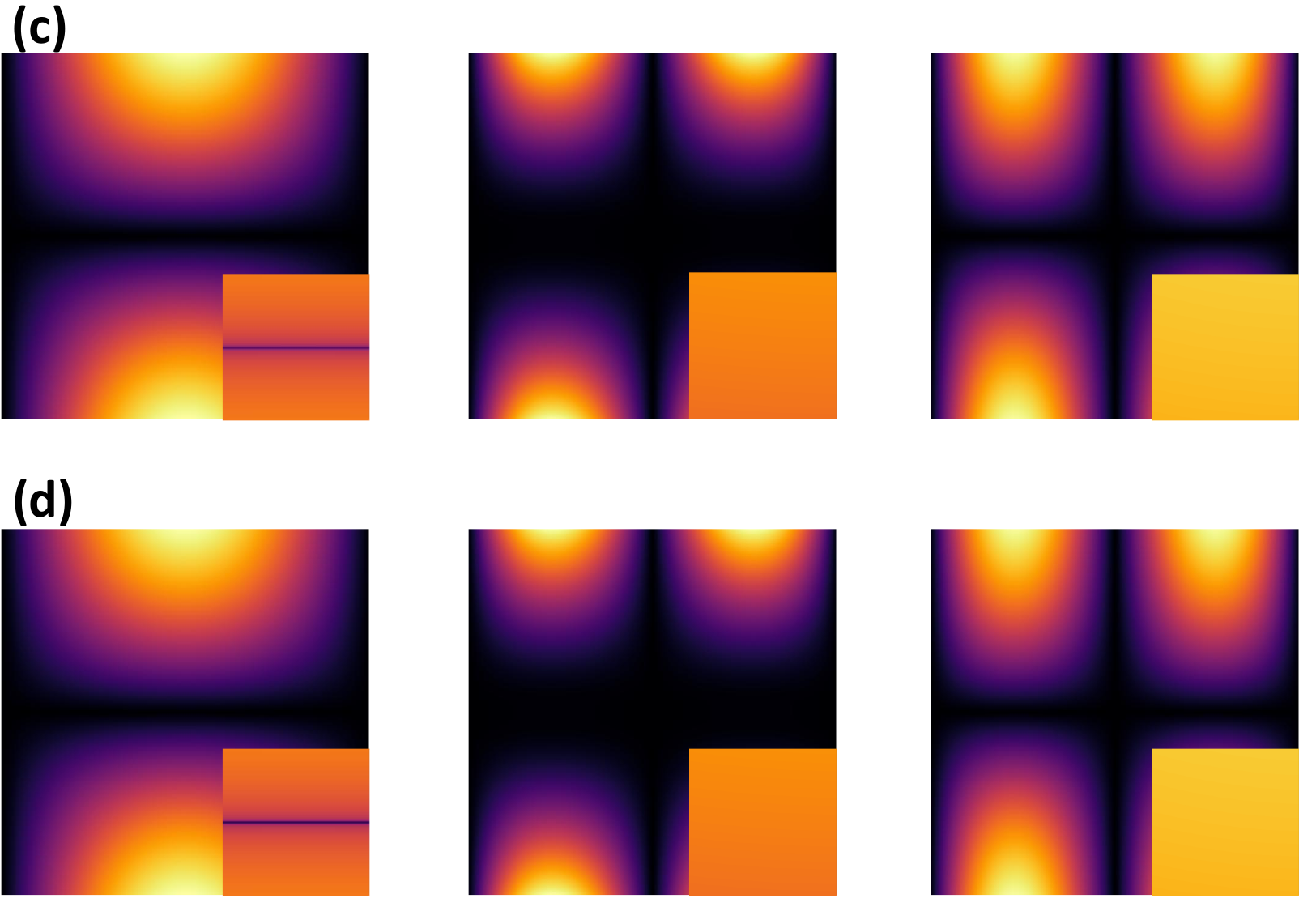}
  \includegraphics[width=0.49\textwidth, page=28, trim = 0cm 7cm 0cm 0cm, clip]{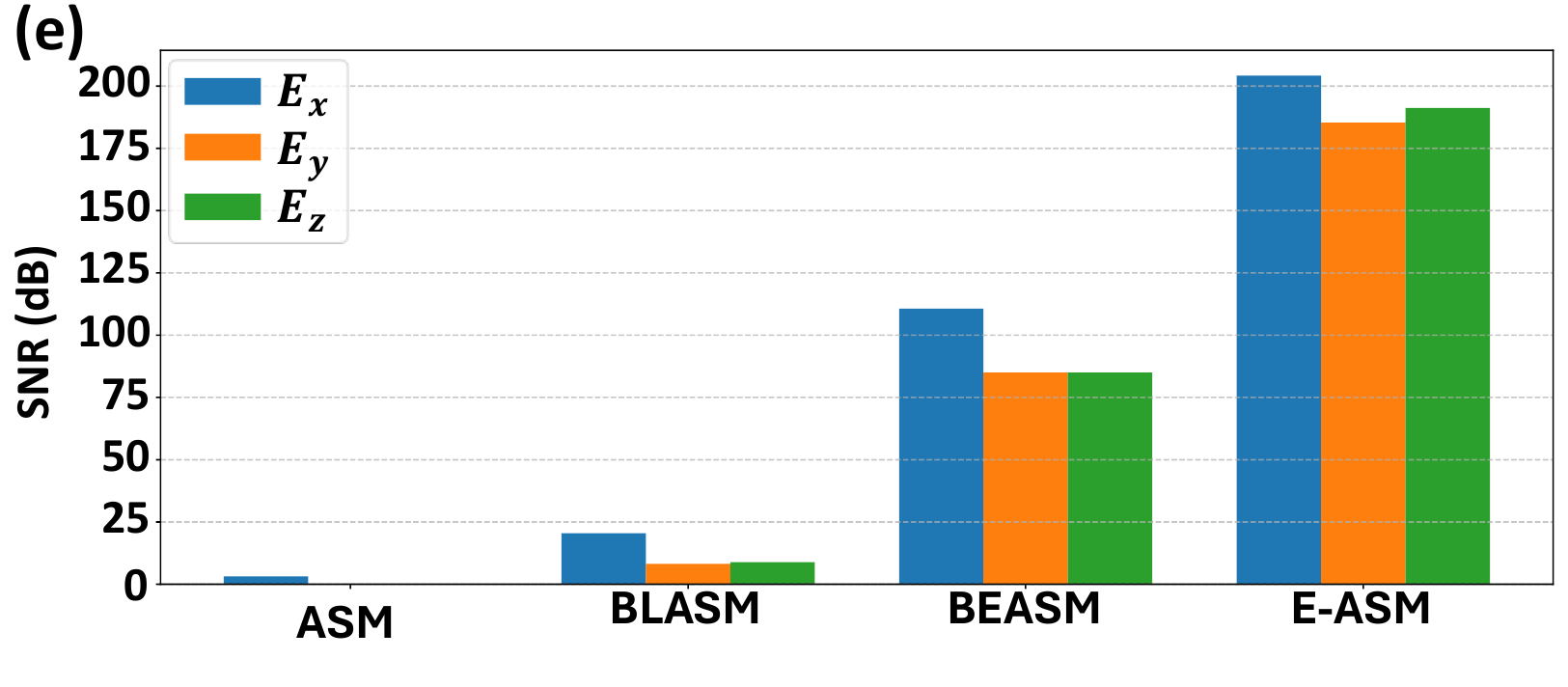}
  \includegraphics[width=0.49\textwidth, page=29, trim = 0cm 7cm 0cm 0cm, clip]{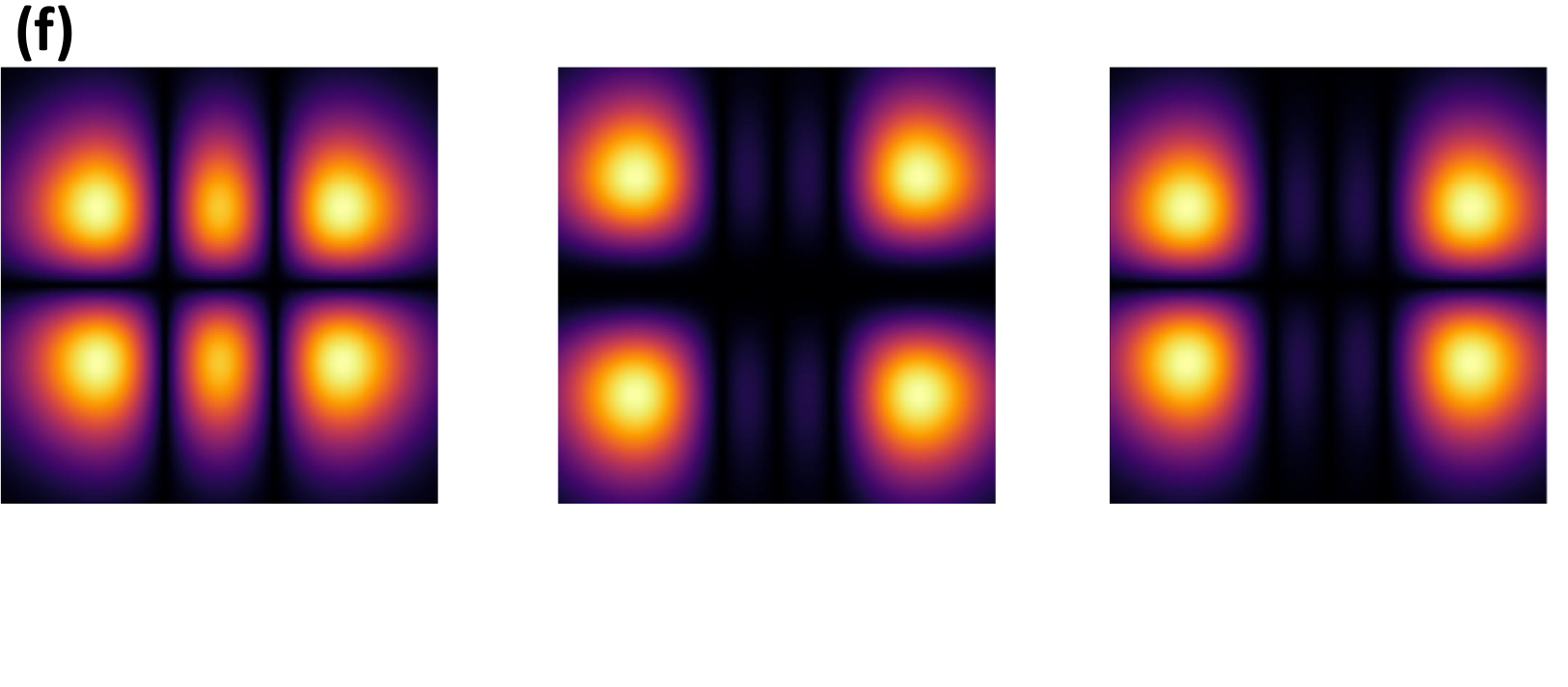}
  \caption{The simulation results of the $E_x, E_y, E_z$ field on the destination plane. (a) ASM, (b) BLASM, (c) BEASM, (d) E-ASM, respectively.
  (e) The SNR (dB) for each propagation algorithm. (f) $4\times$ zoom-out by using E-ASM.}
  \label{figS5}
\end{figure}

\section{The symmetric solutions of the vector beams and the raw 
PEC and PMC solutions at the beam waist}\label{secS12}
This section presents the raw PEC and PMC solutions together with 
the symmetric fundamental Gaussian, $\text{HG}_{21}$, and $\text{LG}_{20}$ (cosine) modes at the beam waist. The computational domain spans $1024\,\mathrm{\mu m} \times 1024\,\mathrm{\mu m}$ and 
consists of $2048 \times 2048$ pixels, with a beam waist of $w_0 = 128\,\mathrm{\mu m}$ for all three modes. In the main text, the fundamental Gaussian beam is propagated over $z = 0.103\,\mathrm{m}$ from its waist plane; the corresponding field distribution at the waist is shown in Fig.~\ref{figS8}.
\clearpage
\begin{figure}[htbp]
  \centering
  \includegraphics[width=\textwidth, page=30]{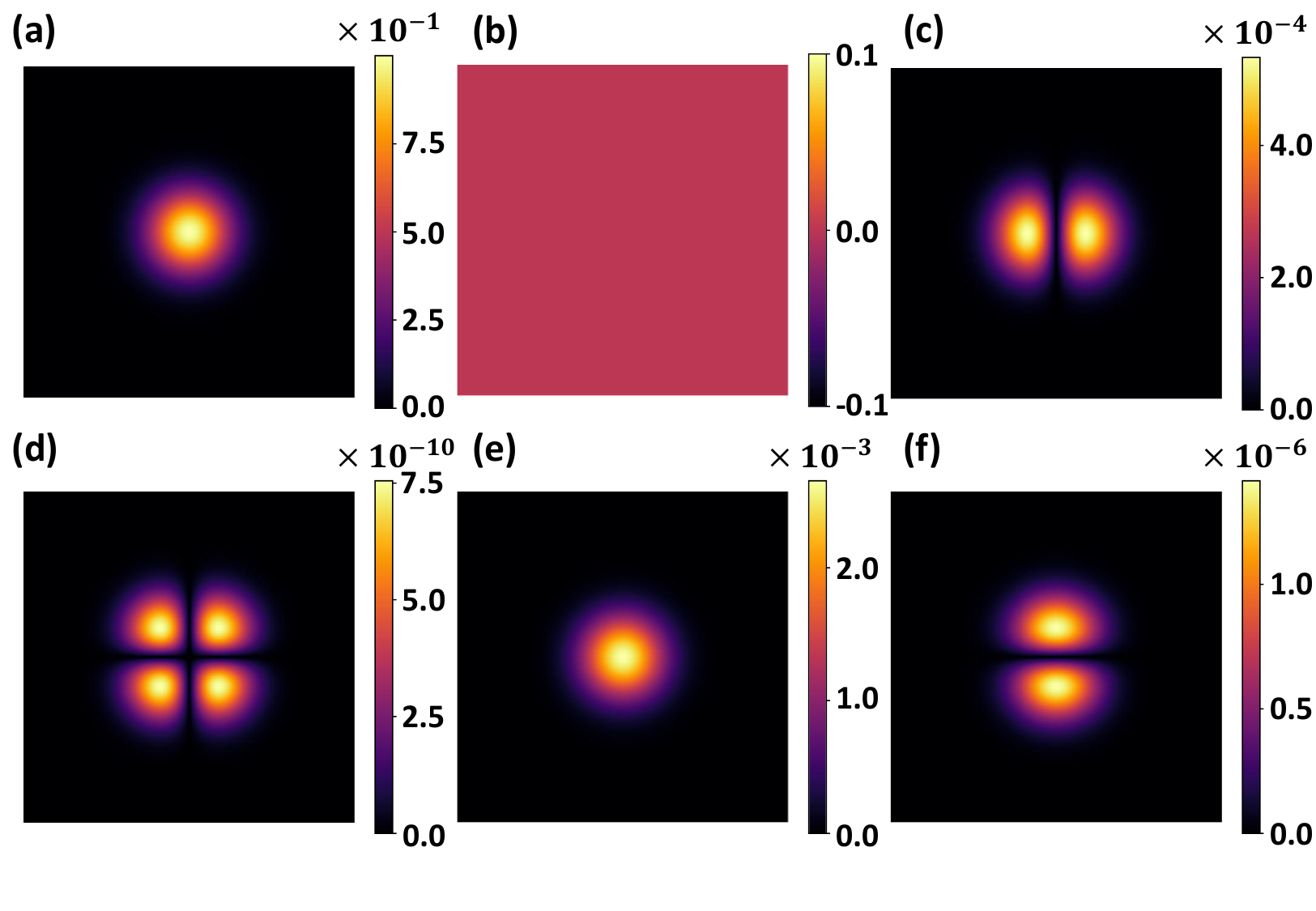}
  \includegraphics[width=\textwidth, page=31]{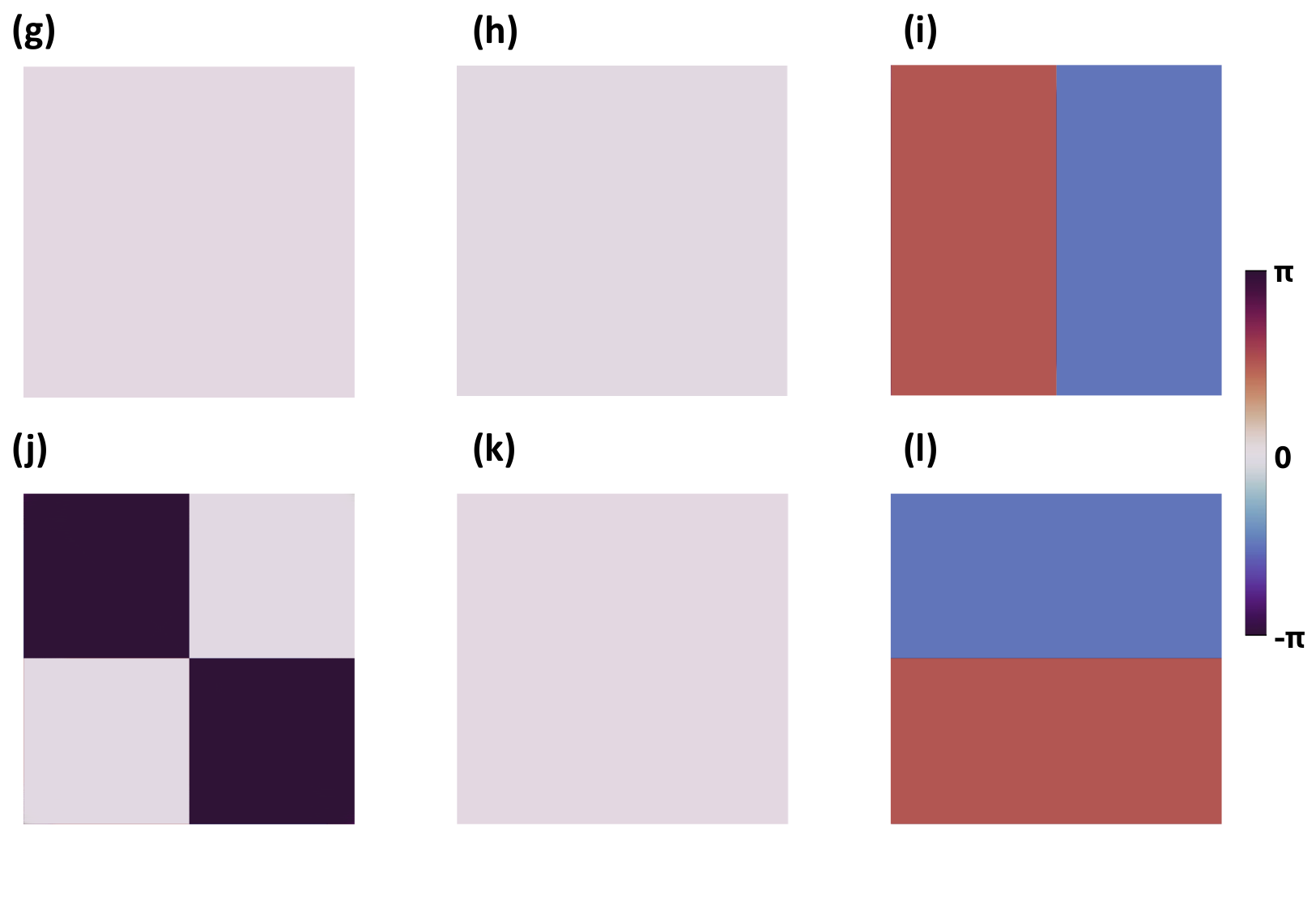}
  \caption{(a--f) The amplitude and (g--l) the phase of the $E_x, E_y,E_z,H_x,H_y, H_z$ components of the H-driven PEC solution of the fundamental Gaussian beam.}
  \label{figS6}
\end{figure}

\begin{figure}[htbp]
  \centering
  \includegraphics[width=\textwidth, page=32]{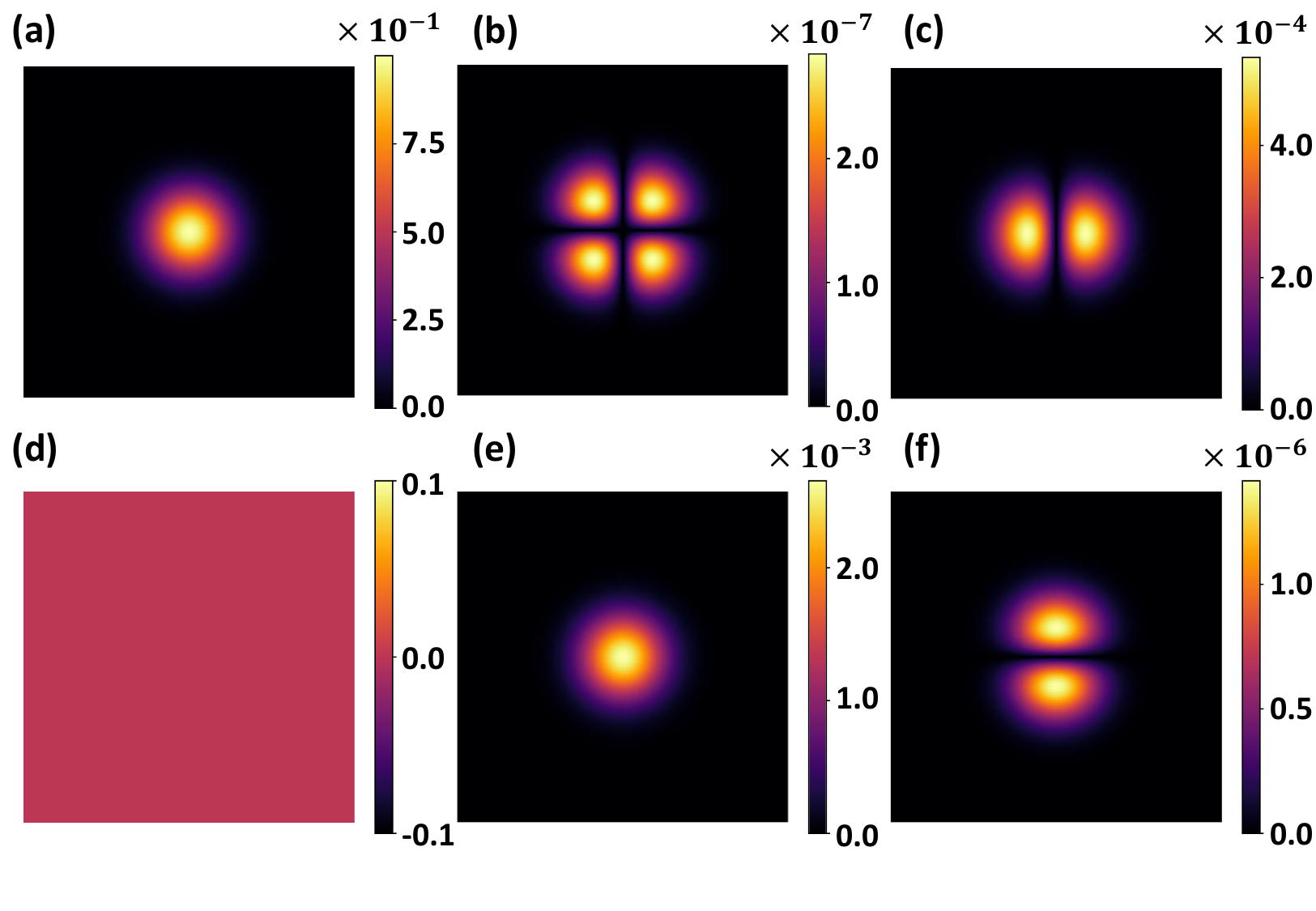}
  \includegraphics[width=\textwidth, page=33]{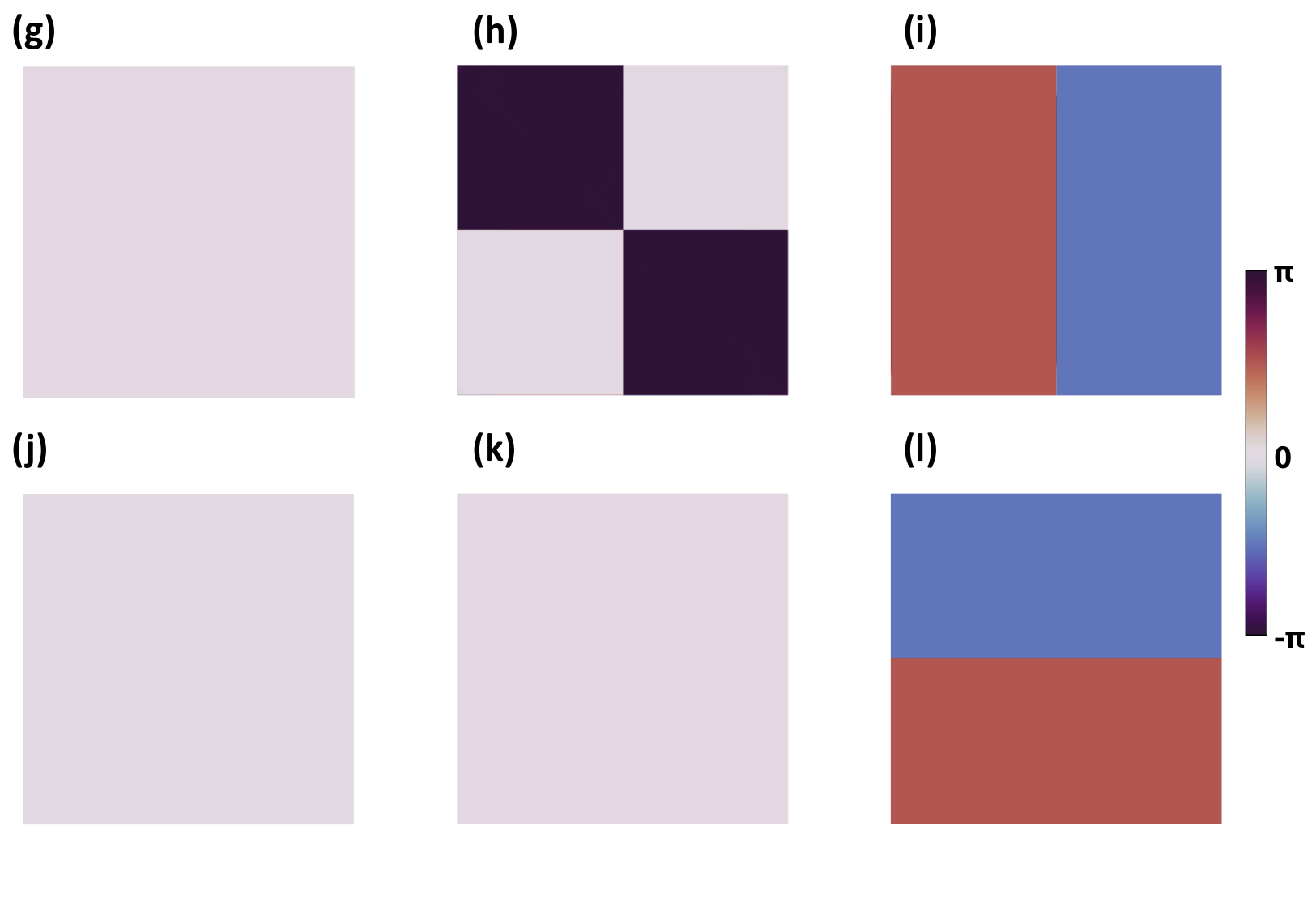}
  \caption{(a--f) The amplitude and (g--l) the phase of the $E_x, E_y,E_z,H_x,H_y, H_z$ components of the E-driven PMC solution of the fundamental Gaussian beam.}
  \label{figS7}
\end{figure}

\begin{figure}[htbp]
  \centering
  \includegraphics[width=\textwidth, page=34]{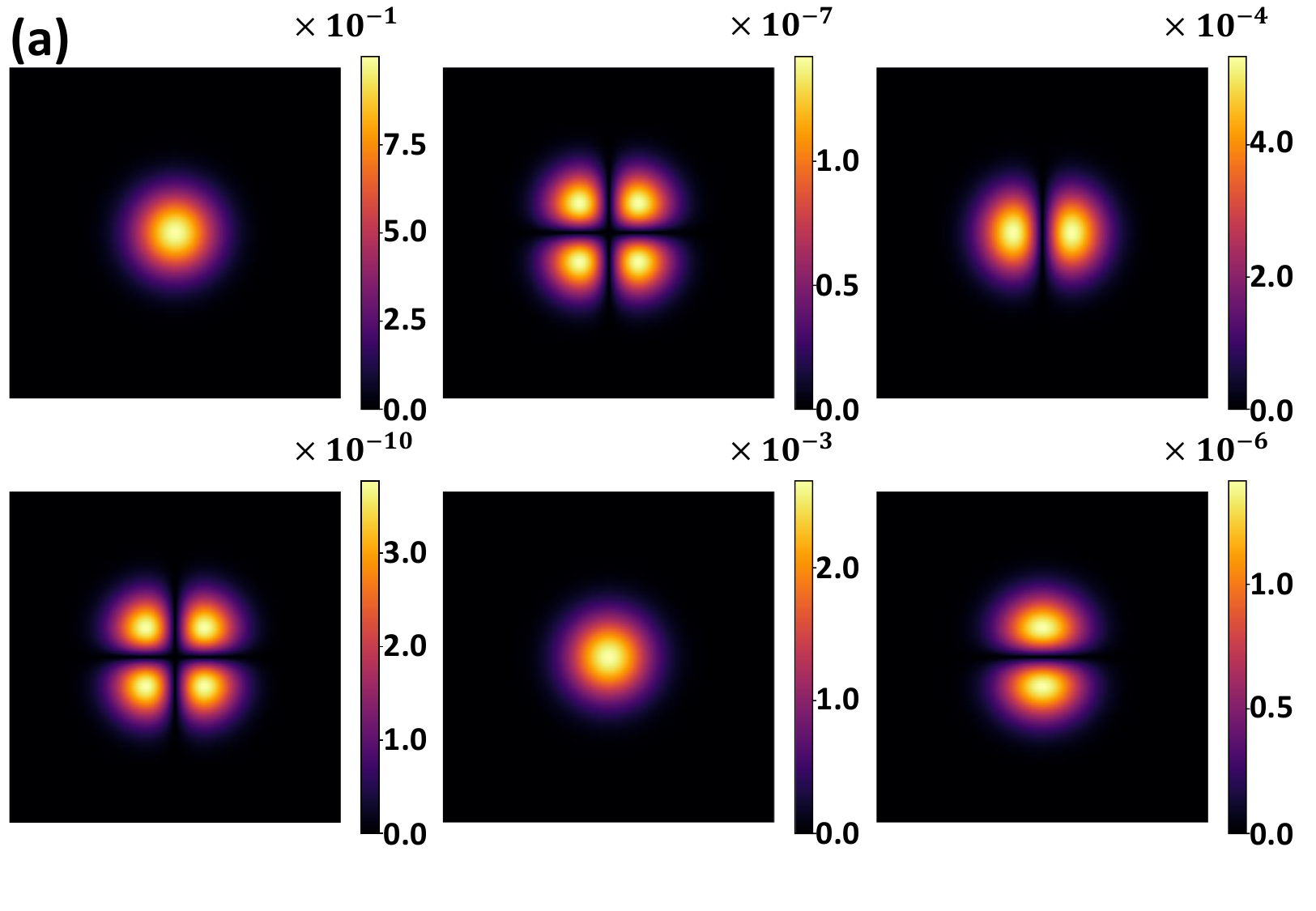}
  \includegraphics[width=\textwidth, page=35]{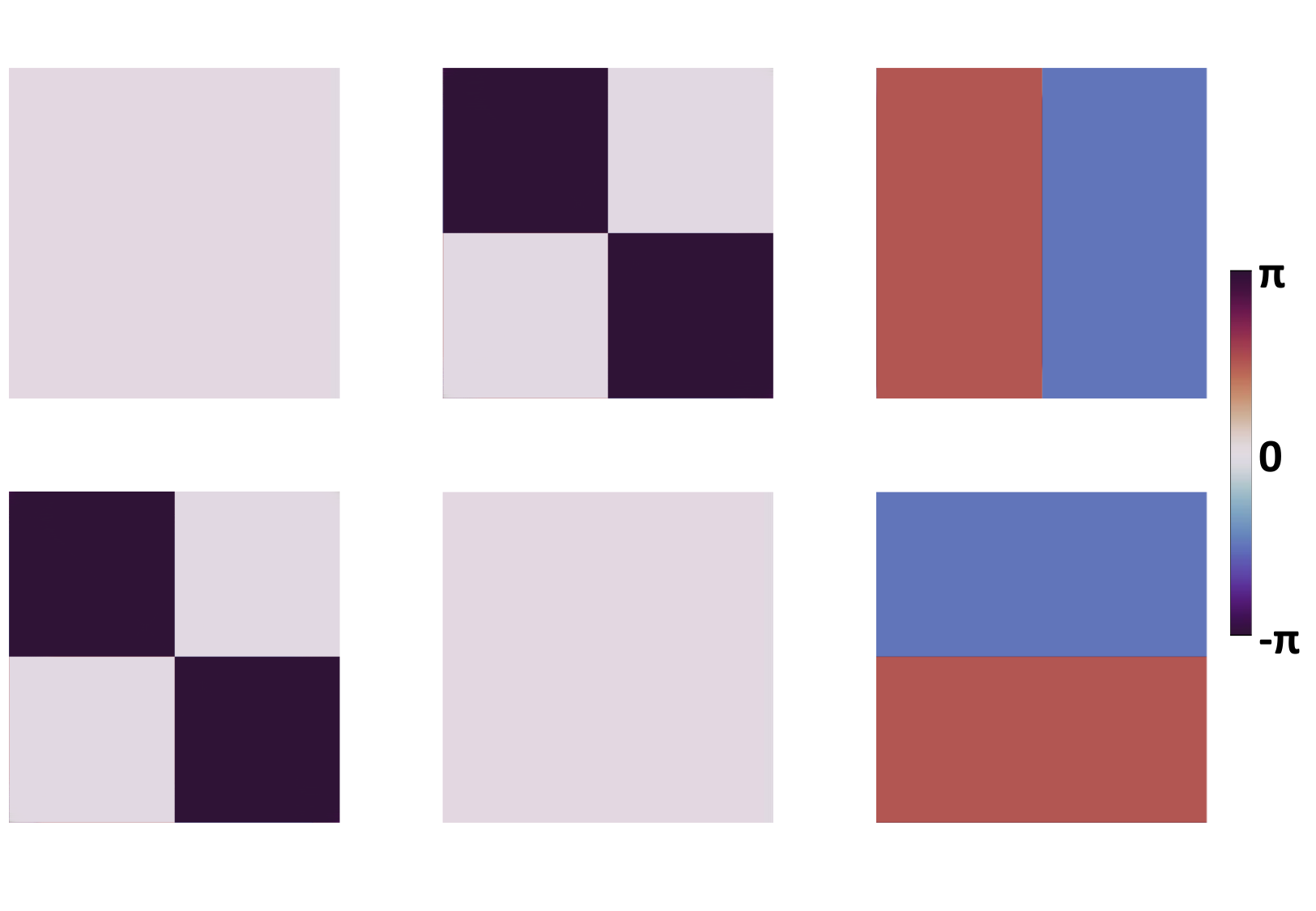}
  \caption{(a--f) The amplitude and (g--l) the phase of the $E_x, E_y,E_z,H_x,H_y, H_z$ components of the symmetric solution of the fundamental Gaussian beam.}
  \label{figS8}
\end{figure}

\begin{figure}[htbp]
  \centering
  \includegraphics[width=\textwidth, page=36]{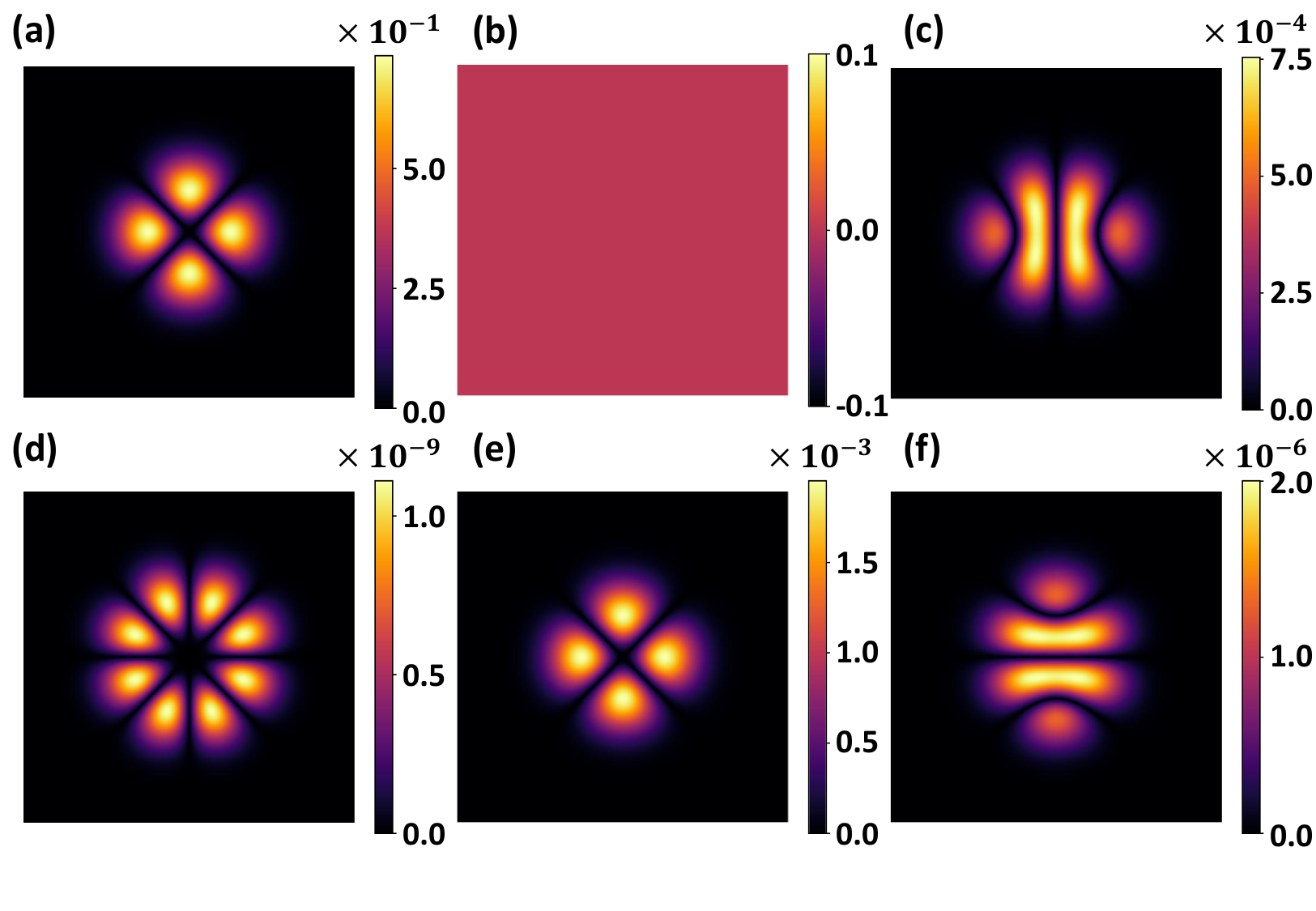}
  \includegraphics[width=\textwidth, page=37]{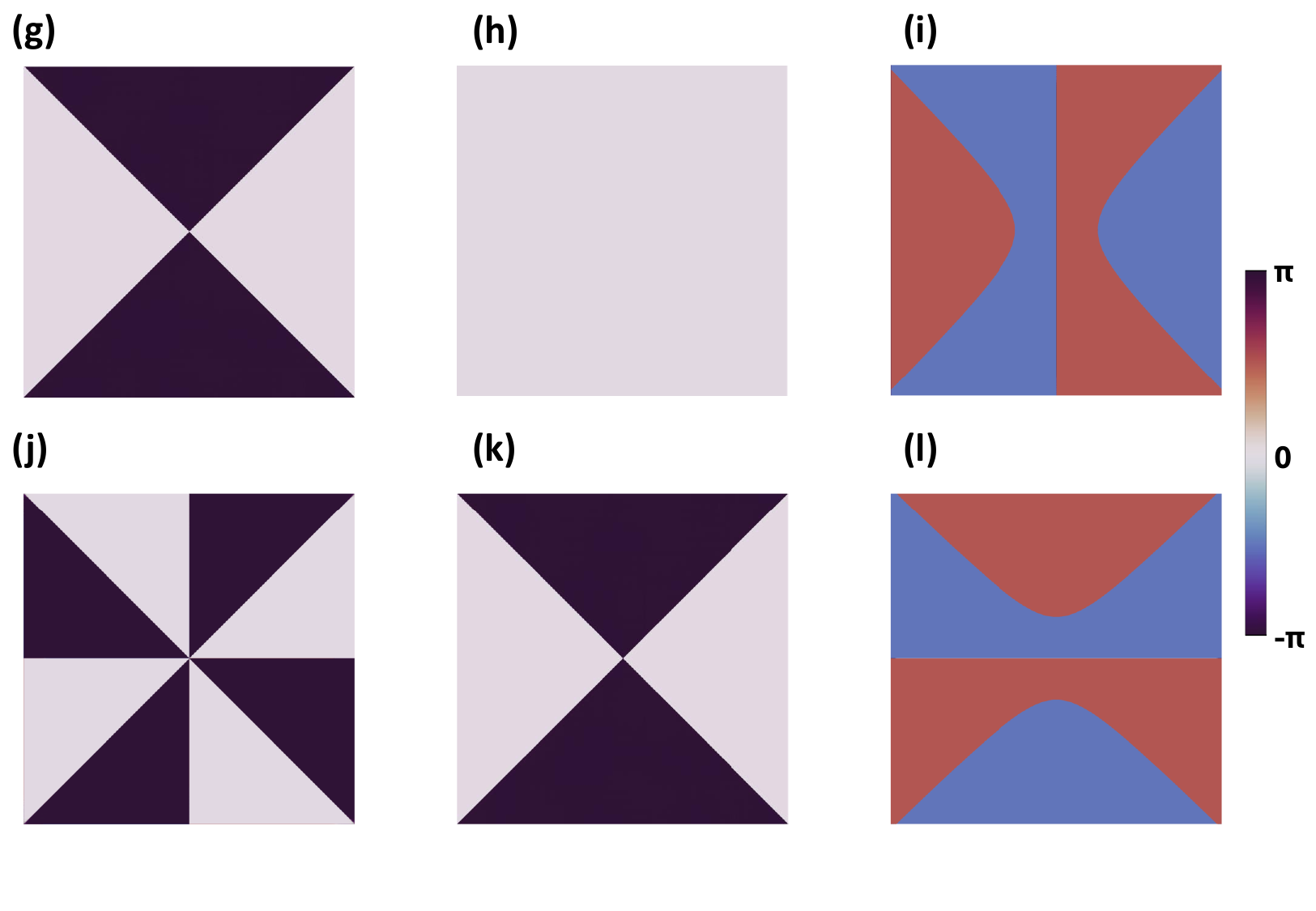}
  \caption{(a--f) The amplitude and (g--l) the phase of the $E_x, E_y,E_z,H_x,H_y, H_z$ components of the H-driven PEC solution of the $\text{LG}_{20}$ cosine beam.}
  \label{figS9}
\end{figure}

\begin{figure}[htbp]
  \centering
  \includegraphics[width=\textwidth, page=38]{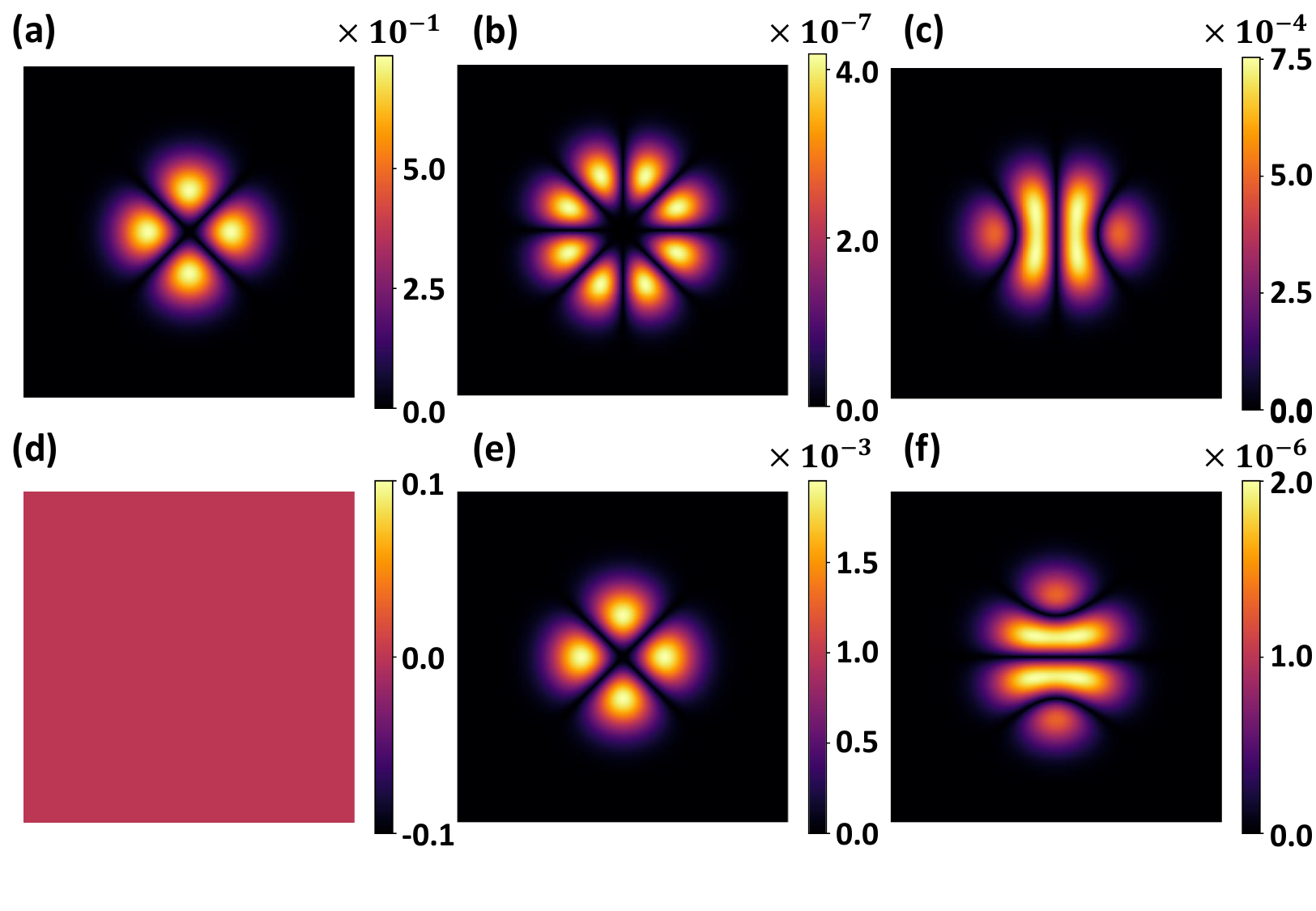}
  \includegraphics[width=\textwidth, page=39]{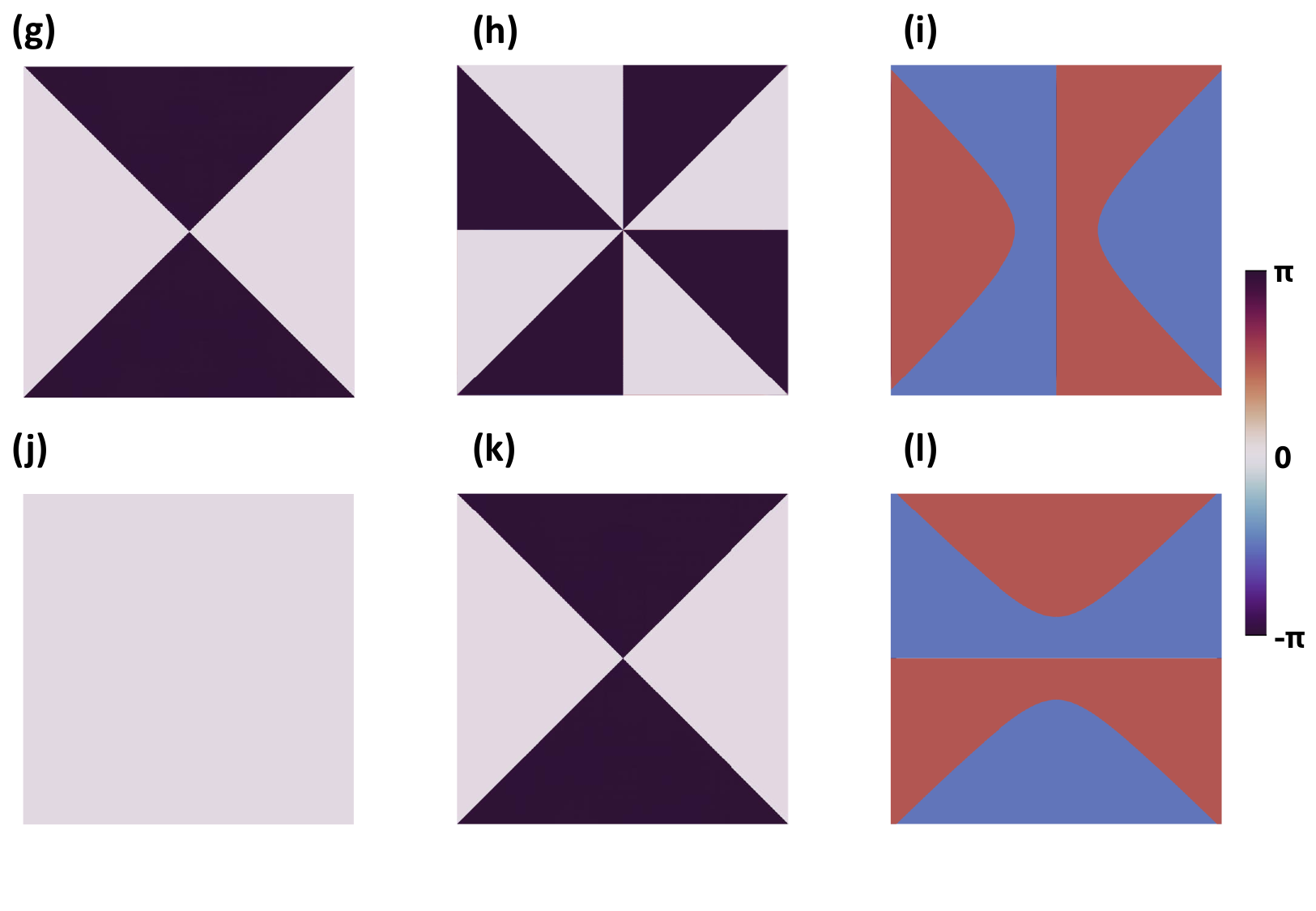}
  \caption{(a--f) The amplitude and (g--l) the phase of the $E_x, E_y,E_z,H_x,H_y, H_z$ components of the E-driven PMC solution of the $\text{LG}_{20}$ cosine beam.}
  \label{figS10}
\end{figure}

\begin{figure}[htbp]
  \centering
  \includegraphics[width=\textwidth, page=40]{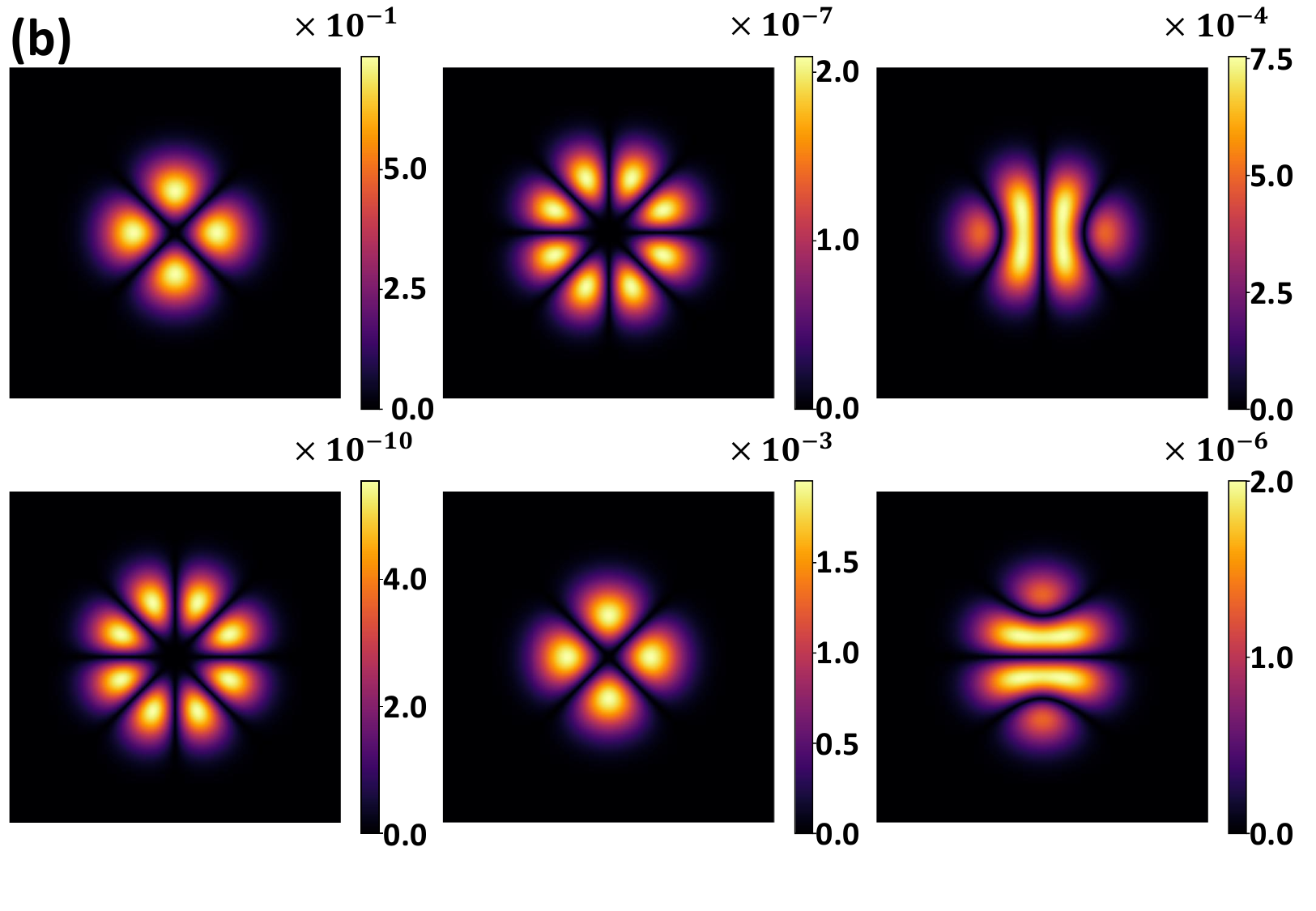}
  \includegraphics[width=\textwidth, page=41]{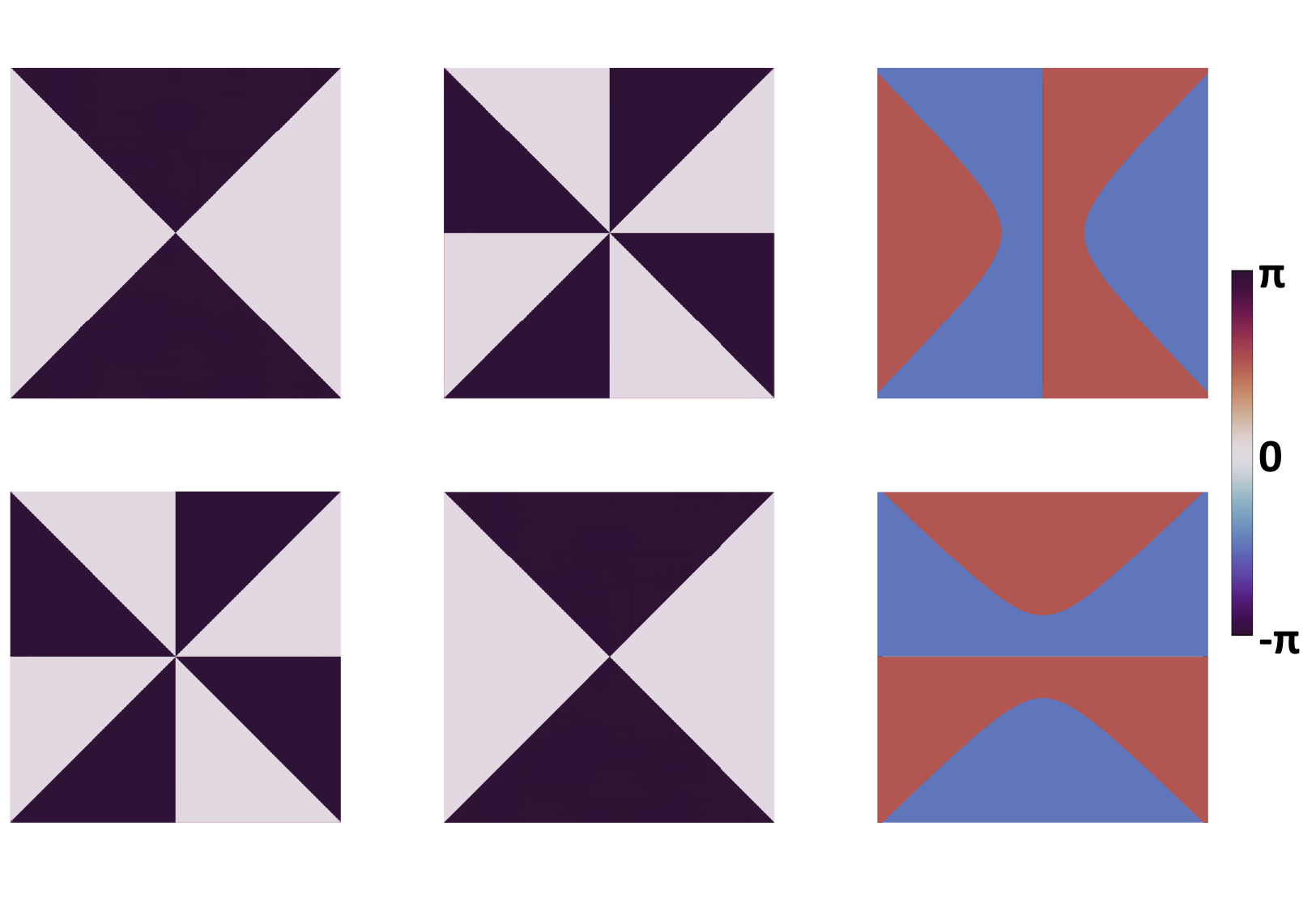}
  \caption{(a--f) The amplitude and (g--l) the phase of the $E_x, E_y,E_z,H_x,H_y, H_z$ components of the symmetric solution of the $\text{LG}_{20}$ cosine beam.}
  \label{figS11}
\end{figure}

\begin{figure}[htbp]
  \centering
  \includegraphics[width=\textwidth, page=42]{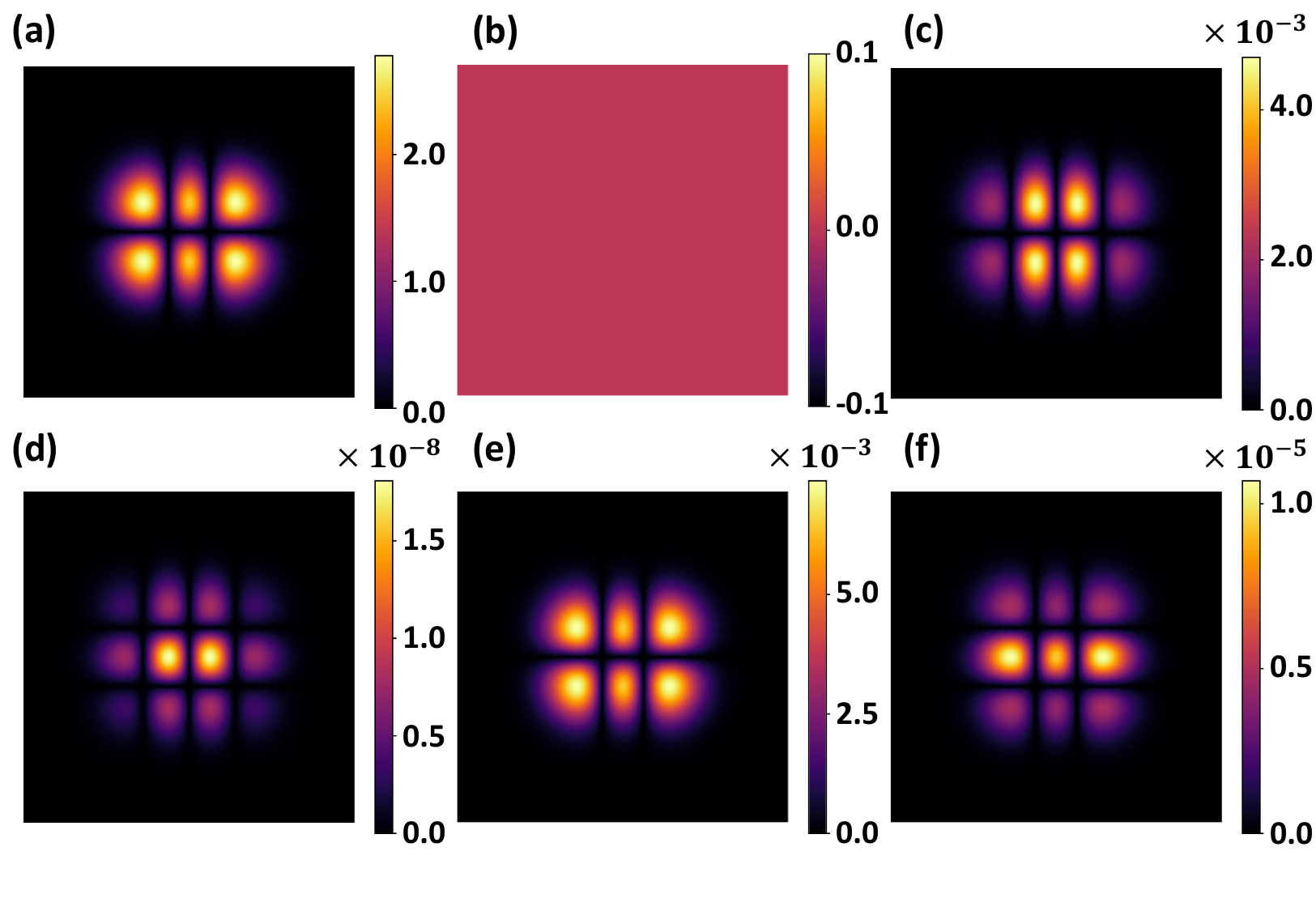}
  \includegraphics[width=\textwidth, page=43]{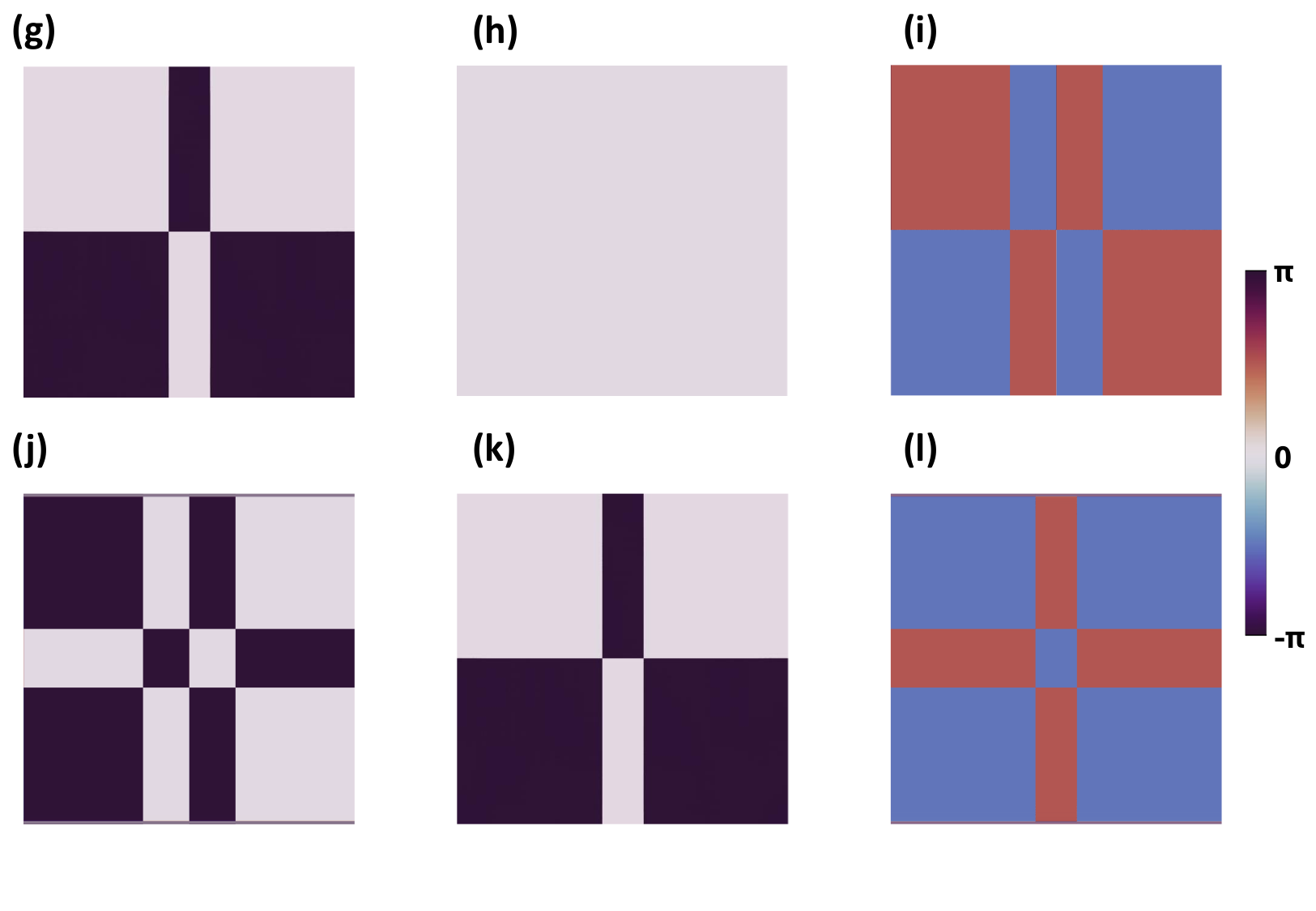}
  \caption{(a--f) The amplitude and (g--l) the phase of the $E_x, E_y,E_z,H_x,H_y, H_z$ components of the H-driven PEC solution of the $\text{HG}_{21}$ mode.}
  \label{figS12}
\end{figure}

\begin{figure}[htbp]
  \centering
  \includegraphics[width=\textwidth, page=44]{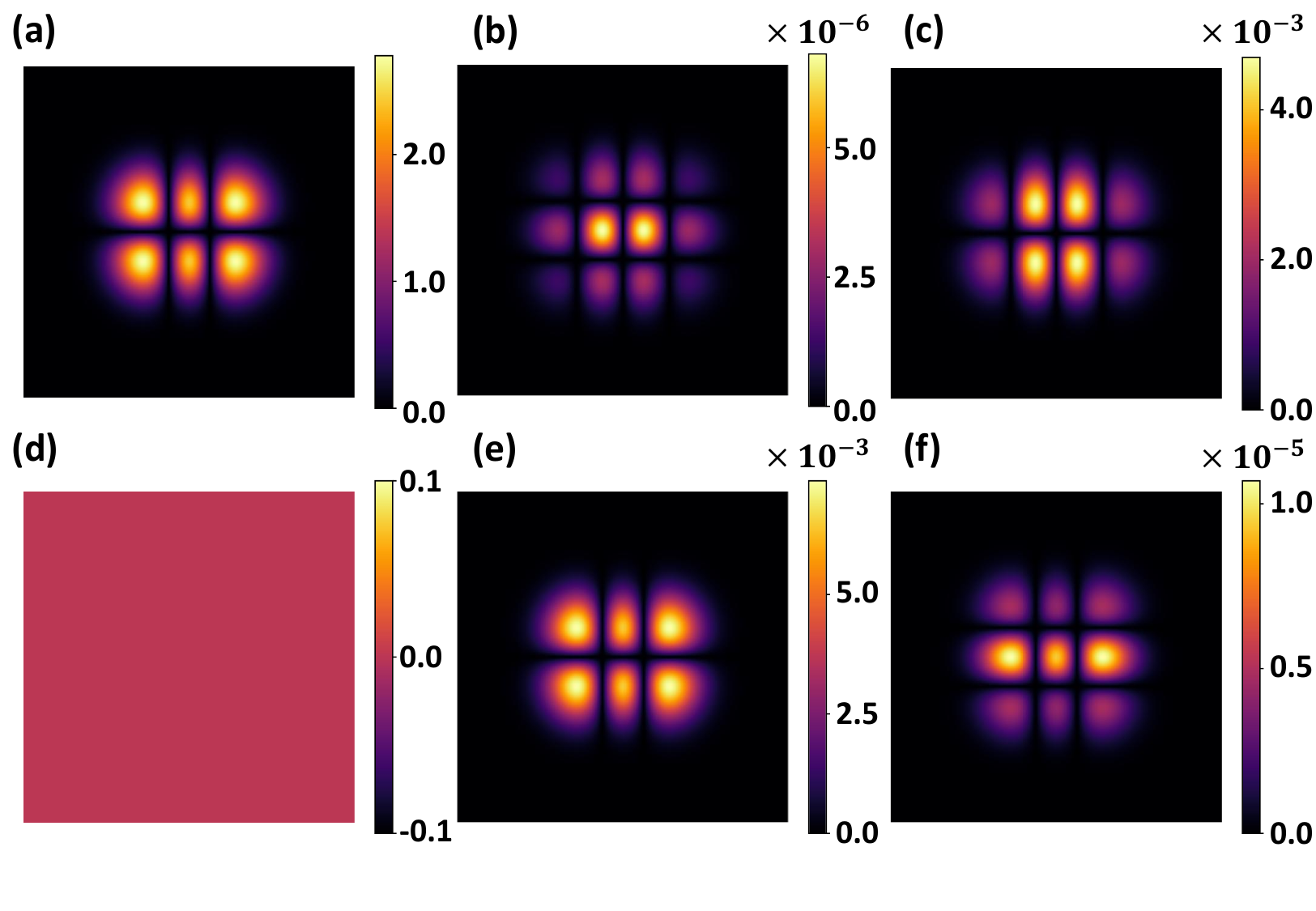}
  \includegraphics[width=\textwidth, page=45]{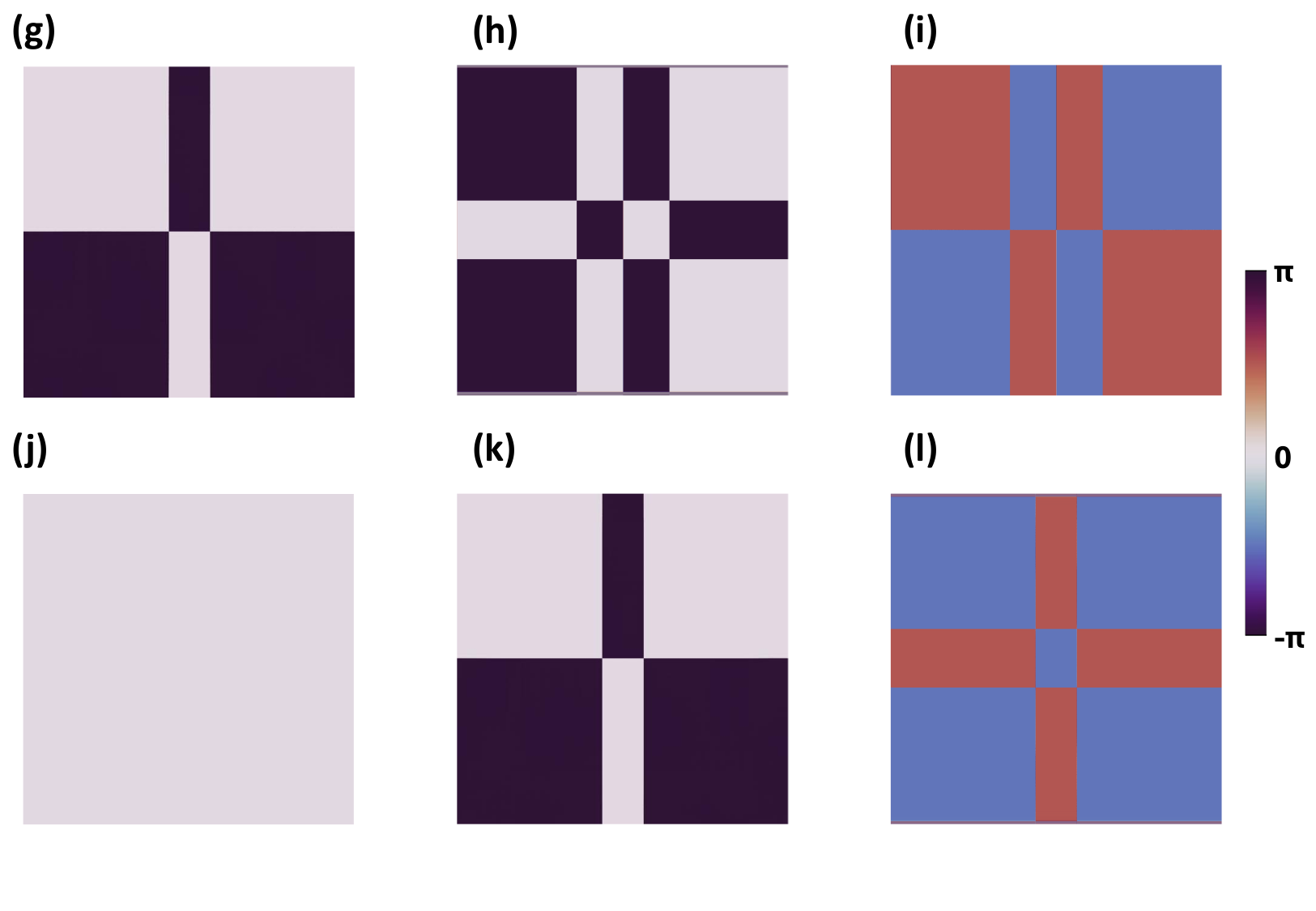}
  \caption{(a--f) The amplitude and (g--l) the phase of the $E_x, E_y,E_z,H_x,H_y, H_z$ components of the E-driven PMC solution of the $\text{HG}_{21}$ mode.}
  \label{figS13}
\end{figure}

\begin{figure}[htbp]
  \centering
  \includegraphics[width=\textwidth, page=46]{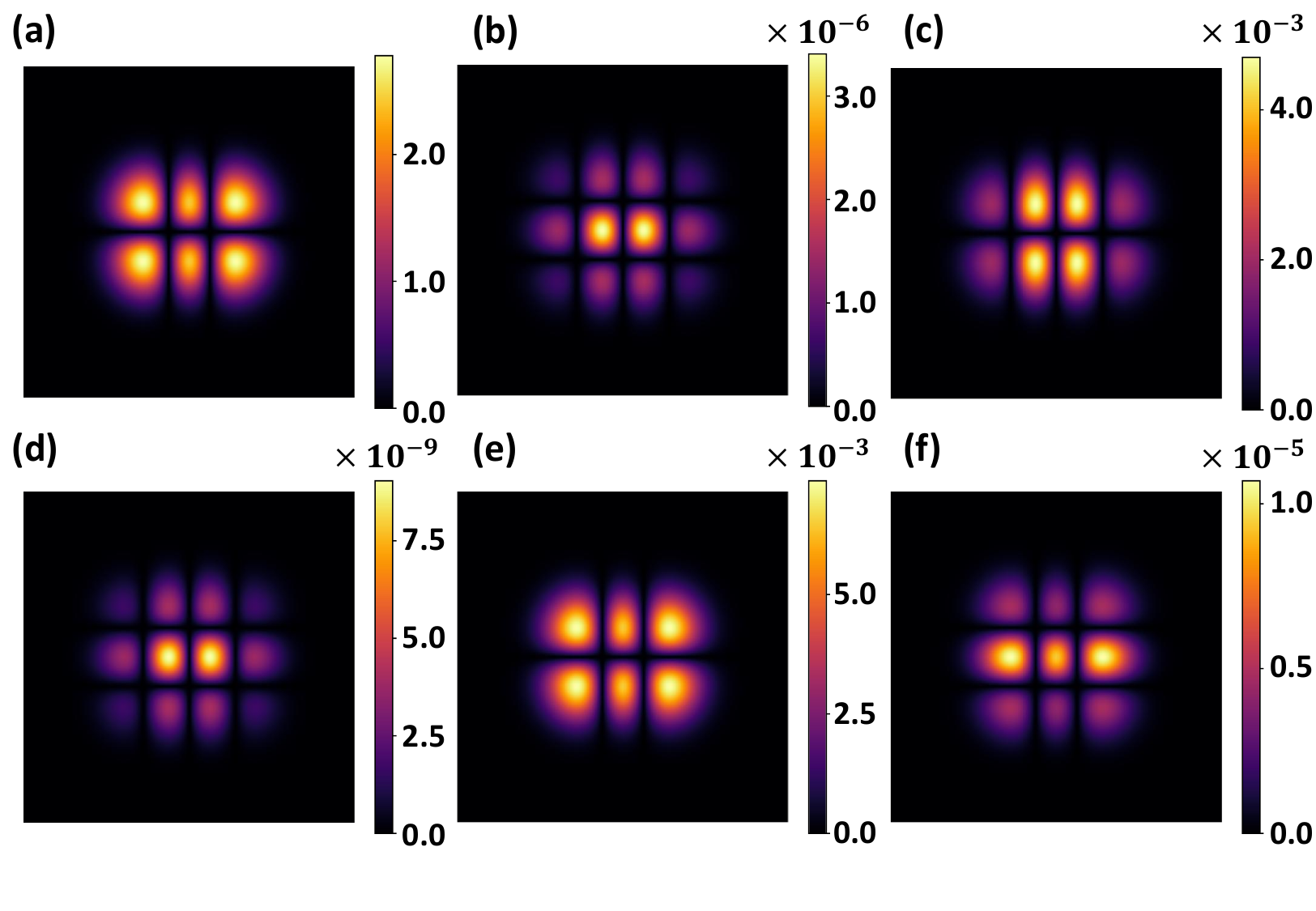}
  \includegraphics[width=\textwidth, page=47]{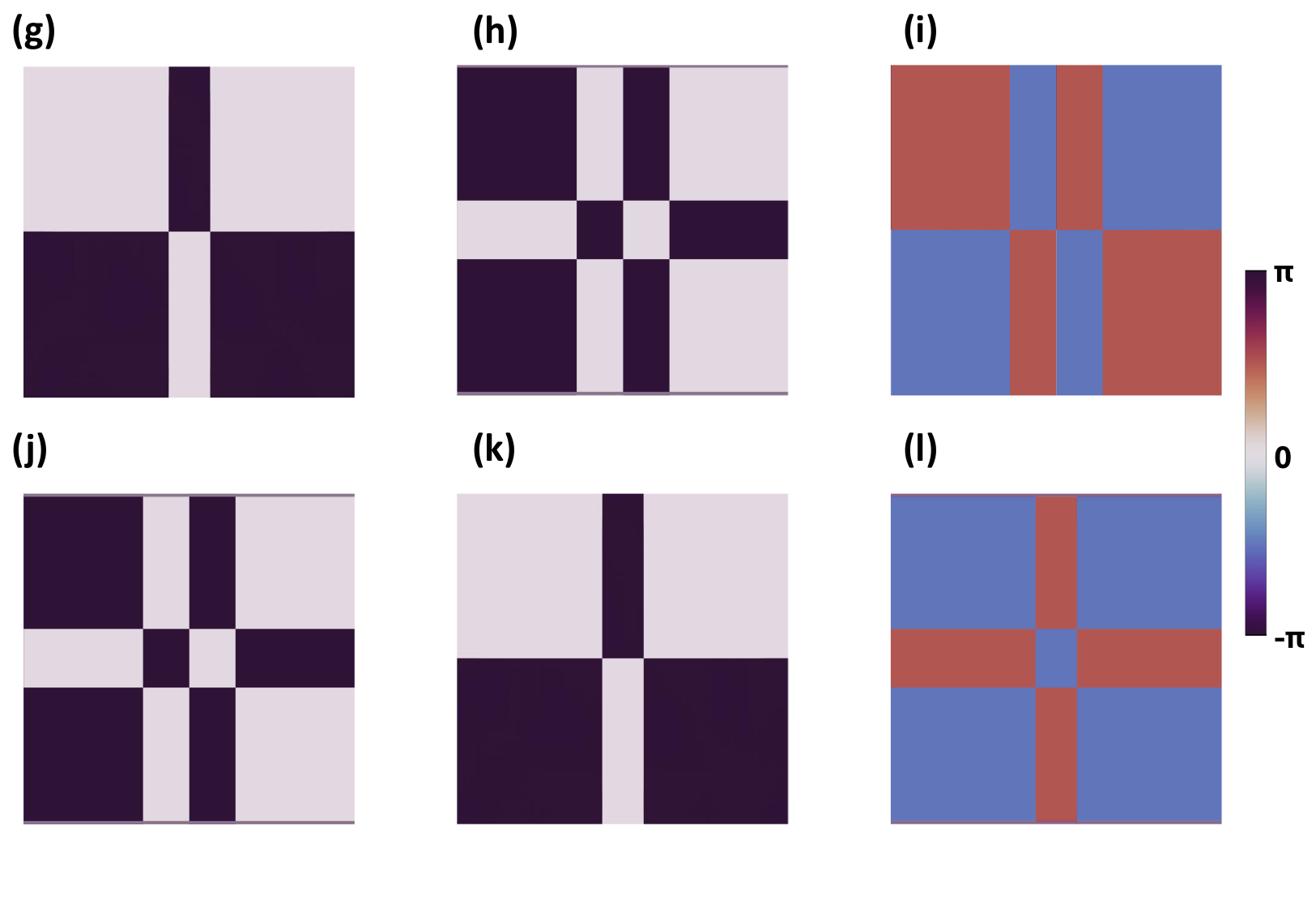}
  \caption{(a--f) The amplitude and (g--l) the phase of the $E_x, E_y,E_z,H_x,H_y, H_z$ components of the symmetric solution of the $\text{HG}_{21}$ mode.}
  \label{figS14}
\end{figure}
\clearpage
\noindent
As illustrated herein, the missing components, i.e., the $E_y$ component of the PEC solution and the $H_x$ component of the PMC solution, are reduced to half of the original PMC / PEC result after the symmetrization process. Although this missing component is relatively small, this symmetrized solution is a good approximation of the real vector beam. 

\section{The PEC and PMC solutions used for the Q-Stratton-Chu propagation}\label{secS13}
This section details the raw PEC and PMC numerical solutions utilized for the Q-Stratton-Chu propagation of the fundamental Gaussian beam presented in the main text. The numerical simulation configurations and boundary conditions are identical to those specified in the main text.

\begin{figure}[htbp]
  \centering
  \includegraphics[width=\textwidth, page=48]{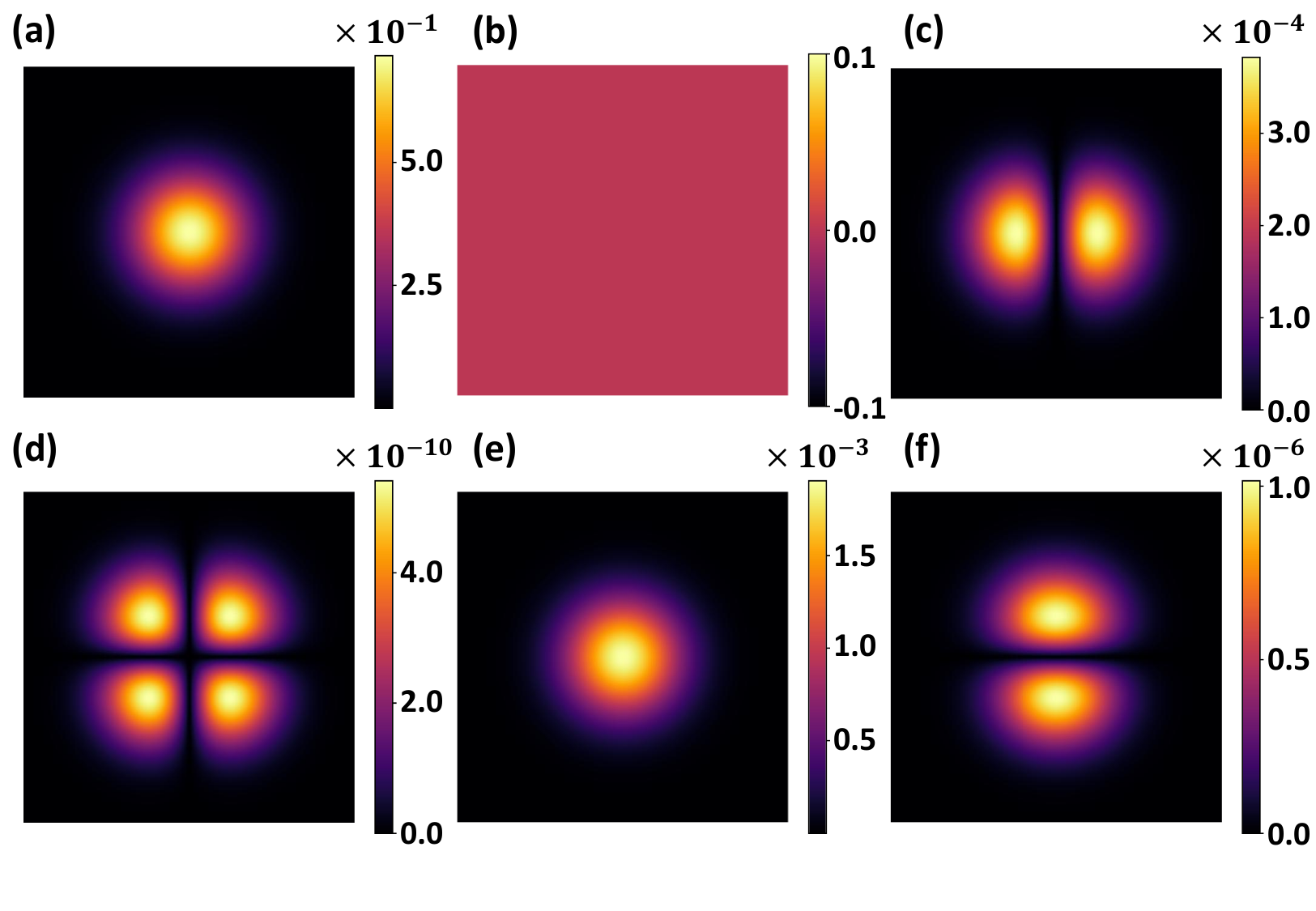}
  \includegraphics[width=\textwidth, page=49]{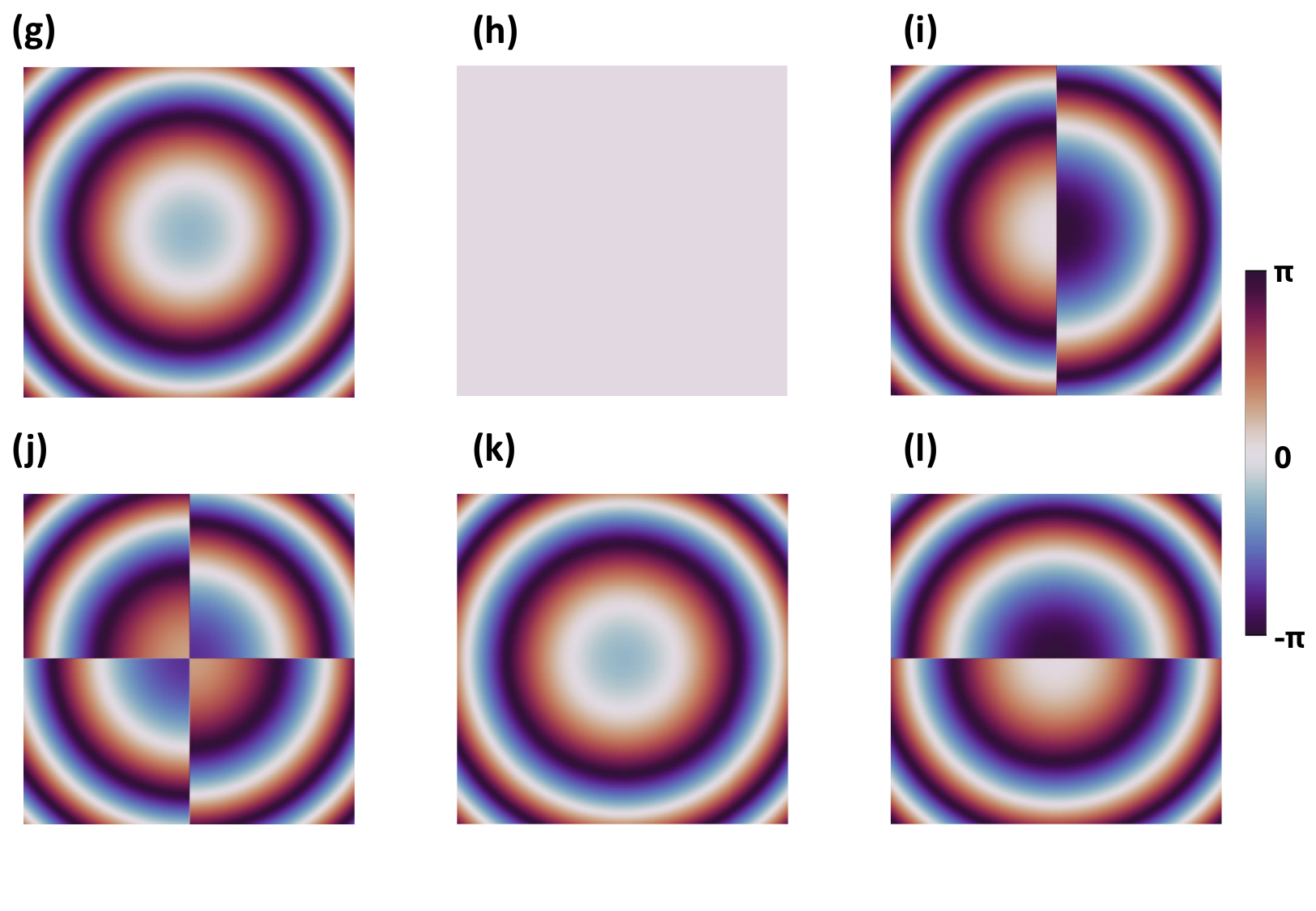}
  \caption{(a--f) The amplitude and (g--l) the phase of the $E_x, E_y,E_z,H_x,H_y, H_z$ components of the H-driven PEC solutions of the fundamental Gaussian beam after the Q-Stratton-Chu propagation.}
  \label{figS15}
\end{figure}

\begin{figure}[htbp]
  \centering
  \includegraphics[width=\textwidth, page=50]{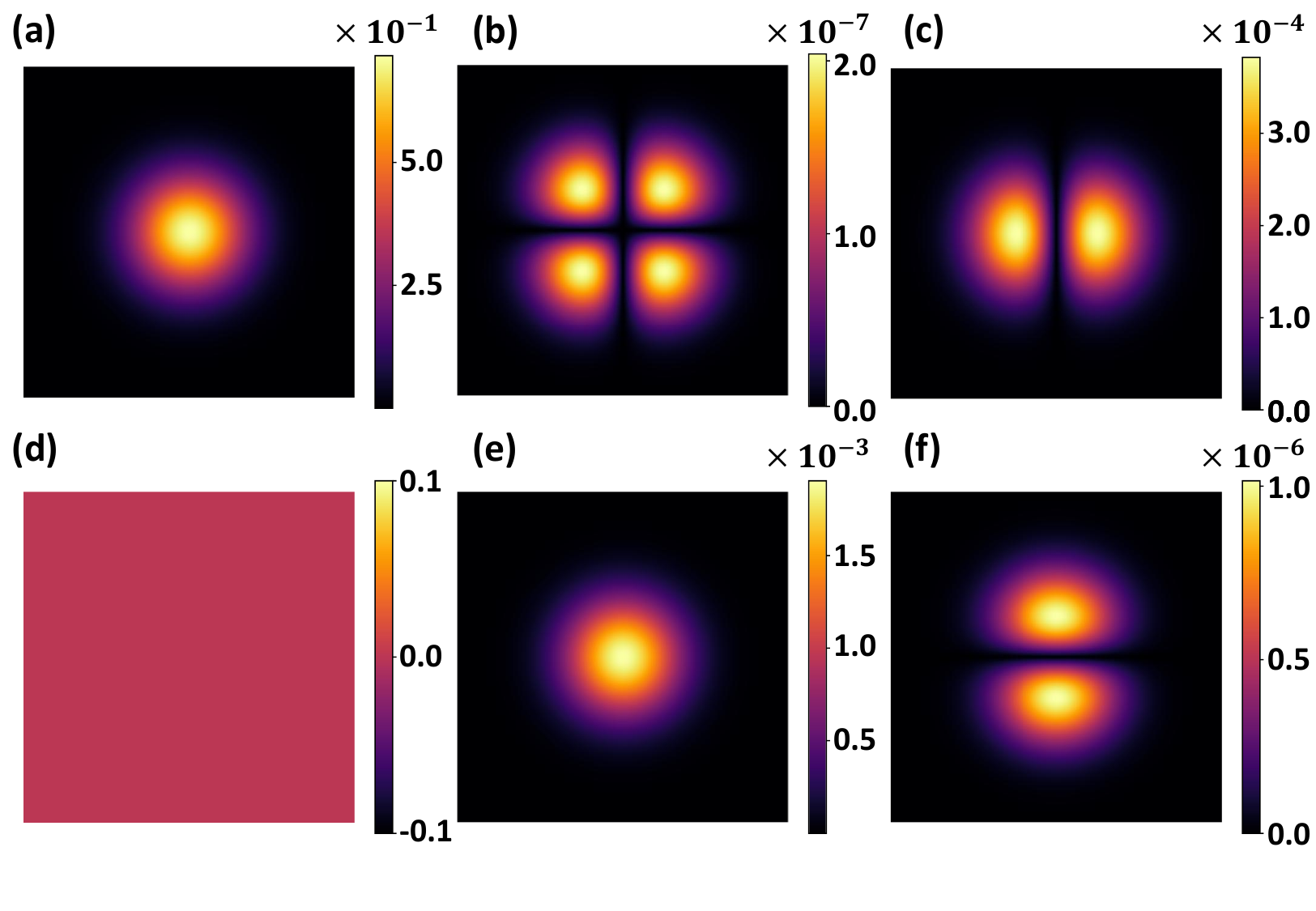}
  \includegraphics[width=\textwidth, page=51]{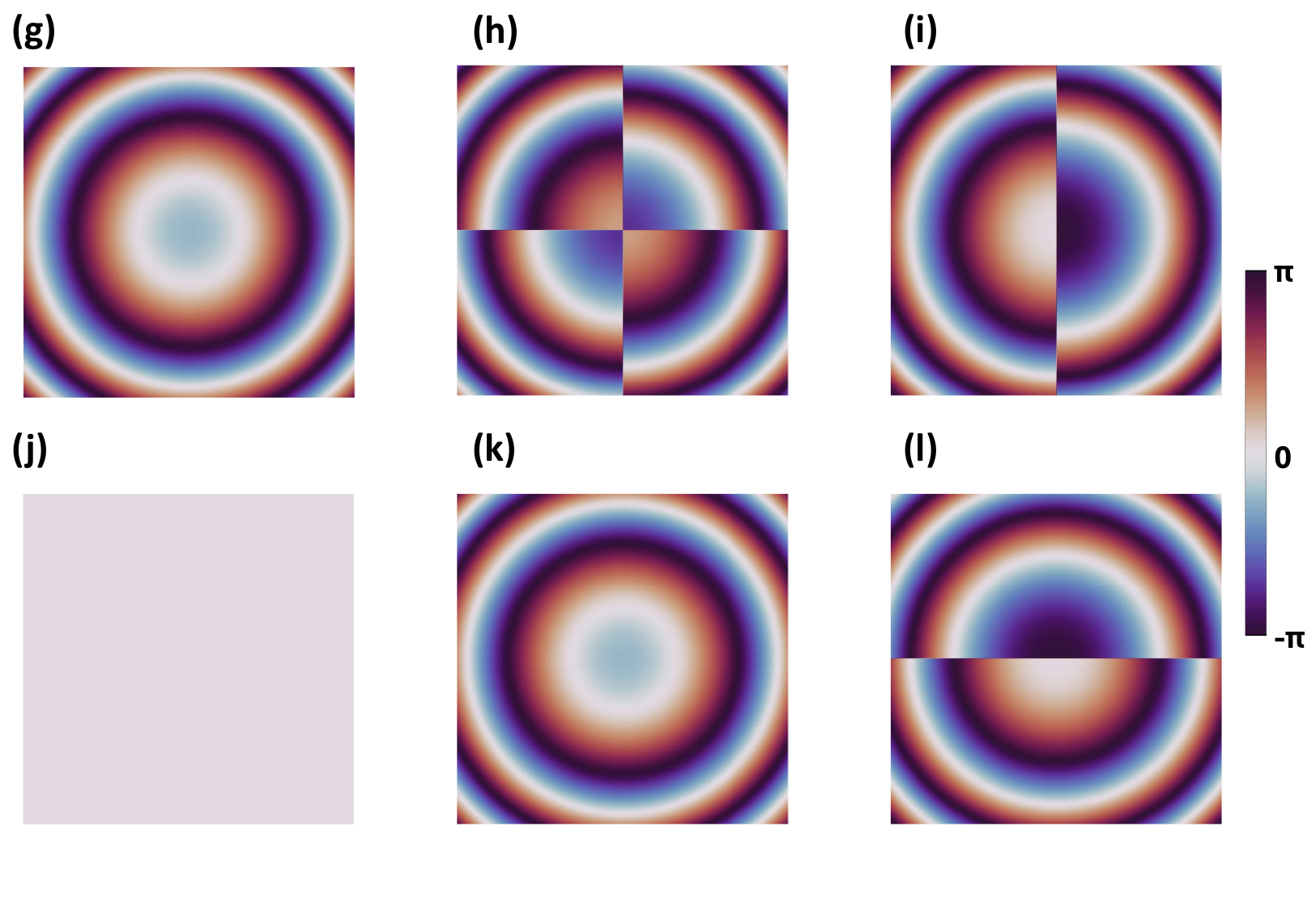}
  \caption{(a--f) The amplitude and (g--l) the phase of the $E_x, E_y,E_z,H_x,H_y, H_z$ components of the E-driven PMC solutions of the fundamental Gaussian beam after the Q-Stratton-Chu propagation.}
  \label{figS16}
\end{figure}
\clearpage
\bibliography{ref}
\clearpage
\end{document}